\documentclass{jfm}

\usepackage{amsmath}
\usepackage{mathtools}

\usepackage{tikz}
\usetikzlibrary{shapes.geometric, arrows}

\usepackage{accents}
\newlength{\dhatheight}

\usepackage{multirow}

\def\KsqBr{\sqrt{K\raisebox{1.ex}{\scriptsize$+$}_{\;}}\raisebox{-0.6ex}{\makebox[0mm][l]{\hspace*{-3.5mm}\scriptsize$Br$}}}
\def\KsqBrp{\sqrt{K'\raisebox{1.ex}{\scriptsize$+$}_{\;}}\raisebox{-0.6ex}{\makebox[0mm][l]{\hspace*{-3.5mm}\scriptsize$Br$}}}
\def\Ksqynp{\sqrt{K_{\; \;}}\raisebox{-0.27ex}{\makebox[0mm][l]{\hspace*{-1.8mm}\scriptsize$y$}}}
\def\Ksqy{\sqrt{K\raisebox{1.05ex}{\scriptsize$+$}}\raisebox{-0.27ex}{\makebox[0mm][l]{\hspace*{-2.4mm}\scriptsize$y$}}}
\def\Ksqx{\sqrt{K\raisebox{1.05ex}{\scriptsize$+$}}\raisebox{-0.27ex}{\makebox[0mm][l]{\hspace*{-2.4mm}\scriptsize$x$}}}
\def\Ksqz{\sqrt{K\raisebox{1.05ex}{\scriptsize$+$}}\raisebox{-0.27ex}{\makebox[0mm][l]{\hspace*{-2.4mm}\scriptsize$z$}}}

\newcommand{\chainII}{\protect\mbox{-- $\cdot$ --}}

\newcommand{\mylab}[3]{\raisebox{#2}[0mm][0mm]{\makebox[0mm][l]{\hspace*{#1}#3}}}

\def\aaa{\textit{a}}
\def\bbb{\textit{b}}
\def\ccc{\textit{c}}
\def\ddd{\textit{d}}
\def\eee{\textit{e}}
\def\fff{\textit{f}}
\def\ggg{\textit{g}}
\def\hhh{\textit{h}}
\def\iii{\textit{i}}

\def\lll{\textit{l}}
\def\mmm{\textit{m}}

\def\ooo{\textit{o}}
\def\ppp{\textit{p}}

\begin{document}

\newcommand{\squaresolidred}{\raisebox{0.5pt}{\tikz{\draw[-,red,solid,line width = 0.5pt](-2.5mm,0) -- (2.5mm,0);\node[draw,scale=0.53,regular polygon, regular polygon sides=4,fill=red,rotate=0](){};}}}

\newcommand{\triangsolidviolet}{\raisebox{0.5pt}{\tikz{\draw[-,violet,solid,line width = 0.5pt](-2.5mm,0) -- (2.5mm,0);\node[draw,scale=0.35,regular polygon, regular polygon sides=3,fill=violet,rotate=0](){};}}}

\newcommand{\circlesolidblue}{\raisebox{0.5pt}{\tikz{\draw[-,black!100!,solid,line width = 0.5pt](-2.5mm,0) -- (2.5mm,0);\node[draw,scale=0.58,circle,fill=blue](){};}}}

\newcommand{\circlesolidblack}{\raisebox{0.5pt}{\tikz{\draw[-,black!100!,solid,line width = 0.5pt](-2.5mm,0) -- (2.5mm,0);\node[draw,scale=0.58,circle,fill=black!100!](){};}}}

\newcommand{\circlesolid}{\raisebox{0.5pt}{\raisebox{0.4pt}{${\mathinner{\cdotp\cdotp}}$}\tikz{\node[draw,scale=0.58,circle,fill=none](){};}\raisebox{0.4pt}{${\mathinner{\cdotp\cdotp}}$}}}

\newcommand{\triangdowndot}{\raisebox{0pt}{\raisebox{0.6pt}{${\mathinner{\cdotp\cdotp}}$}\tikz{\node[draw,scale=0.35,regular polygon, regular polygon sides=3,fill=none,rotate=180](){};}\raisebox{0.6pt}{${\mathinner{\cdotp\cdotp}}$}}}

\newcommand{\triangdowndotblack}{\raisebox{0pt}{\raisebox{0.6pt}{${\mathinner{\cdotp\cdotp}}$}\tikz{\node[draw,scale=0.35,regular polygon, regular polygon sides=3,fill=black!100!,rotate=180](){};}\raisebox{0.6pt}{${\mathinner{\cdotp\cdotp}}$}}}

\newcommand{\dimonddot}{\raisebox{0pt}{\raisebox{0.68pt}{${\mathinner{\cdotp\cdotp}}$}\tikz{\node[draw,scale=0.5,diamond,fill=none](){};}\raisebox{0.68pt}{${\mathinner{\cdotp\cdotp}}$}}}
\newtheorem{lemma}{Lemma}
\newtheorem{corollary}{Corollary}

\shorttitle{Drag reduction by permeable substrates} 
\shortauthor{G. G{\'o}mez-de-Segura and R. Garc{\'i}a-Mayoral} 

\title{Turbulent drag reduction by anisotropic permeable substrates -- analysis and direct numerical simulations}

\author
 {
 G. G{\'o}mez-de-Segura \aff{}
  \and 
  R. Garc{\'i}a-Mayoral \aff{}
  }

\affiliation
{
\aff{}
Department of Engineering, University of Cambridge, Cambridge CB2 1PZ, UK
}

\maketitle

\begin{abstract}

We explore the ability of anisotropic permeable substrates to reduce turbulent skin-friction, studying the influence that these substrates have on the overlying turbulence.
For this, we perform DNSs of channel flows bounded by permeable substrates. 
The results confirm theoretical predictions, and the resulting drag curves are similar to those of riblets.
For small permeabilities, the drag reduction is proportional to the difference between the streamwise and spanwise permeabilities.
This linear regime breaks down for a critical value of the wall-normal permeability, beyond which the performance begins to degrade.
We observe that the degradation is associated with the appearance of spanwise-coherent structures, attributed to a Kelvin-Helmholtz-like instability of the mean flow.
This feature is common to a variety of obstructed flows, and linear stability analysis can be used to predict it.
For large permeabilities, these structures become prevalent in the flow, outweighing the drag-reducing effect of slip and eventually leading to an increase of drag.
For the substrate configurations considered, the largest drag reduction observed is $\approx 20-25\%$ at a friction Reynolds number $\delta^+ = 180$.  

\end{abstract}

\section{Introduction}\label{sec:intro} 

The high skin friction experienced in turbulent flows represents a problem for several engineering applications, such as  pipelines and transportation vehicles. The need is therefore to develop new technologies that reduce turbulent drag,
preferably passive, since in contrast with active technologies, these do not require an energy input and have generally lower manufacturing costs.
In this paper we present the potential of anisotropic permeable substrates, a passive technology, to reduce turbulent skin friction, as has recently been proposed by \cite{Abderrahaman2017}.

Most of the literature in turbulent flows over permeable substrates has focused on isotropic materials, observing a substantial increase in drag with respect to a smooth wall \citep{BreugemBoersma2006,Rosti2015,Kuwata2016}. This increase has often been attributed to the onset of large spanwise-coherent structures, which increase the momentum transfer and thus the Reynolds stresses near the wall.
Here, we study the effect of anisotropy and provide physical insight into the behaviour of anisotropic permeable substrates in turbulent flows for drag-reducing purposes, when the permeability is preferential in the streamwise direction. Recent studies have also covered anisotropic substrates, albeit not considering the case of streamwise-preferential permeability \citep{Kuwata2017,Suga2018}.

Previous studies have shown that streamwise-preferential complex surfaces can reduce drag in turbulent flows \citep{Bechert1997,Luchini1991,Jimenez1994,ggFTaC}. This is indeed the case for some of the most common passive technologies for drag reduction, such as riblets or superhydrophobic surfaces.  
Recently, \cite{Abderrahaman2017} suggested that the drag reduction ability of anisotropic permeable substrates 
is based on the same mechanism. 
The general idea is that complex surfaces can reduce drag if they offer more resistance to the cross flow than to the streamwise mean flow.
When the surface texture is vanishingly small compared to the near-wall turbulent structures, the effect of complex surfaces can be reduced to an apparent slip in the tangential directions. 
\cite{Luchini1991}, \cite{Luchini1996} and \cite{Jimenez1994} showed that the change in drag is proportional to the difference between the streamwise and spanwise slips. 
\cite{Hahn2002} observed this behaviour also in turbulent flows over substrates permeable in the streamwise and spanwise directions only. They observed that the streamwise slip is beneficial for drag reduction, while the spanwise slip has an opposite effect. Their substrates, however, were ideal, in the sense that they were impermeable in the wall-normal direction.
Hence, the work by \cite{Hahn2002} is closely connected to studies where only tangential slips are allowed, while the surface remains impermeable, such as those carried out by \cite{Min2004} or \cite{BusseSandham2012} in the context of superhydrophobic surfaces.
Recently, \cite{Rosti2015} have studied permeable substrates with very low wall-normal permeability, which would also fall under this category.
The analysis by \cite{ggMadrid} shows that the deleterious effect of the spanwise slip saturates if this is not accompanied by a corresponding wall-normal transpiration. 
Therefore, surfaces with isotropic slip can also reduce drag, although suboptimally.

The linear theory of \cite{Luchini1991} and \cite{Jimenez1994} is valid only as long as the texture lengthscales are small compared to the characteristic lengthscales of near-wall turbulence.
As the texture size increases, additional deleterious effects set in, breaking down the drag-reducing performance and eventually leading to an increase of drag. The mechanisms behind these deleterious effects vary from one technology to another. 
In riblets, for instance, the degradation of performance is due to the appearance of spanwise-coherent rollers, which arise from a Kelvin-Helmholtz instability \citep{Garcia-Mayoral2011}.
These structures are in fact a common feature to a variety of obstructed flows \citep{Ghisalberti2009}.

Several studies on permeable substrates have also reported the existence of such structures \citep{BreugemBoersma2006,Kuwata2016,Zampogna2016,Suga2017}. 
In these studies, the large increase of the Reynolds stresses compared to that over a smooth wall and the subsequent increase in drag was associated to the presence of Kelvin-Helmholtz rollers. 
\cite{Abderrahaman2017} suggested the formation of these rollers as a possible drag-degrading mechanism for anisotropic permeable substrates. They proposed a model to bound the maximum achievable drag reduction based on the onset of the Kelvin-Helmholtz-like instability. 
\cite{ggFTaC} extended the analysis and identified the wall-normal permeability as the governing parameter in this instability. 
This result agrees with the work performed by \cite{Jimenez2001}, who observed the formation of Kelvin-Helmholtz rollers over substrates which were permeable in the wall-normal direction only, and inferred that the relaxation of the impermeability condition at the wall was sufficient to elicit the rollers. 

Several drag-reducing surfaces show a linear regime, where the drag reduction increases linearly with a certain characteristic length of the texture, followed by a saturation and an eventual increase of drag \citep{Garcia-Mayoral2011b}.  
Although the same has not been shown for anisotropic permeable substrates, 
the similarities between the drag reduction curves of riblets and those of seal fur by \cite{Itoh2006} suggest a similar behaviour \citep{Abderrahaman2017}.
The effect of the seal fur studied by \cite{Itoh2006} would be to some extend that of an anisotropic permeable material, since it is a layer of hairs preferentially aligned in the streamwise direction.

In the current work, we investigate the drag reduction ability of anisotropic permeable substrates.
The aim of this work is to understand how the overlying turbulent flow is modified by the presence of such substrates and build predictive models to estimate their drag-reducing behaviour. 
For that, we perform a series of DNSs of channel flows bounded by permeable substrates, which are selected using the information obtained from a linear stability theory and the linearised theory of \cite{Luchini1991} and \cite{Jimenez1994} for drag reduction.

The present paper is organised as follows.
In \S\ref{sec:Flow_porous} we discuss several models to characterise the flow within the permeable substrates and present the analytic solution to the model subsequently used, Brinkman's model.
How streamwise-preferential permeable substrates can reduce drag is explained in \S\ref{sec:theory}, where we also discuss the theoretical models derived by \cite{Abderrahaman2017} and \cite{ggFTaC}. The former provides estimates for the expected drag reduction in the linear regime, while the latter bounds the achievable drag reduction based on linear stability theory. 
These models allow us to select particular permeable substrates for the subsequent DNS study.
Details for the DNS setup are presented in \S\ref{sec:numerics}. In \S\ref{sec:results_DNS}, we present the DNS results for the permeable substrates selected
and assess the validity of the theoretical models. Drag reduction curves for different anisotropic permeable substrates are also included, allowing to define design guidelines for optimal substrate configurations.
Finally, conclusions are summarised in \S\ref{sec:conclusions}.

\section{Flow within the permeable substrate}\label{sec:Flow_porous}



Following \cite{Abderrahaman2017} and \cite{ggFTaC}, we focus on permeable materials where the pores are much smaller than any near-wall turbulent lengthscale. We therefore opt for a macroscopic, homogenised approach to model the flow within the permeable medium, due to the high resolution required otherwise to explicitly solve the flow within the pores.
The permeable medium is modelled as homogeneous, by defining a local, instantaneous average solution of the flow within the fluid-solid matrix.

A classical approach to characterise the homogenised flow within a permeable medium is Darcy's equation \citep{Darcy1856}. This is the simplest model amongst the continuum approaches, and results from a volume average of the Stokes equation over many pores/particulate obstacles.
Note that under the assumption of vanishingly small pore size, such averages could still be conducted in small volumes compared to the scales of the overlying flow. Darcy's equation
is a balance between the pressure gradient across the permeable medium and the viscous drag caused by the pressure of the solid matrix. More sophisticated
continuum approaches used in the literature include homogenisation techniques \citep{Zampogna2016,Bagheri2017} or the Volume Averaged Navier-Stokes equations (VANS) \citep{Whitaker1996,OchoaTapiaWhitaker1995a,OchoaTapiaWhitaker1995}.
Several authors have recently used the latter to study flows over permeable substrates \citep{BreugemBoersma2006,TiltonCortelezzi2008,Rosti2015}.

\begin{figure}
		\centering
  		    \includegraphics[width=0.9\textwidth,trim={0cm 5.5cm 0cm 5.5cm},clip]{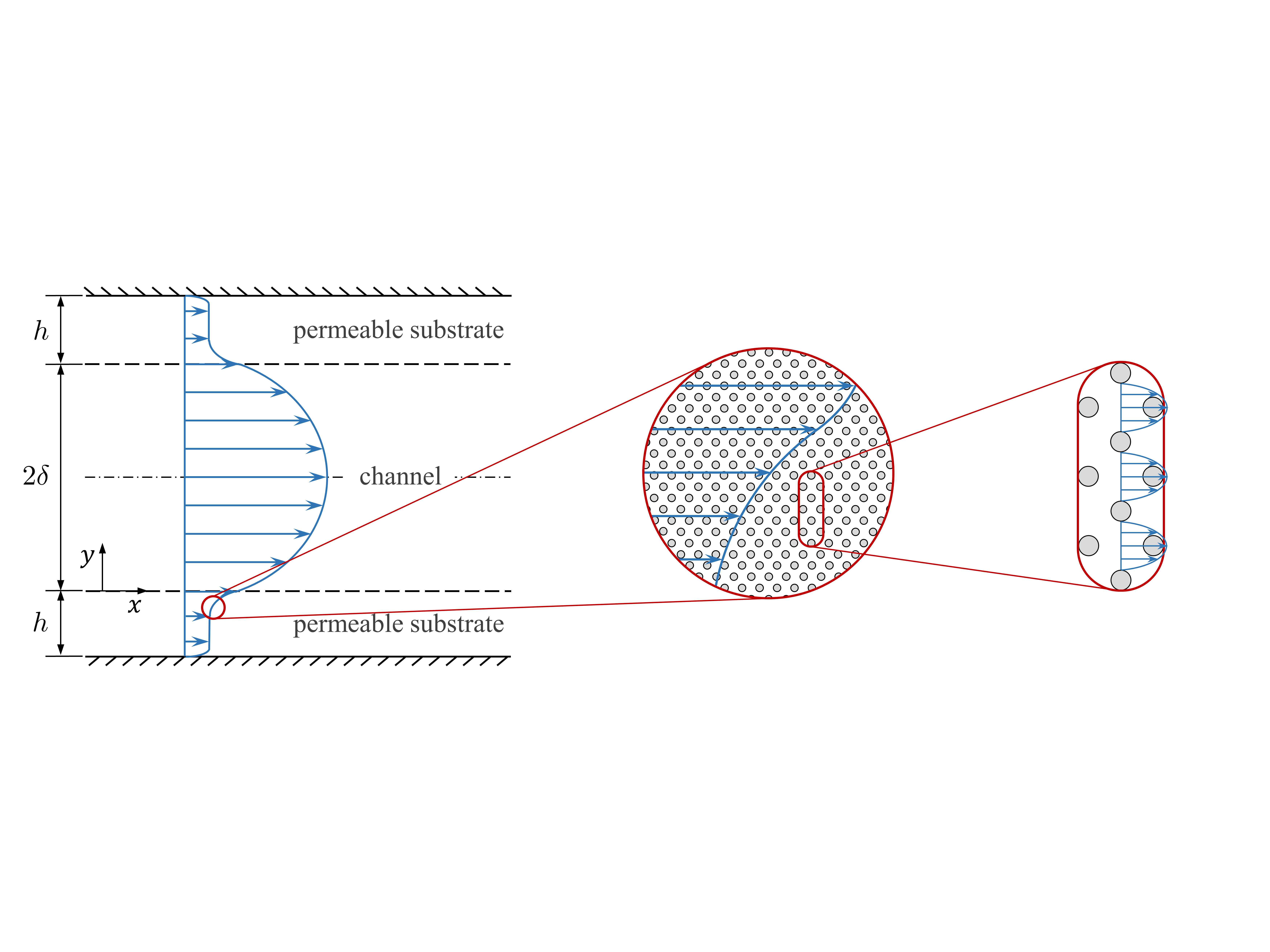}
  		\mylab{-12.3cm}{3.85cm}{(\aaa)}%
  		\mylab{-6.5cm}{2.95cm}{(\bbb)}%
  		\mylab{-2.4cm}{2.95cm}{(\ccc)}%
  		\caption{(\aaa) General layout throughout the present work. (\bbb) Detail of the macroscale flow within the substrate. (\ccc) Detail of the microscale flow within the substrate. }
	    	\label{fig:geometry}	
\end{figure}

\begin{figure}
		\centering
  		\includegraphics[height=0.32\textwidth]{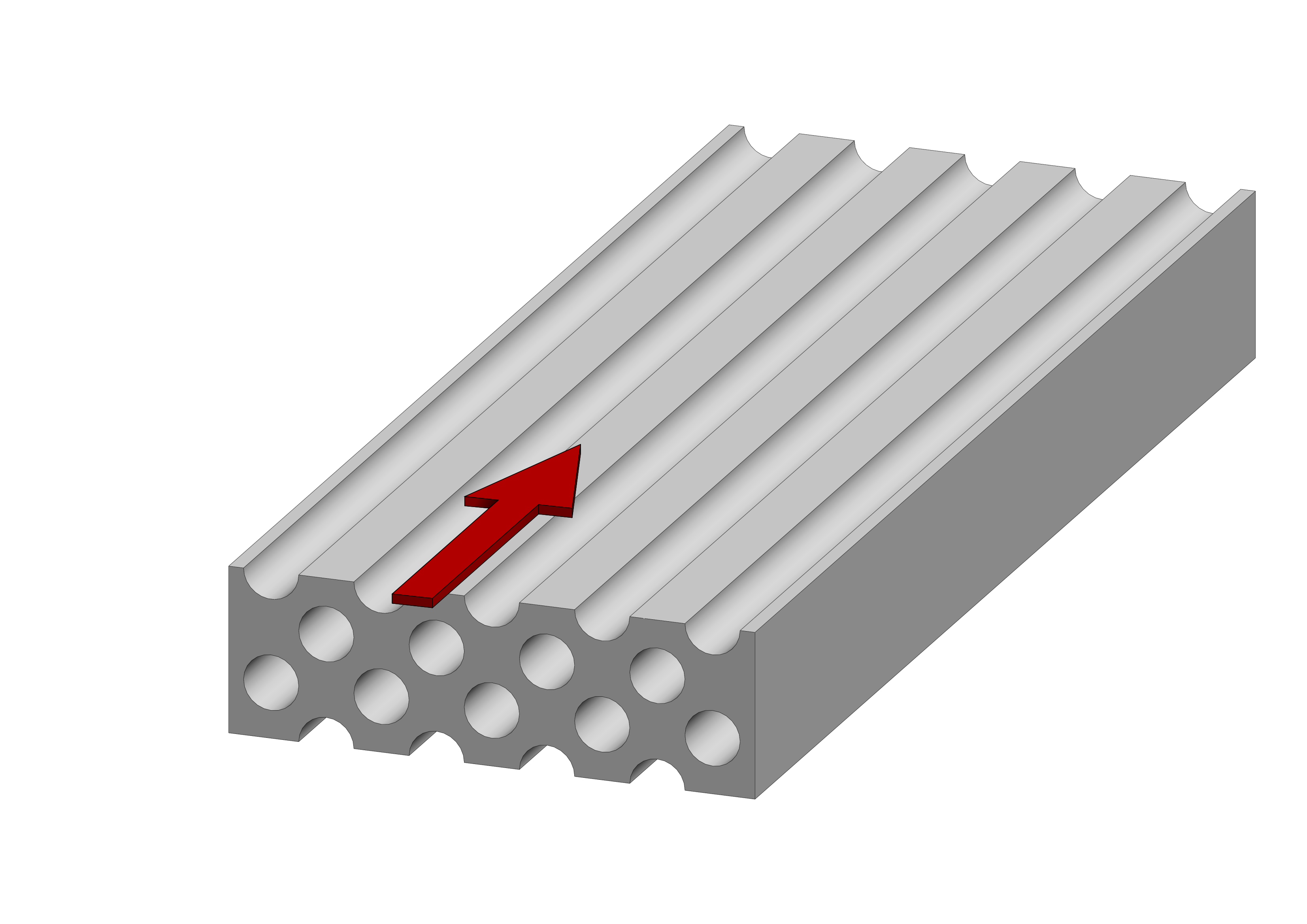}
  		\mylab{-4.3cm}{3.65cm}{(\aaa)}%
  		\hspace{0.5cm}
  		\includegraphics[height=0.32\textwidth,trim={0cm 0cm 0cm 0cm},clip]{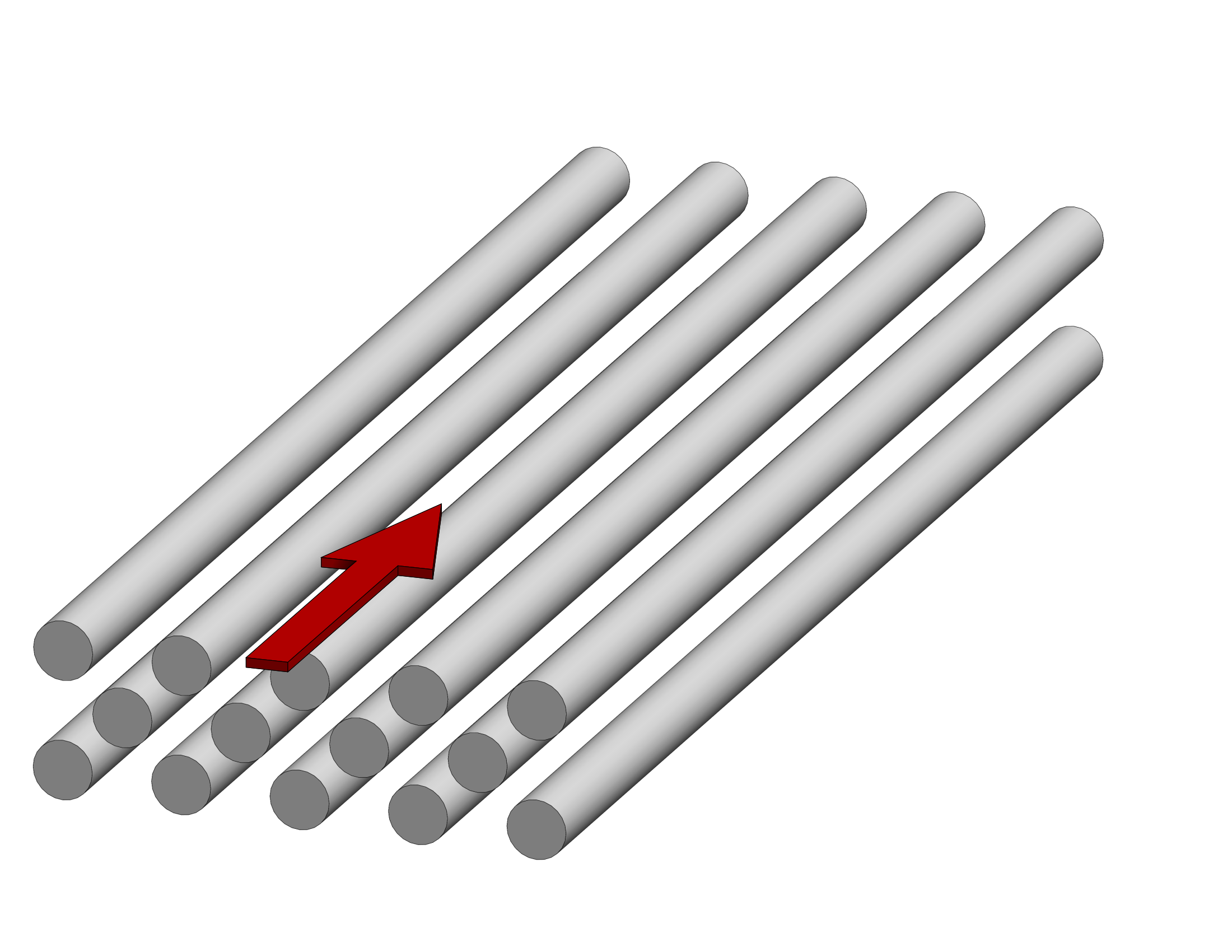}%
  		\mylab{-4.25cm}{3.65cm}{(\bbb)}%
  		\caption{Conceptual sketches of (\aaa) a poorly-connected permeable material, where no diffusive effects connect different pores, and (\bbb) a highly-connected material, where the interstitial flow is well interconnected and diffusion effects can propagate throughout. The red arrow represents the direction of the overlying flow.}\label{fig:materials}
\end{figure}

The volume average, implicit in Darcy's equation, accounts for the viscous stresses caused by velocity gradients over lengths smaller than the averaging one. This effectively filters out diffusive effects acting over larger lengthscales. If the latter are relevant, they can be accounted for by including a macroscopic diffusive term, yielding Brinkman's equation \citep{Brinkman1947},
\begin{equation}
	\nabla p = - \nu \mathbf{K}^{-1} \mathbf{u} + \tilde{\nu} \nabla^2 \mathbf{u}.\label{eq:Br}
\end{equation}%
\noindent
The first two terms in equation~\eqref{eq:Br} constitute Darcy's equation, and the last term, $\tilde{\nu} \nabla^2 \mathbf{u}$, is the Brinkman term,
with $\mathbf{u}$ the velocity vector, p the kinematic pressure, and $\nu$ and $\tilde{\nu}$ the molecular viscosity of the fluid and the effective macrocopic viscosity, respectively.
The homogenised flow within the permeable substrate and the different lengthscales accounted for by the various terms in equation~\eqref{eq:Br} are illustrated in figure~\ref{fig:geometry}.
Panel (\textit{c}) portrays the flow between the obstacles, which results in Darcy's equation when averaged, while 
panel (\textit{b}) portrays the large scale diffusion missed by the volume averaging and captured by the Brinkman term.
Brinkman's model is suitable for substrates made up of open matrices of obstacles,
where fluid regions are significantly interconnected and diffusion can act efficiently over large scales. But,
it does not represent correctly substrates made up of microducts essentially isolated from each other, where diffusion cannot act over scales larger than the pores \citep{Levy1983,Auriault2009}. \cite{Abderrahaman2017} used this distinction to characterise substrates as `highly-connected' or `poorly-connected', and argued that the former offered better properties for drag reduction.
The two types of materials are illustrated in figure~\ref{fig:materials}.

In poorly-connected substrates, Darcy's equation provides a reasonable model for the flow within \citep{Levy1983,Auriault2009}, but it cannot capture the interfacial layer that forms immediately below the substrate-fluid interface, where the velocity transitions from Darcy's velocity deep inside the substrate to a certain slip velocity at the interface plane. For that, the `jump condition' proposed by \cite{Beavers1967} is generally used, which imposes a slip velocity proportional to the external shear at the substrate-fluid interface. 
The constant of proportionality, $\alpha_{BJ}$, accounts for the structure of the permeable material and is determined empirically.

In highly-connected substrates, in contrast, the Brinkman model allows to capture the interfacial region, under certain assumptions.
This equation is also a volume averaging model, so it implicitly assumes that any small volume within the substrate contains a large number of obstacles. However, as the averaging volume approaches the interface with the free flow, this assumption would eventually cease to hold.
The specialised literature shows no general agreement regarding the treatment of the substrate-fluid interface \citep{Bagheri2017,Zampogna2016,OchoaTapiaWhitaker1995,LeBarsWorster2006}. Some studies impose jump conditions, as discussed previously, although these can be of different types, such as a jump in velocity \citep{Beavers1967}, a jump in shear stress but not in velocity \citep{OchoaTapiaWhitaker1995}, or continuity of both velocity and shear stress \citep{VafaiKim1990,LeBarsWorster2006, Battiato2012,Battiato2014}.
Previous studies have shown an analogy between Brinkman's model and Beavers and Joseph's `jump condition' at the substrate-fluid interface \citep{Taylor1971,NealeNader1974,Abderrahaman2017}.
Other studies, in contrast, 
define an adaptation region of certain thickness where the permeability transitions smoothly from its value within the substrate to infinity in the free flow. This is the case of \cite{BreugemBoersma2006}, where they use the more general VANS approach 
with an adaptation region of thickness $\delta_i$. For substrates where the inertial terms are negligible, this approach would be analogous to using Brinkman's model and `blurring' the solution with a moving average of thickness $\delta_i$.

The analysis of \cite{Abderrahaman2017} and \cite{ggFTaC} suggested that highly-connected materials would yield greater drag reduction.
Furthermore, for the small values of permeabilities considered in this study, the flow within the substrate would be dominantly viscous. In this scenario, the Brinkman model provides a simple but reasonable approximation. We therefore follow the above works and use Brinkman's equation to model the flow within the substrate.
For simplicity, we assume that pores are infinitely small, so the continuum hypothesis would hold for any vanishingly small volume, and Brinkman's equation remains valid near the interface \citep{VafaiKim1990}.
For larger permeabilities, the inertial terms might also become important and they need to be considered by including an additional Forchheimer term \citep{Forchheimer1901,Joseph1982,Whitaker1996}.


\subsection{Analytic solution of Brinkman's equation}\label{sec:Flow_Br}

In the present work we consider channels of height $2\delta$ delimited by two identical anisotropic permeable substrates of thickness $h$, as sketched in figure~\ref{fig:geometry}. The substrate-channel interfaces are located at $y=0$ and $y=2 \delta$, and the substrates are bounded by impermeable walls at $y=-h$ and $y=2 \delta + h$.
Throughout the paper we will refer to the free-flow region between $y=0$ and $y=2 \delta$ as `channel' and to the permeable region below $y=0$ (or above $y= 2 \delta$) as `substrate'.
The flow within the permeable substrates is modelled using equation~\eqref{eq:Br}, where the fluid density $\rho$ is assumed to be unity for convenience. The simplicity of Brinkman's equation allows to solve it analytically, and the particularised solution at the substrate-channel interface can be implemented as boundary condition for the DNS of the channel, fully coupling the flow in both regions. 
The procedure to solve Brinkman's equation is detailed in Appendix~\ref{sec:appA}. Here only the problem formulation and its solution are presented.   

As discussed above, poorly-connected substrates have negligible macroscale viscous effects, which in equation~\eqref{eq:Br} can be interpreted as having $\tilde{\nu} = 0$, recovering Darcy's equation. Highly-connected media, in turn, would asymptotically tend to have macroscale diffusion as efficient as a free flow, so $\tilde{\nu} \approx \nu$ \citep{Tam1969, Levy1983, NealeNader1974, Abderrahaman2017}. 
\cite{Abderrahaman2017} and \cite{ggFTaC} suggested that such materials would have a better potential for drag reduction, as it will be discussed in \S\ref{sec:theory}. Here we follow them and assume $\tilde{\nu} = \nu$.
The permeable substrates are characterised then by their thickness, $h$, and their permeabilities $K_x$, $K_y$ and $K_z$ in the streamwise, $x$, wall-normal, $y$, and spanwise, $z$, directions, respectively,
which are considered to be the principal directions of the permeability tensor $\mathbf{K}$ in equation~\eqref{eq:Br}.
The tensor has dimensions of length squared, and is a measure of the ability of the fluid to flow through a permeable medium. 
When $\mathbf{K} \rightarrow \infty$ the medium offers no resistance to the flow, and when $\mathbf{K} = 0$ an impermeable medium is recovered. 

Let us consider the lower substrate between $y=-h$ and $y=0$. To solve equation~\eqref{eq:Br}, we impose no slip and impermeability at $y=-h$, and continuity of the tangential and normal stresses at the substrate-channel interface, i.e. at $y=0$. The solution within the substrate is coupled to the flow within the channel by imposing the continuity of the three velocity components.
The resulting boundary conditions at $y=0$ are then

\begin{subequations}\label{eq:BC_interface}
	\begin{align}
		\nu \left[ \frac{\partial u}{\partial y} + \frac{\partial v}{\partial x} \right]_{y=0^+} &= \tilde{\nu} \left[ \frac{\partial u}{\partial y} + \frac{\partial v}{\partial x} \right]_{y=0^-},\label{eq:BC_interface1}\\
		\nu \left[ \frac{\partial w}{\partial y} + \frac{\partial v}{\partial z} \right]_{y=0^+} &= \tilde{\nu} \left[ \frac{\partial w}{\partial y} + \frac{\partial v}{\partial z} \right]_{y=0^-},\label{eq:BC_interface2}\\
		\left[ -p + 2 \nu \frac{\partial v}{\partial y} \right]_{y=0^+} &= \left[ -p + 2 \tilde{\nu} \frac{\partial v}{\partial y} \right]_{y=0^-},\label{eq:BC_interface3}	
	\end{align}	
\end{subequations}

\noindent where $y=0^+$ and $y=0^-$ correspond to the channel and the substrate sides of the interface, respectively. 
Under the above assumptions, 
the boundary conditions~\eqref{eq:BC_interface} can be further simplified. 
The continuity of tangential stresses becomes that of $\partial u/ \partial y$ and $ \partial w / \partial y$, and the continuity of normal stresses that of $p$.
Equation~\eqref{eq:Br} is then solved by taking Fourier transforms in the tangential directions $(x,z)$. Following the derivations presented in Appendix~\ref{sec:appA}, the analytic solution particularised at $y=0$ provides the following expressions for the velocities,

\begin{subequations}\label{eq:BC3D}
	\begin{align}
		\left. \hat{u} \right|_{y=0^+} &= \left. \hat{u} \right|_{y=0^{-}} = \left. \mathcal{C}_{uu} \frac{d \hat{u}}{dy} \right|_{y=0^{+}} + \left. \mathcal{C}_{uw} \frac{d \hat{w}}{dy} \right|_{y=0^{+}} + \left. \mathcal{C}_{up} \hat{p}\right|_{y=0^{+}},\label{eq:BC3D1}\\
		\left. \hat{w} \right|_{y=0^+} &= \left. \hat{w} \right|_{y=0^{-}} = \left. \mathcal{C}_{wu} \frac{d \hat{u}}{dy} \right|_{y=0^{+}} + \left. \mathcal{C}_{ww} \frac{d \hat{w}}{dy} \right|_{y=0^{+}} + \left. \mathcal{C}_{wp} \hat{p}\right|_{y=0^{+}},\label{eq:BC3D2}\\
		\left. \hat{v} \right|_{y=0^+} &= \left. \hat{v} \right|_{y=0^{-}} = \left. \mathcal{C}_{vu} \frac{d \hat{u}}{dy} \right|_{y=0^{+}} + \left. \mathcal{C}_{vw} \frac{d \hat{w}}{dy} \right|_{y=0^{+}} + \left. \mathcal{C}_{vp} \hat{p}\right|_{y=0^{+}}, \label{eq:BC3D3}
	\end{align}	
\end{subequations}
\noindent where the hat denotes variables in Fourier space. The coefficients $\mathcal{C}_{ij}$ are complex and depend on the structure of the permeable substrate through $K_x$, $K_y$, $K_z$ and $h$, as well as on the overlying flow through the streamwise and spanwise wavenumbers, $\alpha_x$ and $\alpha_z$,
or the corresponding wavelengths, $\lambda_x = 2 \pi / \alpha_x$ and $\lambda_z = 2 \pi / \alpha_z$.
The same procedure can be used to obtain a symmetric solution for the upper substrate, and the resulting expressions for the interface at $y=2 \delta$ can be found in Appendix~\ref{sec:appA}.
The effect of the permeable substrates on the channel flow is introduced through equations~\eqref{eq:BC3D} and the corresponding equations at $y=2\delta$, which serve as boundary conditions.

\begin{figure}
		\centering
  		\includegraphics[height=0.29\textwidth]{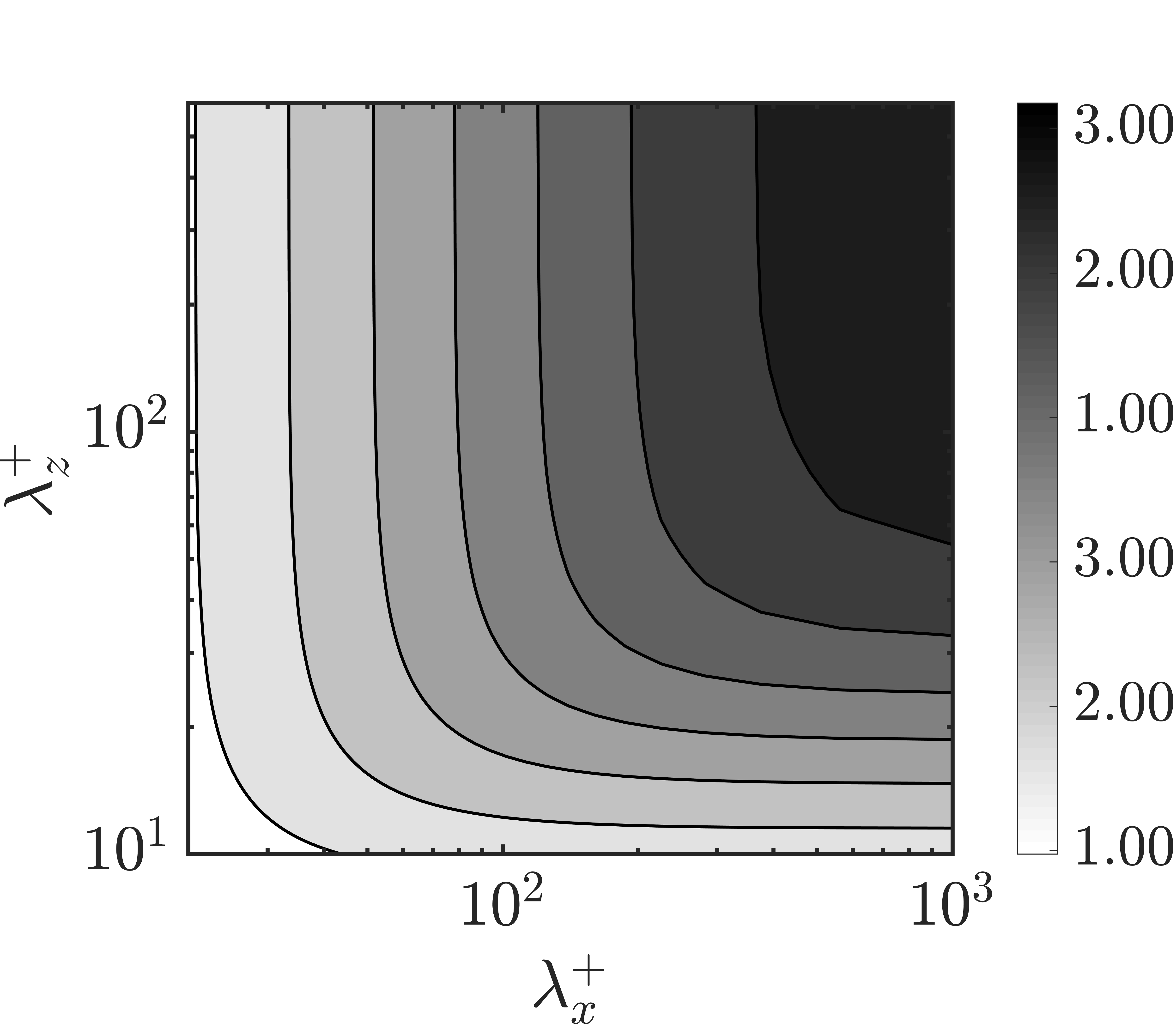}%
  		\mylab{-4.3cm}{3.65cm}{(\aaa)}%
  		\hspace{0.2cm}
  		\includegraphics[height=0.29\textwidth,trim={3cm 0cm 0cm 0cm},clip]{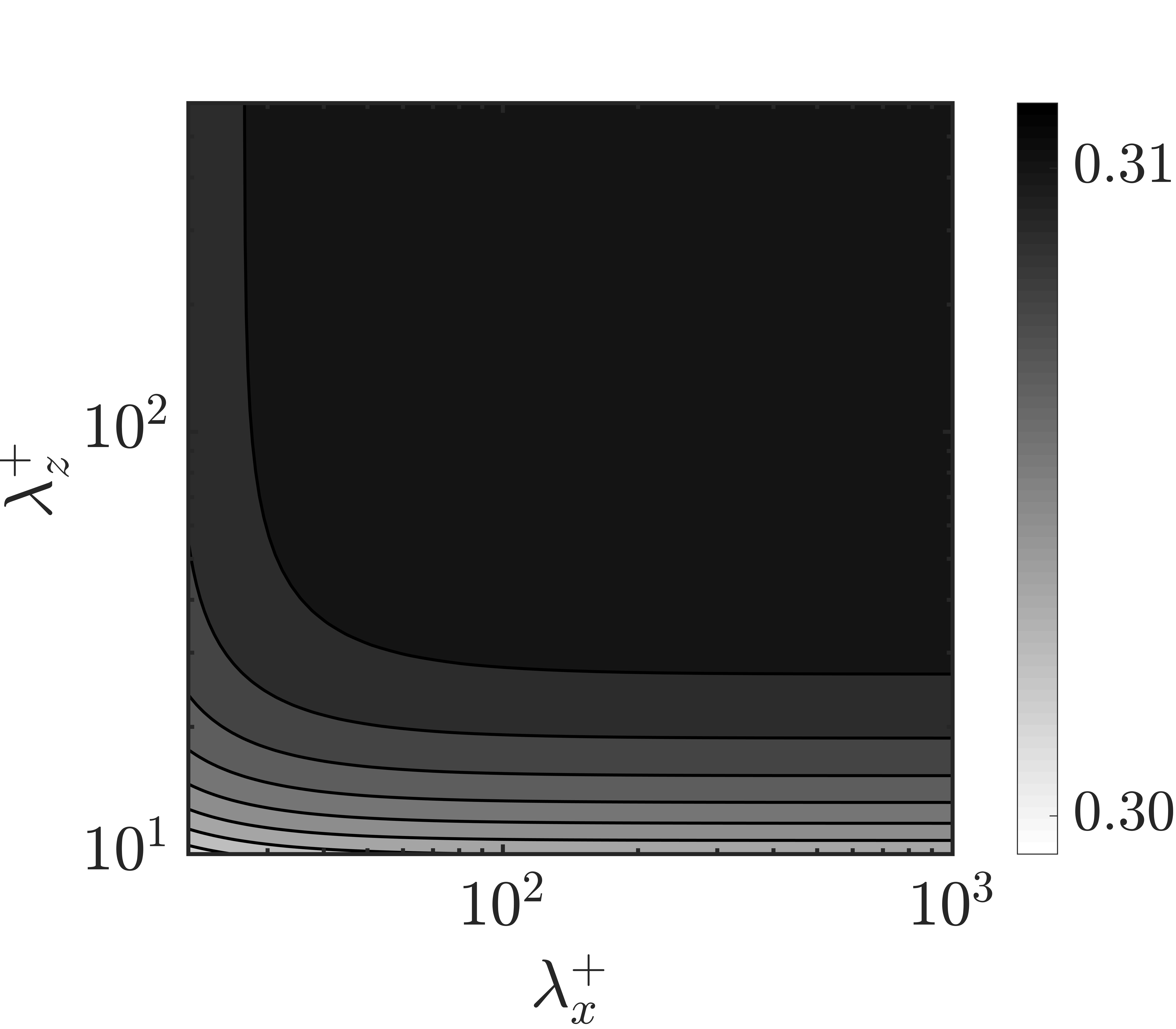}%
  		\mylab{-4.25cm}{3.65cm}{(\bbb)}%
  		\hspace{0.2cm}
  		\includegraphics[height=0.29\textwidth,trim={3cm 0cm 0cm 0cm},clip]{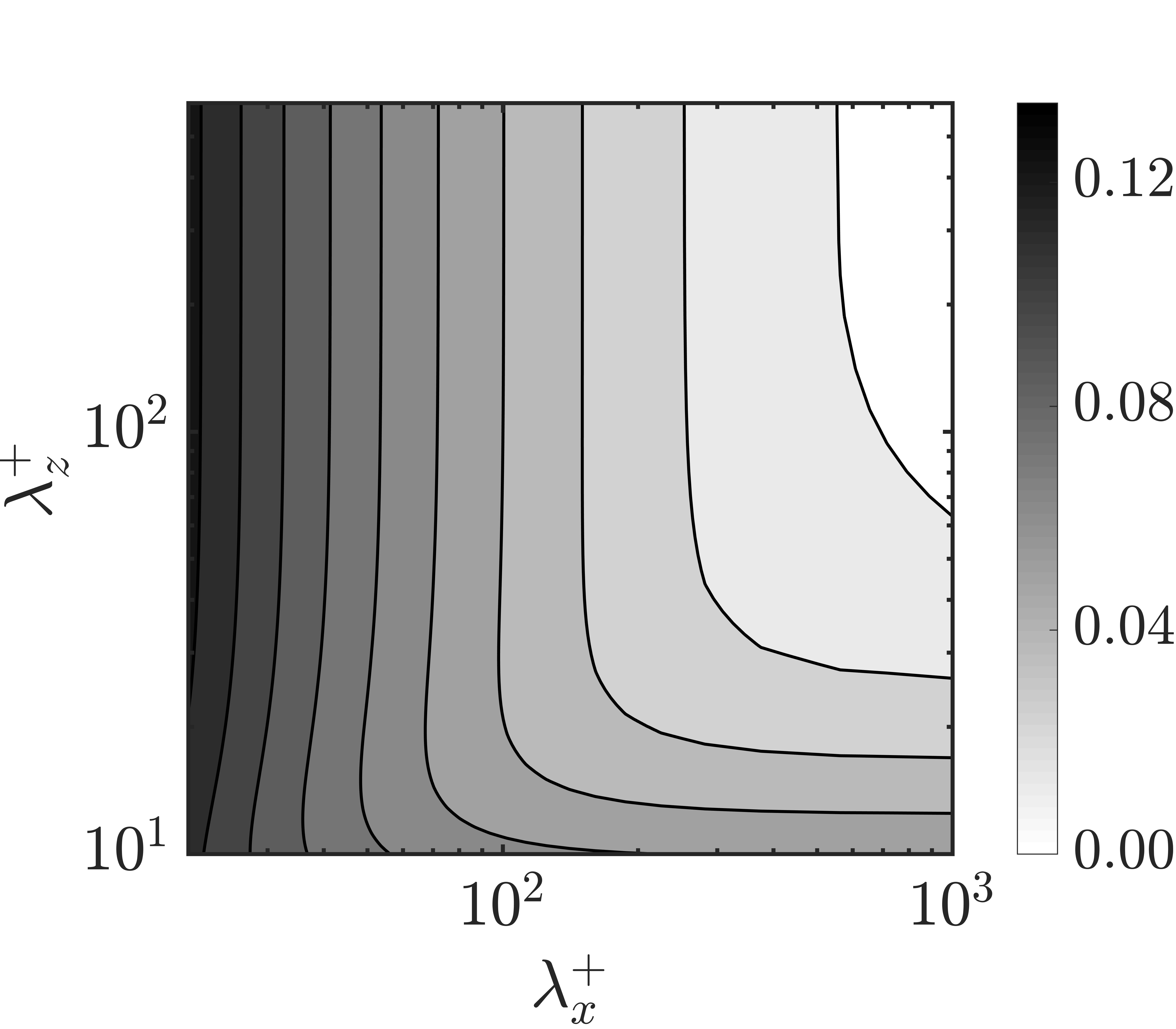}
  		\mylab{-4.25cm}{3.65cm}{(\ccc)}%
  		\caption{Maps of (\aaa) $\mathcal{C}_{uu}^+$, (\bbb) $\mathcal{C}_{ww}^+$ and (\ccc) $-\mathcal{C}_{vp}^+$, from equation~\eqref{eq:BC3D}, as a function of the wavelengths $\lambda_x^+$ and $\lambda_z^+$ for substrate C4 in table~\ref{tab:cases}.}\label{fig:coeffs}
\end{figure}

To illustrate how the coefficients in equation~\eqref{eq:BC3D} vary with the wavelengths, figure~\ref{fig:coeffs} shows maps of $\mathcal{C}_{uu}^+$, $\mathcal{C}_{ww}^+$ and $\mathcal{C}_{vp}^+$, which have zero imaginary part, as a function of $\lambda_x^+$ and $\lambda_z^+$ for a particular substrate.
The superscript `$+$' denotes viscous units, where magnitudes are normalised using the kinematic viscosity, $\nu$, and the friction velocity at the substrate-channel interface,
$u_{\tau} = \sqrt{\tau_w}$. Note that the total stress at that location, $\tau_w$, accounts for both the viscous and the Reynolds stresses.
$\mathcal{C}_{uu}^+$ and $\mathcal{C}_{ww}^+$ relate the streamwise and spanwise velocities with their corresponding wall-normal gradients, respectively, and are connected to the slip boundary conditions typically used in slip-only simulations \citep{Hahn2002,Min2004,BusseSandham2012}. $\mathcal{C}_{vp}^+$ represents an impedance relating the wall-normal velocity and the pressure \citep{Jimenez2001}. 
The slip coefficients  $\mathcal{C}_{uu}^+$ and $\mathcal{C}_{ww}^+$ are purely real, so the tangential velocity is in phase with the tangential shear. The transpiration coefficient $\mathcal{C}_{vp}^+$ is also real but negative, so the wall-normal velocity is in anti-phase with the pressure. 
For the mean flow, i.e $\alpha_x^+ = 0$ and $\alpha_z^+ = 0$ (or alternatively $\lambda_x^+ \rightarrow \infty$ and $\lambda_z^+ \rightarrow \infty$), out of the 9 coefficients from equation~\eqref{eq:BC3D} only $\mathcal{C}_{uu}^+$ and $\mathcal{C}_{ww}^+$ are non-zero 
and their value decreases as the wavenumbers increase, as shown in figure~\ref{fig:coeffs}. In contrast, the transpiration coefficient $\mathcal{C}_{vp}^+$ is zero for the mean flow and becomes increasingly negative as the wavenumbers increase, since short wavelengths penetrate more easily through the substrate.
In \cite{ggFTaC}, we conducted preliminary DNSs of channel flows with permeable substrates where only these three coefficients from equation~\eqref{eq:BC3D} were included.
The DNSs presented here in \S\ref{sec:results_DNS} show that the other coefficients modulate the results, and this modulation can become significant as the permeability increases.

\section{Theoretical models}\label{sec:theory} 

In this section, we present the theoretical models introduced by \cite{Abderrahaman2017} and \cite{ggFTaC} to estimate the drag reduction that permeable substrates can achieve. We also discuss the effect on internal and external flows and how they relate.

\subsection{Drag reduction from surface manipulations}\label{subsec:DU}

\begin{figure}
		\centering
		\includegraphics{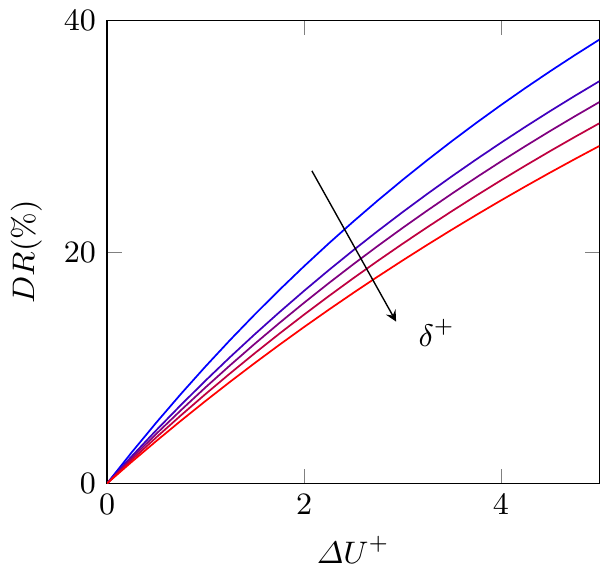}	  
  		\caption{Drag reduction, $DR$, as a function of $\Delta U^+$, as given by equation~\eqref{eq:DR}, for different friction Reynolds numbers. $DR$ has been calculated using the centreline velocities of the smooth channels in \cite{LeeMoser2015}, \cite{Hoyas2006} and \cite{Lozano2014}. Blue to red, $\delta^+ \approx 180,$ $540$, $1000$, $1990$, $5180$. The arrow indicates increasing friction Reynolds number.}
	    \label{fig:DR_vs_DU}	
\end{figure}

The friction coefficient, $c_f$, can be defined as
\begin{equation}
	c_f = 2 \frac{\tau_{w}}{U_{\delta}^2}= 2 \frac{1}{U_{\delta}^{+2}},
	\label{eq:cf}
\end{equation}
\noindent where the density is assumed to be unity. The choice on the reference velocity $U_{\delta}$ depends on the type of flow studied. In external flows, the free stream velocity is typically used, while in internal flows the bulk velocity is more common.
The substrates studied here would mainly be aimed at external flow applications, for instance as coatings in vehicle surfaces.
The simulations, however, have been conducted in channels for simplicity.
In this framework, \cite{Garcia-Mayoral2011} argued that choosing the centreline velocity as the reference for $c_f$ permitted a closer comparison with external-flow friction coefficients.

In the case of small surface textures, their effect is confined to the near-wall region.
According to the classical theory of wall turbulence,  
sufficiently far away from the wall, the only effect of any surface manipulation is to modify the intercept of the logarithmic law, while the K\'arm\'an constant and the wake function remain unaltered \citep{Clauser1956}. The centreline velocity is then $U_{\delta}^{+} = U_{\delta 0}^{+} + \Delta U^+$, where the subscript `$0$' indicates values for a reference smooth channel and $\Delta U^+$ is the shift of the logarithmic velocity profile with respect to the smooth wall. The drag reduction ($DR$) can then be expressed in terms of $\Delta U^+$,
\begin{equation}
	DR = - \frac{c_f - c_{f0}}{c_{f0}} = 1 - \frac{1}{\left( 1 + \Delta U^+/U_{\delta 0}^{+} \right)^2}.
	\label{eq:DR}
\end{equation}
\noindent If $\Delta U^+ > 0$, the logarithmic region is shifted upwards and drag is reduced. Conversely, if $\Delta U^+ < 0$, the logarithmic region is shifted downwards and drag is increased. Note that $DR$ depends on the friction Reynolds number, $\delta^+$, through $U_{\delta 0}^+$, while $\Delta U^+$ does not. The latter therefore provides a more universal measure, as it can be extrapolated to higher $\delta^+$ \citep{Garcia-Mayoral2011b,SpalartMcLean2011,Gatti2016,RGM2018}.
The change of $DR$ with $\Delta U^+$ given by equation~\eqref{eq:DR} and its dependence with $\delta^+$ are depicted in figure~\ref{fig:DR_vs_DU}. This figure shows a decrease of $DR$ with the Reynolds number, due to larger values of $U_{\delta 0}^+$. This can be expected to lead to discrepancies in $DR$ between simulations and experiments at low Reynolds numbers, and industrial applications at high Reynolds numbers. To circumvent this, in the present paper we quantify drag reduction in terms of $\Delta U^+$.

\subsection{Drag reduction from virtual origins}\label{subsec:DRgeneral}

Drag reduction from non-smooth, passive surfaces has recently been reviewed in \cite{RGM2018} as a virtual-origin effect, 
where the reduction of drag is essentially caused by an offset between the positions of the virtual, equivalent smooth walls perceived by the mean flow and the overlying turbulent flow.
For vanishingly small surface textures, \cite{Luchini1991} proposed that $\Delta U^+$ produced by any complex surface is given by
\begin{equation}
	\Delta U^+ \approx \ell_{U}^+ - \ell_{T}^+,
\label{eq:DR_gg}
\end{equation}
\noindent where $\ell_U^+$ refers to the virtual origin experienced by the mean flow, defined as the depth below a reference plane where the mean flow would perceive a non-slipping wall; and $\ell_T^+$ refers to the virtual origin experienced by turbulence. 
\cite{Luchini1996} suggested that the latter could be identified as the origin experienced by the quasi-streamwise vortices.
These virtual origins are measured from a reference plane, often taken at the top plane of the surface geometry, for instance at the riblet tips \citep{Luchini1991} or at the substrate-fluid interface plane for permeable substrates \citep{Abderrahaman2017}. This is where we set $y=0$.
The virtual origins perceived by the mean flow and the vortices are therefore at $y^+=-\ell_U^+$ and $y^+=-\ell_T^+$, respectively.
As discussed below, these are directly connected to the concepts of `slip lengths' and `protrusion heights' typically used in the literature.

\begin{figure}
		\centering
		\vspace{0.1cm}
  		\includegraphics[width=0.7\textwidth,trim={0 3cm 0 2.6cm},clip]{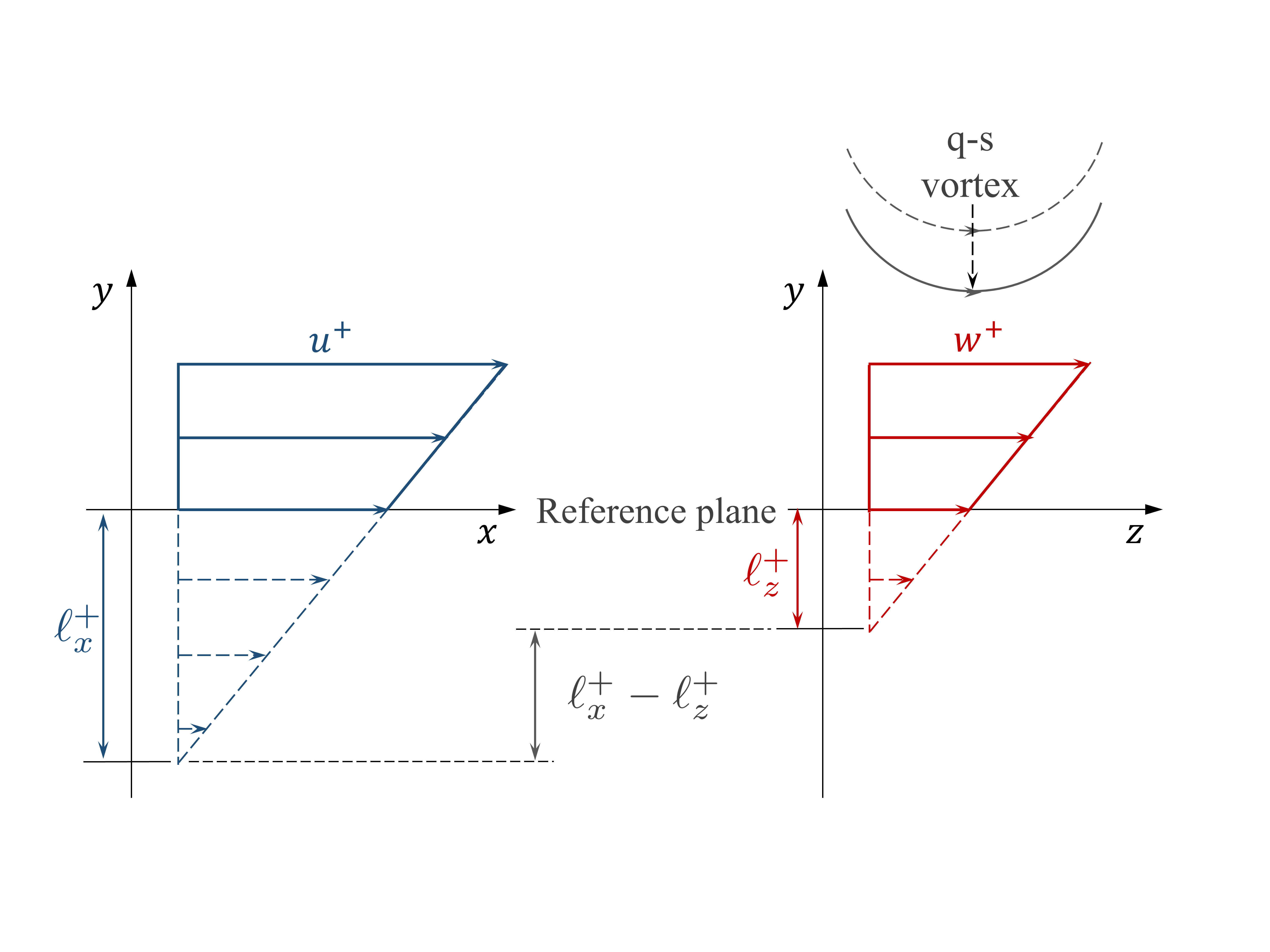}
  		\mylab{-9.3cm}{4cm}{(\aaa)}%
  		\mylab{-4.3cm}{4cm}{(\bbb)}%
  		\caption{Sketch of the (\aaa) streamwise and (\bbb) spanwise slip lengths, $\ell_x^+$ and $\ell_z^+$, and the corresponding virtual origins at $y^+ = -\ell_x^+$ and $y^+ = -\ell_z^+$. A quasi-streamwise vortex (q-s vortex), inducing cross-flow $w^+$, is sketched in (\bbb).}
  		\label{fig:VOs}
\end{figure}

If the surface texture is small, the overlying flow does not perceive the detail of the texture, but a homogenised effect, quantified by these virtual origins.
Complex surfaces therefore change drag by causing a relative y-displacement of turbulence with respect to the mean flow, as depicted in figure~\ref{fig:VOs}, but turbulence remains otherwise smooth wall-like \citep{Luchini1991,Jimenez1994,Luchini1996, RGM2018}.
If $\ell_T^+ < \ell_U^+$, 
quasi-streamwise vortices are, compared to a smooth wall, shifted farther away from the origin of the mean flow.
As a result, the local momentum flux close to the surface decreases, thereby reducing the shear and the skin friction. 
Conversely, if $\ell_U^+ < \ell_T^+$, the vortices perceive a deeper origin than the mean flow and friction drag increases. 

\cite{Luchini1991} and \cite{Luchini1996} proposed that the virtual origin of turbulence is given by that of the spanwise velocity.
Given that for small surface textures the velocity profile near the surface is linear, the concept of virtual origins can be represented by Robin boundary conditions at the reference plane $y^+=0$,
\begin{subequations} \label{eq:Slip_BC}
    \begin{gather}
       \left. u^+ \right|_{y^+=0}= \ell_{x}^+ \left. \frac{\partial u}{\partial y}^+ \right|_{y^+=0}, \label{subeq:Slip_BCa}\\
       \left. w^+ \right|_{y^+=0}= \ell_{z}^+ \left. \frac{\partial w}{\partial y}^+ \right|_{y^+=0} \label{subeq:Slip_BCb},
    \end{gather}
\end{subequations}
\noindent 
where the Robin coefficients $\ell_x^+$ and $\ell_z^+$, typically referred to as the streamwise and spanwise slip lengths, are roughly equal to the depths of the virtual origins, i.e. $\ell_{U}^+ \approx \ell_{x}^+$ and $\ell_{T}^+ \approx \ell_{z}^+$.
In addition, the mean streamwise shear is $dU^+/dy^+|_{y^+=0} \approx 1$ and the slip length, $\ell_x^+$, is interchangeable with the slip velocity, $U_{slip}^+$. 

Boundary conditions of the form of equation~\eqref{eq:Slip_BC} are generally used in slip-only simulations, such as in \cite{Min2004} or \cite{BusseSandham2012}.
In these simulations, however, the effect of $\ell_z^+$ on $\Delta U^+$ saturates, that is the linear expression $\Delta U^+ \approx \ell_{x}^+ - \ell_{z}^+$ is valid only for $\ell_z^+ \lesssim 1$  and increasing $\ell_z^+$ beyond $\approx 4$ has only a negligible effect on $\Delta U^+$.
\cite{ggMadrid} noted that this saturation effect is a result of the impermeability condition imposed at the interface, $v = 0$, as the imposed impermeability impedes the displacement of the quasi-streamwise vortices further towards the interface. 
This effect would be present in the drag-reducing simulations of \cite{Hahn2002} and \cite{Rosti2018}, which considered zero or very low values of wall-normal permeabilities, so that $v \approx 0$ at the interface.
This would not be the case for the permeable substrates in general, or for those studied in this paper in particular.
For the DNSs presented in \S\ref{sec:results_DNS}, we consider equal wall-normal and spanwise permeabilities, $K_y^+ = K_z^+$. The slip in the spanwise direction is then always accompanied by a corresponding wall-normal transpiration, and the virtual origin perceived by turbulence is roughly given by $\ell_T^+ \approx \ell_z^+$ with no saturation.
For a more general case where $K_z^+ \neq K_y^+$, however, the virtual origin of turbulence would deviate from $\ell_z^+$ \citep{ggMadrid}.

\subsection{Virtual origins for anisotropic permeable substrates}\label{subsec:DR}

\cite{Abderrahaman2017} derived the streamwise and spanwise slip lengths, as well as $\Delta U^+$, for a permeable substrate.
The authors calculated $\ell_x^+$ and $\ell_z^+$ by solving the flow within the permeable medium in response to an overlying shear, obtaining a solution of the form of equation~\eqref{eq:Slip_BC}, a procedure that has also been followed for riblets or superhydrophobic textures \citep{Luchini1991,Ybert2007}.   
\cite{Abderrahaman2017} solved equation~\eqref{eq:Br} for $u$ and $w$ under homogeneous shear, for which the pressure terms zero out. This is actually the solution for mode zero, i.e. $\alpha_x = 0$ and $\alpha_z = 0$, in Appendix~\ref{sec:appA}.
Obtaining the relationships between the velocities and their corresponding shears at the interface, $\ell_x^+$ and $\ell_z^+$ are
\begin{subequations} \label{eq:SL}
    \begin{gather}
    \ell_x^+ = \xi \sqrt{K_x^+} \tanh \left( \frac{h^+}{\sqrt{K_x^+}} \right), \label{subeq:SLa}\\
    \ell_z^+ = \xi \sqrt{K_z^+} \tanh \left( \frac{h^+}{\sqrt{K_z^+}} \right), \label{subeq:SLb}
    \end{gather}
\end{subequations}
\noindent where $\xi$ is the ratio between the molecular and effective viscosities of the permeable substrate, and would be $\xi \approx 1$ for highly-connected substrates with $\tilde{\nu} \approx \nu$.
Note that $\ell_x^+$ and $\ell_z^+$ in equation~\eqref{eq:SL} are the coefficients $\mathcal{C}_{uu}^+$ and $\mathcal{C}_{ww}^+$ for mode $(0,0)$ in equation~\eqref{eq:BC3D}.
For poorly-connected substrates, \cite{Abderrahaman2017} obtained the same solution using Darcy's equation with Beavers \& Joseph's jump conditions at the interface, in which case $\xi$ would be the inverse of Beavers \& Joseph's constant of proportionality, $\alpha_{BJ}$.

\cite{Abderrahaman2017} concluded that the highest performance for a given anisotropic material would be achieved for sufficiently deep substrates, where $h^+ \gtrsim \Ksqx, \Ksqz$. In this case, both hyperbolic tangents in equation~\eqref{eq:SL} tend to unity and the slip lengths become $\ell_x^+ \approx \xi \Ksqx$ and $\ell_z^+ \approx \xi \Ksqz$.
Introducing these results into equation~\eqref{eq:DR_gg}, $\Delta U^+$ becomes

\begin{equation} \label{eq:DR_K}
    \Delta U^+ \approx \xi \left( \sqrt{K_x^+} - \sqrt{K_z^+} \right).
\end{equation}
 
The microstructure of the substrate, represented by $\xi$, has therefore an important effect on the drag-reducing performance of the substrate. The optimum configuration would be obtained for highly-connected materials with $\xi \approx 1$ (i.e. $\tilde{\nu} \approx \nu$), which supports our previous assumption in \S\ref{sec:Flow_Br}. 
Furthermore, to maximise drag reduction, we seek highly anisotropic materials, maximising the streamwise permeability, $K_x^+$, while minimising the spanwise one, $K_z^+$.
Note that equation~\eqref{eq:DR_K} considers deep substrates, so that the flow near the substrate-channel interface does not perceive the bottom no-slipping wall. This assumption eliminates the substrate thickness, $h^+$, from the parameter space under consideration.

The linear theory that results in equation~\eqref{eq:DR_K} is valid only if the texture lengthscales are small compared to the characteristic lengthscales of near-wall turbulence, so that the near-wall cycle is not altered.
For a given permeable material (i.e. with fixed permeability values $K_x$, $K_z$ and $K_y$), the permeabilities $K_x^+$ and $K_z^+$ in viscous units would increase as the friction Reynolds number increases, thereby increasing $\Delta U^+$.
As $K_x^+$ and $K_z^+$ increase, equation~\eqref{eq:DR_K} would eventually stop holding, as other mechanisms set in, degrading the drag-reducing performance.

\subsection{Onset of Kelvin-Helmholtz rollers}\label{subsec:LSA}

Equation~\eqref{eq:DR_K} does not explicitly include the wall-normal permeability, or transpiration in general.
However, most complex surfaces that produce slip produce also a non-zero wall-normal velocity at the reference plane, such as permeable substrates \citep{BreugemBoersma2005}, riblets \citep{Garcia-Mayoral2011} or superhydrophobic surfaces \citep{Seo2018}, and this effect induces generally a degradation in drag.
\cite{Abderrahaman2017} and \cite{ggFTaC} argued that the onset of the Kelvin-Helmholtz instability discussed in the introduction would disrupt the linear regime of equation~\eqref{eq:DR_K}, and could therefore be used to establish an \textit{a priori} limit for its range of validity.

Kelvin-Helmholtz rollers are ubiquitous over permeable substrates and are known to increase drag \citep{BreugemBoersma2006,Kuwata2016,Suga2017}.
\cite{Abderrahaman2017} developed a model based on Darcy's equation to characterise the onset of such structures, which is well-suited for poorly-connected permeable media. 
\cite{ggFTaC} extended their analysis for highly-connected permeable substrates, which have a greater potential for drag reduction, and showed that the latter exhibit a different behaviour for the onset of the instability.

In this section, we summarise the procedure and results from \cite{ggFTaC}.
The procedure is based on a linear stability analysis on the mean turbulent profile to capture the onset of Kelvin-Helmholtz rollers, as in \cite{Jimenez2001}, \cite{Garcia-Mayoral2011} and \cite{Abderrahaman2017}.
The analysis is restricted to spanwise-homogeneous modes, as Kelvin-Helmholtz rollers are predominantly spanwise coherent.
Considering normal-mode solutions of the form $v' = \hat{v}(y) \exp (i(\alpha x - \omega t))$, where the wavenumber $\alpha$ is real and the angular frequency $\omega= \omega_r + i \omega_i$ is complex,
the Orr-Sommerfeld equation with a variable eddy viscosity in $y$ \citep{Cess1958} can be solved.
Modes are unstable if $\omega_i$ is positive. 
At the substrate-channel interface, equation~\eqref{eq:BC3D} for $\hat{u}$ and $\hat{v}$ was imposed, 
particularising for $\alpha_z = 0$ and $\hat{w} = 0$.

\begin{figure}
		\centering
  		    \includegraphics[width=0.55\textwidth]{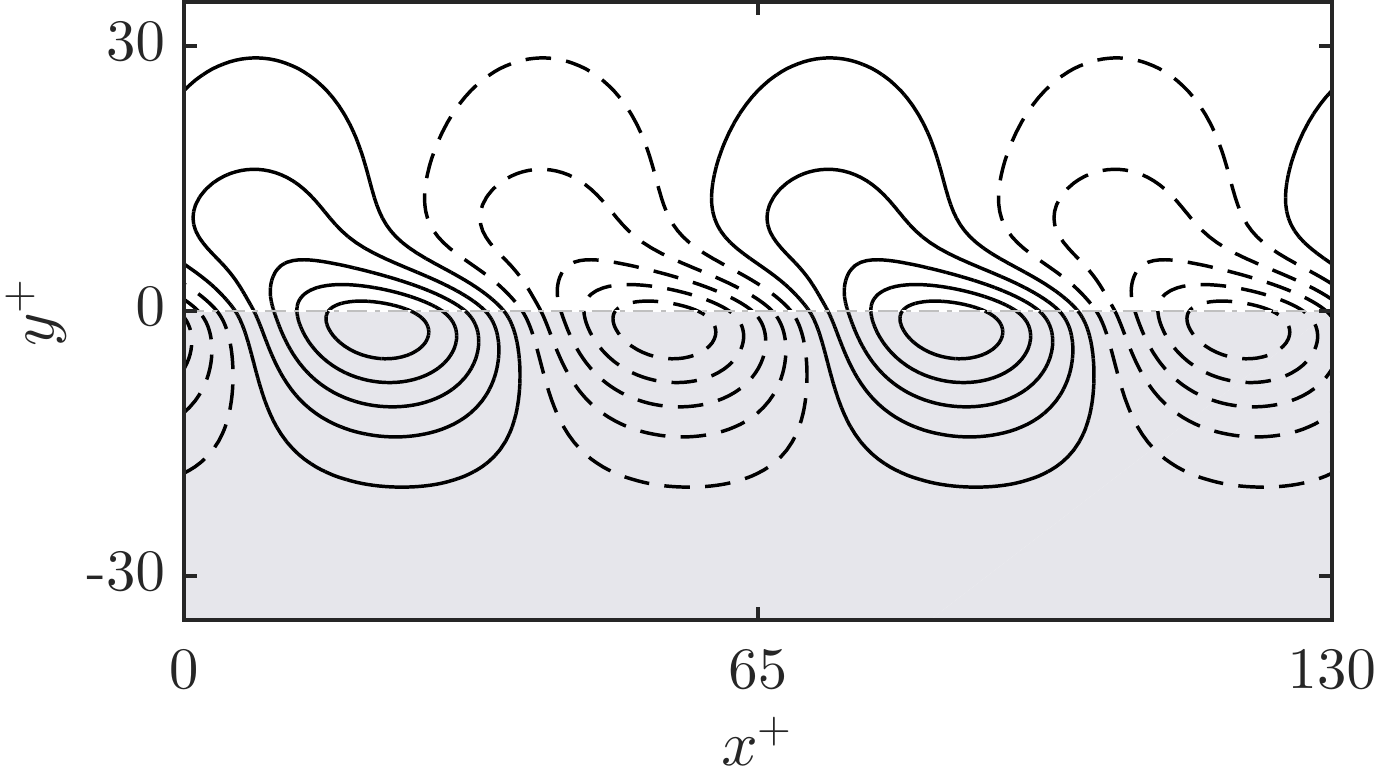}
  		\caption{Isocontours of the streamfunction for the most unstable mode, with $\lambda^+ \approx 70$, for a substrate with $K_x^+ = 100$, $K_y^+ = 10$ and $h^+ = 100$. The solid and dashed lines correspond to clockwise and counter-clockwise rotation, respectively. }
	    	\label{fig:LSA_KH}	
\end{figure}

The presence of permeable substrates at the boundaries of the channel destabilises the otherwise stable mean flow, in agreement with \cite{Jimenez2001}, \cite{TiltonCortelezzi2008} and \cite{Abderrahaman2017}. 
The most unstable mode forms counterrotating rollers separated by a wavelength $\lambda^+ \approx 70$, as shown in figure~\ref{fig:LSA_KH}, which resemble Kelvin-Helmholtz rollers. 
\cite{Garcia-Mayoral2011} and \cite{Abderrahaman2017} argued that the height at which the energy-producing term of the Orr-Sommerfeld equation, $d^2 U / dy^2$, concentrates, $y_{c}^{+} \simeq 9$, sets the lengthscale for the instability.
This height is essentially independent of the Reynolds number when measured in wall units, resulting in the optimum $\lambda^+ \sim 2 \pi y_c^+$ regardless of the topology of the substrate.

\begin{figure}
		\centering
		\includegraphics{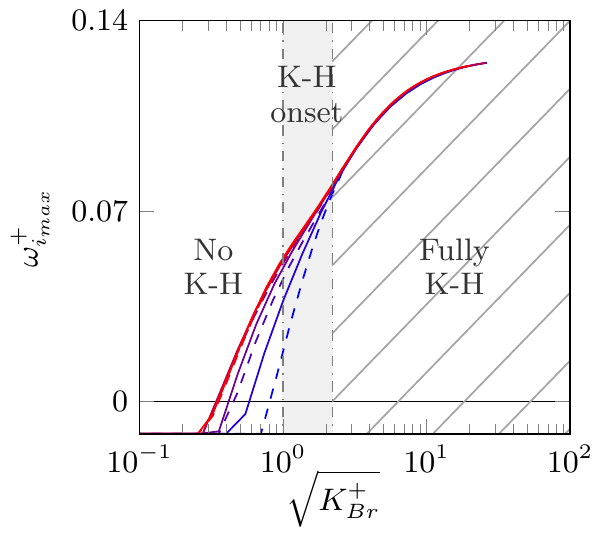} 		  
  		\caption{Maximum amplification, $\omega_{i_{max}}^+$, versus the permeability lengthscale, $\protect\KsqBr$, for different permeable substrates. \broken,  $h^+=10$;  \full, $h^+=100$; from blue to red, anisotropy ratios $\phi_{xy} = \sqrt{K_x/K_y} \approx 1$, $3$, $10$, $30$. The shaded region corresponds to the estimated range for the onset of Kelvin-Helmholtz rollers (K-H), with the dashed-dotted lines corresponding to $\protect\KsqBr \approx 1$ and $2.2$.}
	    	\label{fig:sigmai}	
\end{figure}

While the substrate topology does not significantly alter the wavelength of the most amplified mode, it determines its amplification. \cite{ggFTaC} proposed a single, empirically-fitted parameter to capture the effect of the topology on the amplification, given for streamwise-preferential substrates by

\begin{equation}
	K_{Br}^{+} = K_{y}^+ \tanh \left( \frac{\sqrt{2 K_{x}^{+}}}{y_{c}^{+}}  \right) \tanh^2 \left( \frac{h^+}{\sqrt{12 K_{y}^{+}}} \right).
	\label{eq:Br_parameter_plus}
\end{equation}%
\noindent Figure~\ref{fig:sigmai} shows how the amplification for different substrates is essentially a function of $K_{Br}^{+}$ only. 
From an application point of view, we are interested in sufficiently deep and streamwise-preferential substrates, $h^+ \gtrsim \Ksqx \gg \Ksqy$. If this is the case, the second hyperbolic tangent in equation~\eqref{eq:Br_parameter_plus} is approximately $1$. For $\Ksqx \gtrsim 5$, the first hyperbolic tangent is also approximately $1$, and $K_{Br}^{+}$ becomes
\begin{equation}
    K_{Br}^{+} \approx K_{y}^{+}.
    \label{eq:Br_Ky}
\end{equation}%
\noindent Hence, the amplification of the instability is mainly determined by $K_y^+$, and $h^+$ and $K_x^+$ have only a secondary effect. 

Depending on the value of $K_{Br}^{+}$, \cite{ggFTaC} hypothesised three regimes for the instability, as shown in figure~\ref{fig:sigmai}:
a low-permeability regime, $\KsqBr \lesssim 1$, where the instability would be weak and not expected to emerge in the flow; a high-permeability regime, $\KsqBr \gtrsim 2.2$, where the amplification approaches an asymptote and the instability would be fully developed; and an intermediate regime, where the instability would set in. 
\cite{Garcia-Mayoral2011} found that, for riblets, the instability sets in for amplifications of approximately half the maximum. Following this, \cite{ggFTaC} hypothesised that the intermediate regime would occur for $\Ksqy \approx \KsqBr = 1-2.2$, and the linear drag reduction of equation~\eqref{eq:DR_K} could only hold for lower values of $\Ksqy$.
This hypothesis will be re-assessed based on the present DNS results in \S\ref{subsec:DNS_LSA}.

\subsection{Theoretical prediction of drag-reducing curves}

Combining the information on the linear drag reduction of equation~\eqref{eq:DR_K} and the range of $\Ksqy$ for the onset of Kelvin-Helmholtz rollers, the trend of the drag reduction curves for anisotropic permeable substrates can be estimated \citep{Abderrahaman2017,ggFTaC}.
An optimal substrate should seek to maximise the difference $\Ksqx - \Ksqz$ to obtain a large slip effect, while maintaining $\Ksqy$ as low as possible to inhibit the appearance of Kelvin-Helmholtz rollers.
Fibrous substrates as those proposed in figure~\ref{fig:materials}(\textit{b}) would conform such a material, for instance.

In this study, a substrate configuration will be represented by three dimensionless parameters; the anisotropy ratios $\phi_{xy} = \sqrt{K_x/K_y}$ and $\phi_{zy} = \sqrt{K_z/K_y}$, and the dimensionless thickness, $h/\Ksqynp$.
Given that both $K_y^+$ and $K_z^+$ have an adverse effect on the drag, in what follows we consider materials with preferential permeability in $x$ and equally low permeabilities in $y$ and $z$, $K_x^+ > K_z^+ = K_y^+$.
This implies $\phi_{xy} > 1$ and $\phi_{zy}=1$. In addition, we consider deep substrates with large $h/\Ksqynp$, so that the substrate thickness does not affect the overlying flow. In \S\ref{sec:results_DNS}, we study
substrates with $\Ksqx \lesssim 10$, so a thickness $h^+ \gtrsim 50$ would suffice. In practical aeronautic applications, for instance, this would imply permeable layers with sub-millimetre thickness.

For substrates with $\phi_{zy} = 1$, the expression for $\Delta U^+$ in equation~\eqref{eq:DR_K} becomes
\begin{equation} \label{eq:DR_K2}
    \Delta U^+ = \left( \phi_{xy} - 1 \right) \sqrt{K_y^+}.
\end{equation}

\begin{figure}
		\centering
		    \includegraphics{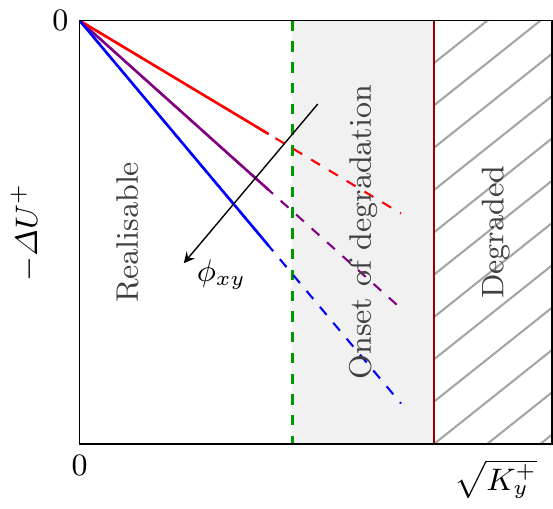}
  		    \mylab{-5.9cm}{5.cm}{(\aaa)}%
  		    \hspace{0.3cm}%
  		    \includegraphics{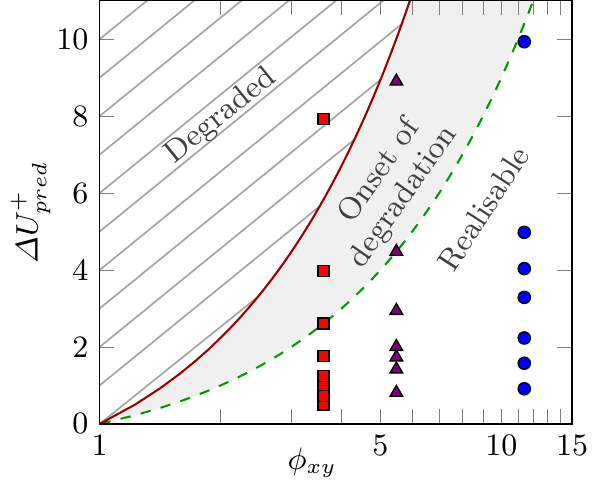}
  			\mylab{-6.cm}{5.cm}{(\bbb)}%
  		\caption{(\aaa) Sketch of the predicted $\Delta U^+$ as a function of $\protect\Ksqy$. Each line corresponds to a substrate with a different anisotropy ratio, $\phi_{xy} = \sqrt{K_x^+/K_y^+}$, and follows the behaviour of the linear expression~\eqref{eq:DR_K2}. The shaded region corresponds to the permeability values for which the Kelvin-Helmholtz rollers would be expected to develop, as in figure~\ref{fig:sigmai}. (\bbb) Predicted values of $\Delta U^+$ using the linear expression~\eqref{eq:DR_K2} as a function of the anisotropy ratio $\phi_{xy}$. In both panels, the dashed-green and solid-red lines define the limits for the onset of Kelvin-Helmholtz rollers estimated at $\protect\KsqBr |_{lim} \approx \protect\Ksqy \approx 1$ and $2.2$, and they separate three regions: the empty-colored one, where no Kelvin-Helmholtz instability would be expected; the shaded one, where the instability would set in; and the hatched one, where the instability would be fully developed. Symbols represent the DNS cases studied in \S\ref{sec:results_DNS} for three substrates with different anisotropies, from red to blue $\phi_{xy} \approx 3.6$, $5.5$ and $11$.}
	    	\label{fig:LSA}	
\end{figure}

\noindent The drag reduction for a given substrate configuration (i.e. for a fixed $\phi_{xy}$) can then be expressed solely as a function of the wall-normal permeability lengthscale, $\Ksqy$, which can be interpreted as a substrate Reynolds number, as sketched in figure~\ref{fig:LSA}(\textit{a}).
In a wind-tunnel experiment, for instance, $\Ksqy$ could be changed by changing the friction velocity, while $\phi_{xy}$ remained unaltered for a given substrate.

From the present analysis, the resulting drag reduction curves would be analogous to those for riblets \citep{Bechert1997,Garcia-Mayoral2011b}.
The curves would exhibit a linear increase of $\Delta U^+$ with $\Ksqy$,
breaking down no later than in the shaded region in the figure, 
where the onset of the Kelvin-Helmholtz instability would be expected. 
According to equation~\eqref{eq:DR_K2}, the slope in the linear regime is predicted to depend on $\phi_{xy}$, and the maximum $\Delta U^+$ for a given $\phi_{xy}$ would be determined by the intercept of the corresponding curve with the shaded region. 
The exact value of $\Ksqy$ for the breakdown, as well as the form of the curves in its proximity and for larger values will be obtained from the DNSs presented in \S\ref{sec:results_DNS}.

The ideas illustrated in figure~\ref{fig:LSA}(\textit{a}) for a few substrate configurations can be summarised for a wide range of anisotropy ratios, as is done in figure~\ref{fig:LSA}(\textit{b}). Following a drag reduction curve as $\Ksqy$ increases in figure~\ref{fig:LSA}(\textit{a}) would be equivalent to ascending vertically along a constant-$\phi_{xy}$ line in figure~\ref{fig:LSA}(\textit{b}).
The linear drag-reducing behaviour of equation~\eqref{eq:DR_K2} is expected to begin to fail in the shaded region.
This shaded region represents the permeability values for which the drag-degrading Kelvin-Helmholtz rollers are expected to appear and is the same as in figure~\ref{fig:LSA}(\textit{a}).
It is determined by introducing the limiting values of $\Ksqy$ specified in \S\ref{subsec:LSA}, $\Ksqy |_{lim}  \approx 1-2.2$, into equation~\eqref{eq:DR_K2}.
Although additional adverse phenomena cannot be ruled out, figure~\ref{fig:LSA}(\textit{b}) allows us to bound the parameter space for realisable drag reduction.
This figure was used to select the region in the parametric space subsequently investigated in \S\ref{sec:results_DNS}.
Most cases studied are in the drag-reducing region, where equation~\eqref{eq:DR_K2} is expected to hold, and a few cases have been selected in the shaded region, to capture the breakdown. 
We have considered three substrate configurations, $\phi_{xy}=\sqrt{13}$, $\sqrt{30}$ and $\sqrt{130}$, represented by the three vertical lines of symbols in figure~\ref{fig:LSA}(\textit{b}), and simulated them at different substrate Reynolds numbers, $\Ksqy$, so that complete drag reduction curves could be obtained.


\subsection{Change in drag in internal and external flows} \label{subsec:Cf}

The expressions for $\Delta U^+$ of equations~\eqref{eq:DR_gg} and \eqref{eq:DR_K} are valid only for external flows with mild or zero pressure gradients, where the flow near the wall is essentially driven by the overlying shear and the effect of the mean pressure gradient within the permeable substrate is negligible. 
We are mainly interested in vehicular applications, where the flow falls into that category, but for completion let us discuss the differences with internal flows. In the latter,
the effect of the mean pressure gradient could be significant.
This effect is essentially additive, so in the following discussion we will leave out turbulence for simplicity, and consider the laminar case.

\begin{figure}
		\centering
	\includegraphics[width=0.99\textwidth]{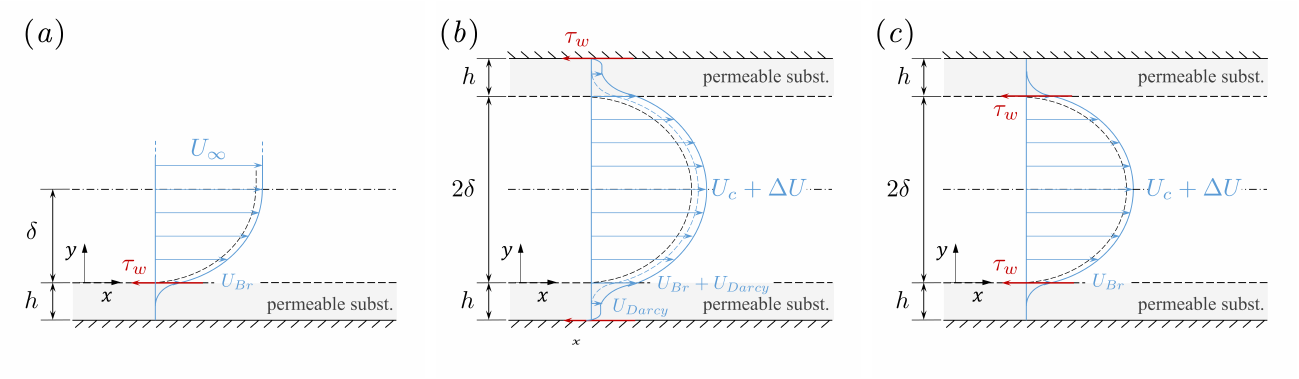}
  		\caption{Mean velocity in internal and external flows. The black-dashed line represents the mean velocity profiles for smooth walls. (\aaa) Boundary layer. (\bbb) Channel flow, where the mean pressure gradient is applied through the whole section of height $2 (\delta + h)$, including the permeable substrates. (\ccc) Artificial internal setup to produce only slip that appears in an external flow, by not applying the mean pressure gradient in the substrate regions.}\label{fig:Px}
\end{figure}

Sketches of the mean velocity profiles in a boundary layer and in a channel are depicted in figures~\ref{fig:Px}(\textit{a}) and \ref{fig:Px}(\textit{b}), respectively.
In a boundary layer over a permeable substrate, there would be a slip velocity at the interface, $U_{Br}$, due solely to the formation of a Brinkman layer within the substrate. 
It follows from equation~\eqref{subeq:SLa} that, provided that the substrate is sufficiently deep, this slip velocity is $U_{Br}^+ \approx \Ksqx$. Compared with a smooth wall, the only change in the mean velocity profile would be a shift by $U_{Br}$, that is $\Delta U^+ \approx \Ksqx$, and the drag reduction experienced would arise entirely from this slip effect.


In channel flows, there are two limiting forms of applying the permeable substrates to the reference smooth channel of height $2 \delta$. They can substitute a layer of solid material, increasing the height to $2(\delta +h)$, or they can be placed on top of the reference smooth channel, reducing the free flow area.
In the first case, depicted in figure~\ref{fig:Px}(\textit{b}),
the mean pressure gradient acts on the region $2 (\delta + h)$, which includes the permeable substrates.
This produces two opposite effects on the drag: a positive effect due to an increment in the flow rate, not only within the substrate but also in the channel core, and a negative effect due to the pressure gradient being applied across a larger cross-section.
In order to evaluate these two effects, we compare the friction coefficient for the permeable and the smooth channel under equal mean pressure gradient $P_x$.
The integral force balance yields 
\begin{equation}
	\tau_{w} = - P_x \left( \delta + h \right),
	\label{eq:stream_momen}
\end{equation} 
\noindent where $\tau_w$ accounts for the net force applied on the substrates. 
As we are now solely considering internal flows, we use the conventional bulk velocity, $U_b$, to define $c_f$, 
\begin{equation}
	c_f = 2 \frac{\tau_{w}}{U_{b}^2} = 2 \frac{\tau_{w}}{\left( U_{b0} + \Delta U_{b} \right)^2} = c_{f0} \frac{1 + h/ \delta}{\left( 1 + \Delta U_{b}/U_{b0}\right)^2} = c_{f0} \frac{\left( 1 + h/ \delta \right)^3}{\left( 1 + \Delta q/q_{0}\right)^2},
	\label{eq:cf}
\end{equation}
\noindent where the subscript `0' refers to the smooth channel. The friction coefficient of the smooth channel is therefore defined as $c_{f0} = - 2 P_x \delta / U_{b0}$, and $q= 2(\delta + h) U_b$ is the mass flow rate. The opposing effects of the increase in cross-section where the pressure gradient acts, $2h$, and the extra flow rate, $\Delta q$, are evidenced in equation~\eqref{eq:cf}. 
The result can be either a drag reduction or a drag increase depending on the values of $\Delta q$ and $h$. 

\begin{figure}
		\centering
  		    \includegraphics[width=0.55\textwidth]{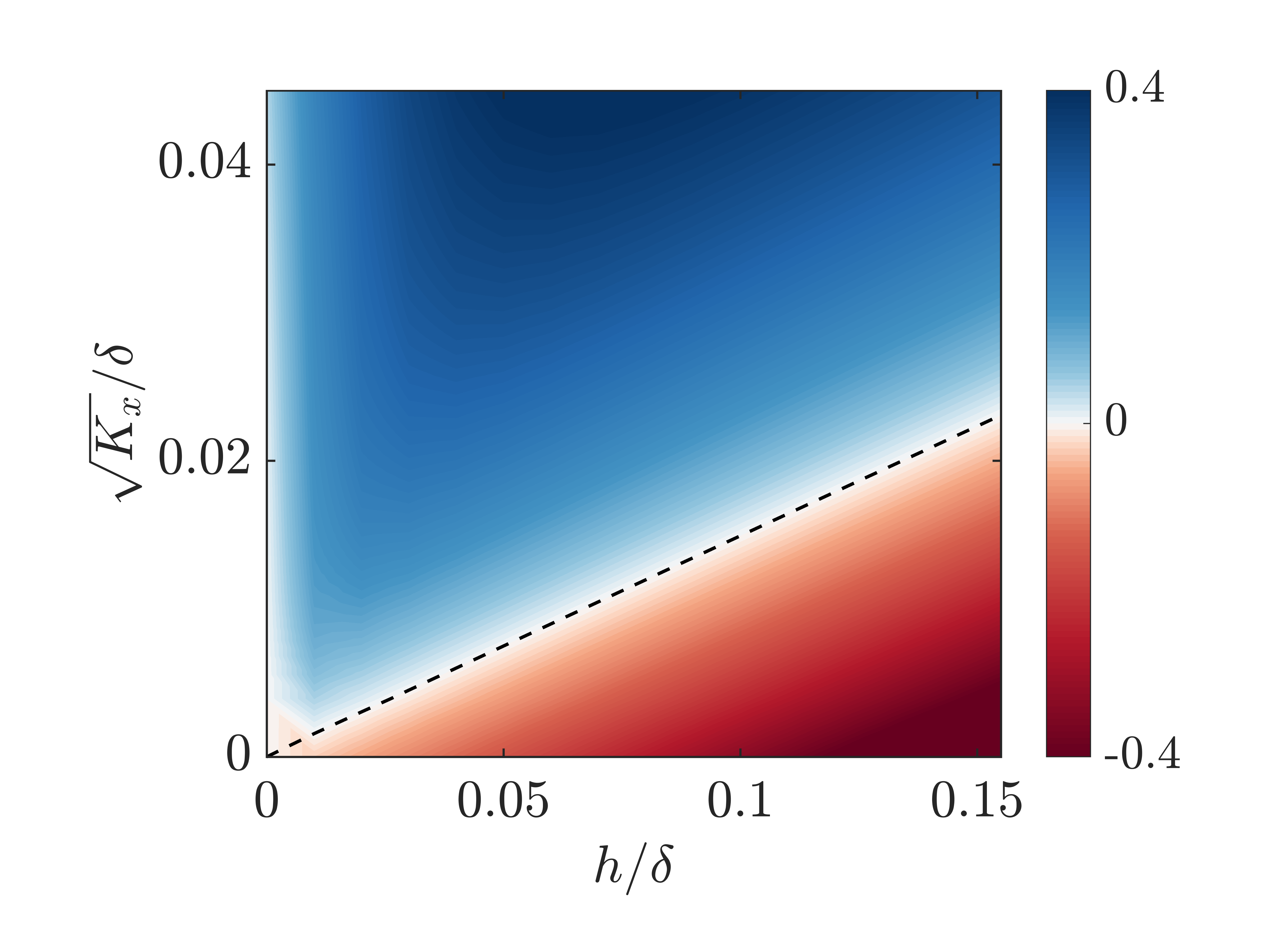}
  		    \mylab{-5cm}{4cm}{\textcolor{white}{drag reduction}}
  		    \mylab{-3.83cm}{1.4cm}{\textcolor{white}{drag increase}}
  		\caption{Map of $DR= - \Delta c_f / c_{f0}$ in an internal channel flow with permeable substrates as a function of the permeability length, $\sqrt{K_x}$, and the thickness of the substrate, $h$, for a friction Reynolds number $\delta^+ = 180$. The channel with substrates has a total height of $2(h+\delta)$, and is compared to a smooth channel of height $2 \delta$, as in figure~\ref{fig:Px}(\bbb). \broken, the first order approximation of zero drag reduction line, with a slope of $0.15$ obtained from equation~\eqref{eq:cf_linear} (valid for $h/\delta > 0.01$).}
	    	\label{fig:internalDR}	
\end{figure}

The extra flow rate, $\Delta q$, can be expressed in terms of $K_x$. From figure~\ref{fig:Px}(\textit{b}), $ \Delta q$ is
\begin{equation}
	\Delta q \approx 2 \delta \left( U_{Br} + U_{Darcy} \right) + 2 q_{substrate},
	\label{eq:q_int}
\end{equation}
\noindent where, in addition to the slip velocity caused by the overlying shear, $U_{Br}$, as in figure~\ref{fig:Px}(\textit{a}), there is an extra slip velocity caused by the mean pressure gradient, $U_{Darcy}$, and a resulting extra flow rate within the substrate, $q_{substrate}$.
The former is obtained from Darcy's law, $U_{Darcy} = - P_{x} K_x / \nu$, and $q_{substrate}$ is obtained using $U_{Darcy}$ and Brinkman's velocity within the substrate, as defined by expression~\eqref{subeq:Br_0_0_1}.
Substituting expression~\eqref{eq:q_int} into equation~\eqref{eq:cf}, the resulting change in $c_f$ depends on $\sqrt{K_x}/h$, $h/ \delta$ and the Reynolds number. This dependency for a friction Reynolds number $\delta^+ = 180$ is illustrated in figure~\ref{fig:internalDR}. 
The figure shows how the beneficial effect of adding a streamwise permeability is opposite to the deleterious effect of the increased area and, for certain substrate geometries, can even outweigh it, resulting in a net drag reduction.
Note that in the turbulent case the effect of the spanwise, Brinkman contribution would also need to be included, as given by equation~\eqref{eq:DR_K}.

To better understand the relationship between $K_x$ and $h$, expression~\eqref{eq:cf} can be simplified further for $\delta \gg h \gtrsim \sqrt{K_x}$.
The extra flow rate is then dominated by $\Delta q \approx 2 \delta U_{Br} \approx 2 \delta \sqrt{K_x} \left. dU/dy \right|_{y=0}$, and in a first order approximation, equation~\eqref{eq:cf} simplifies to 

\begin{equation}
	c_f \approx c_{f0} \frac{ 1 + 3 h/ \delta}{ 1 + \dfrac{2 \sqrt{K_x}}{ U_{b0}} \left. \dfrac{dU}{dy} \right|_{y=0} }.
	\label{eq:cf_linear}
\end{equation}
\noindent It follows from this equation that, in $(h,\sqrt{K_x})$ parameter space, the isocontours of $DR$ are approximately oblique straight lines, as observed in figure~\ref{fig:internalDR}, and specifically the neutral drag curve is $\sqrt{K_x} = 3/2 \, \, U_{b0} / (dU/dy |_{y=0}) \, \, h$, which depends on the friction Reynolds number through $U_{b0}$ and $dU/dy |_{y=0}$.
For $\delta^+ = 180$, the zero drag reduction line is given by $\sqrt{K_x} \approx 0.15 h$, as indicated in figure~\ref{fig:internalDR}, while for $\delta^+ = 5000$, $\sqrt{K_x} \approx 0.007 h$.

The above analysis applies to channel flows where the permeable substrates substitute a layer of solid material and shows that, in this case, the drag can be either reduced or increased. If, on the other hand, the permeable coating was added on top of an existing smooth channel, 
the pressure gradient would still be applied over the whole height of $2 \delta$, which includes the permeable coatings, and the resulting friction coefficient would be $c_f = c_{f0} / (1 + \Delta q/q_{0})^{2}$.
In this case, the flow rate would always decrease, $\Delta q <0$, resulting in an increase of drag independently of the values of $\sqrt{K_x}$ and $h$.

\section{DNS setup}\label{sec:numerics} 
In this section, we present the numerical setup for direct numerical simulations of the domain sketched in figure~\ref{fig:Px}(\ccc), a channel of height $2\delta$ delimited by two identical anisotropic permeable substrates.
The presence of the substrates is taken into account through the boundary conditions defined by equation~\eqref{eq:BC3D}, as in the stability analysis in \S\ref{subsec:LSA}. 


\subsection{The numerical method} \label{subsec:num_method} 
 
The channel flow is governed by the incompressible Navier-Stokes equations,

\begin{subequations}\label{eq:ch_NS}
	\begin{gather}
		\bnabla \bcdot \boldsymbol{u} =0, \label{eq:ch_NS_1}\\
		\frac{\partial \boldsymbol{u}}{\partial t} + \boldsymbol{u} \bcdot \bnabla \boldsymbol{u} = - \nabla p + \frac{1}{Re} \nabla^2 \boldsymbol{u},	
		\label{eq:ch_NS_2}
	\end{gather}	
\end{subequations}
\noindent where the density has been assumed to be unity for simplicity,
$p$ is the pressure, $\boldsymbol{u}= (u,v,w)$ the velocity vector and $Re$ the bulk Reynolds number defined as $Re = U_b \delta / \nu$, with $U_b$ being the bulk velocity in the channel region. 
The DNS code is adapted from \cite{Garcia-Mayoral2011} and \cite{Fairhall2018} and was already used in \cite{ggFTaC}. It solves the incompressible Navier-Stokes equations~\eqref{eq:ch_NS} in a doubly-periodic channel of height $2 \delta$, where $\delta=1$ is the distance between the substrate-channel interface and the centre of the channel.
All simulations are conducted at a fixed friction Reynolds number $Re_{\tau}= u_{\tau} \delta / \nu = 180$ by imposing a constant mean pressure gradient in $y \in [0, 2 \delta]$. 
The kinematic viscosity is $\nu = 1/2870$ and we use a smooth-wall channel with the same mean pressure gradient as reference.

Although for convenience the present DNSs are conducted in channels, our scope of application is mainly external flows with mild pressure gradients. In a channel, in comparison, there would be an additional flow rate from Darcy's contribution discussed in \S\ref{subsec:Cf}. To allow direct extrapolation to external flows, we simply do not include this contribution when implementing the boundary conditions on the mean flow, that is, mode $(0,0)$, which would be the only Fourier mode affected. This numerical artefact would be equivalent to applying the mean pressure gradient in the channel region only, as depicted in figure~\ref{fig:Px}(\textit{c}). The drag reduction in the present simulations results then entirely from the slip velocity due to an overlying shear, as in external flows.


The spatial discretisation is spectral in the wall-parallel directions $x$ and $z$, with $2/3$ rule de-aliasing, and uses second-order centred finite differences on a staggered grid in the wall-normal direction.  
The computational domain is of size $2\pi \times \pi \times 2$ in the streamwise, spanwise and wall-normal directions, respectively. A grid with $192 \times 192 \times 153$ collocation points with grid stretching is used, which in viscous units gives a resolution of $\Delta x^{+} \approx 5.9$, $\Delta z^{+} \approx 2.9$, and $\Delta y^{+} \simeq 0.3$ near the wall and $\Delta y^{+} \simeq 3$ in the centre of the channel.
For the temporal integration, 
a Runge-Kutta discretisation is used, where 
every time-step is divided into three substeps,
each of which uses a semi-implicit scheme for the viscous terms and an explicit scheme for the advective terms.
Discretised this way, the Navier-Stokes equations in~\eqref{eq:ch_NS} result in
	\begin{equation}
	\begin{split}
		\left[ I - \Delta t \frac{\beta_k}{\mathrm{Re} \mathrm{L}} \right] \boldsymbol{u}^{k} = \boldsymbol{u}^{k-1} + \Delta t \Big[ & \frac{\alpha_k}{\mathrm{Re}} \mathrm{L}(\boldsymbol{u}^{k-1}) - \gamma_k \mathrm{N}(\boldsymbol{u}^{k-1}) \\
		& - \zeta_k \mathrm{N}(\boldsymbol{u}^{k-2}) - \left( \alpha_k + \beta_k \right) \mathrm{G}(p^{k}) \Big] 
	\end{split}
	\label{eq:NS_RKl}
	\end{equation}	
\noindent where $\mathrm{L}$, $\mathrm{G}$ and $\mathrm{D}$ represent the discretised laplacian, gradient and divergence operators, respectively, and $\mathrm{N}$ represents the nonlinear, advective operator. The superscript $k=1,2,3$ denotes the Runge-Kutta substep. Hence, the velocities $\boldsymbol{u}^{0}$ and $\boldsymbol{u}^{3}$ correspond to the velocities at time-step $n$ and $n+1$, respectively. Additionally, $\alpha_k$, $\beta_k$, $\gamma_k$ and $\zeta_k$ are the Runge-Kutta coefficients for substep $k$ from \cite{Le1991}.
In equation~\eqref{eq:NS_RKl}, the velocity at substep $k$ is expressed in terms of the velocities at the previous substeps, as well as the pressure at that same substep $k$.
To solve it, a fractional step method is integrated in each substep \citep{Le1991}.

The presence of the permeable substrates is accounted for by the boundary conditions~\eqref{eq:BC3D}, and the coupling between the velocities and the pressure at the interface is implemented implicitly. 
Following \cite{Perot1993,Perot1995}, the discretised incompressible Navier-Stokes equations from \eqref{eq:NS_RKl} can be represented in matrix form,
\begin{equation}
\left[
\begin{array}{cc}
  \mathrm{A} & \mathrm{G}  \\
  \displaystyle
  \mathrm{D} & 0 \\
\end{array}  \right]
\left(
\begin{array}{c}
  \mathbf{u}^{k}   \\
  \displaystyle
  p^{k}  \\
\end{array}  
\right)
= 
\left(
\begin{array}{c}
  r^{k-1}   \\
  \displaystyle
  0  \\
\end{array}  
\right),
\label{eq:FM_1}
\end{equation}
\noindent where $\mathbf{u}^{k}$ and $p^{k}$ are the discrete velocity and pressure unknowns, respectively. $\mathrm{A}$ is the operator containing the implicit part of the diffusive terms, which for the internal points of the domain equation~\eqref{eq:NS_RKl} gives $\mathrm{A} = [ I - \Delta t \frac{\beta_k}{\mathrm{Re} \mathrm{L}} ]$, and
the vector $r^n$ is the explicit right-hand side, which contains all the quantities from previous time-steps.
The boundary conditions given by equations~\eqref{eq:BC3D} are embedded in the block matrices in equation~\eqref{eq:FM_1}. The relationships between the three velocities and the shears $d \hat{u}/dy$ and $d \hat{w}/dy$ are embedded in $\mathrm{A}$, while the coupling between the velocities and pressure is embedded in $\mathrm{A}$ and $\mathrm{G}$.
Taking then the LU decomposition of system~\eqref{eq:FM_1} results in
\begin{equation}
\left[
\begin{array}{cc}
  \mathrm{A} & 0  \\
  \displaystyle
  \mathrm{D} & -\mathrm{D}\mathrm{A^{-1}}\mathrm{G} \\
\end{array}  \right]
\left[
\begin{array}{cc}
  \mathrm{I} & \mathrm{A^{-1}} \mathrm{G}  \\
  \displaystyle
  0 & \mathrm{I} \\
\end{array}  \right]
\left(
\begin{array}{c}
  \mathbf{u}^{k}   \\
  \displaystyle
  p^{k}  \\
\end{array}  
\right)
= 
\left(
\begin{array}{c}
  r^{k-1}   \\
  \displaystyle
  0  \\
\end{array}  
\right)
\label{eq:FM_2}
\end{equation}
\noindent and the operations are solved in the following order
\begin{subequations}\label{eq:FM_eqs}
	\begin{gather}
		\mathrm{A} \mathbf{u^*} =  r^{k-1}, \label{eq:FM_eqs_1}\\
		\mathrm{D} \mathrm{A^{-1}} \mathrm{G} p^{n+1} = \mathrm{D} \mathbf{u^*}, \label{eq:FM_eqs_2}\\
		\mathbf{u}^{k} = \mathbf{u^*} - \mathrm{A^{-1}} \mathrm{G} p^{k}, \label{eq:FM_eqs_3}
	\end{gather}	
\end{subequations}
\noindent where the variable $\mathbf{u^*}$ is an intermediate, non-solenoidal velocity.
The Poisson equation in equation~\eqref{eq:FM_eqs_2} is computationally expensive, as it requires the inversion of matrix $\mathrm{A}$.
For efficiency, $\mathrm{A^{-1}}$ is generally approximated to its first order term, $ \approx \mathrm{I}$ \citep{Perot1993}.
In the present work, we approximate the internal points in $\mathrm{A}$ by $\approx  \mathrm{I}$, while keeping the rows of $\mathrm{A}$ that contain the boundary conditions unchanged, and then invert the resulting matrix to obtain $\mathrm{A}^{-1}$. 

Statistics are obtained by averaging over approximately $100$ eddy-turnovers, once the statistically steady state has been reached.
Statistical convergence was verified using the criterion of \cite{Hoyas2008}.

\subsection{Validation} \label{subsec:validation} 

\begin{table}
 \begin{center}
  \begin{tabular}{lccccccccc}
    Cases  & $\epsilon$ & $\tilde{\nu}/\nu$ & $\delta^+$   &   $K^+$ & $\delta_w / \delta$ & $\delta_i / \delta$ & $c_{f} (\times 10^{-3})$ & $c_{f_{0}} (\times 10^{-3})$ & $\Delta D$ \\[3pt]
       BB\_E80    & 0.8 & -   & 203 & 1.14 & 1.12 & 0.04 & 10.41 & 8.19 & 27.15\\
       BB\_Br     & -   & 1.0 & 204 & 1.19 & 1.11 &  -   & 10.34 & 8.07 & 28.06\\
  \end{tabular}
  \caption{Characteristics of the simulations for VANS approach (BB\_E80) and Brinkman's (BB\_Br). Porosity, $\epsilon$; viscosity ratio $\tilde{\nu} / \nu$; friction Reynolds number, $\delta^+ = u_{\tau} \delta / \nu$; permeability, $K^+ = K u_{\tau}^{2} / \nu^2$; location of zero total stress, $\delta_w$; friction coefficient, $c_f = 2 (u_{\tau} / U_b)^2$; friction coefficient of the corresponding smooth channel, $c_{f_{0}}$; and change of drag defined as $\Delta D  = (c_{f} - c_{f_{0}})/c_{f_{0}}$. Viscous units are defined with $u_{\tau}$ measured at the interface plane $y=0$.}\label{tab:BB}
 \end{center}
\end{table}

\begin{figure}
		\centering
  		    \includegraphics[width=1.0\textwidth,trim={0 5.5cm 0 5.8cm},clip]{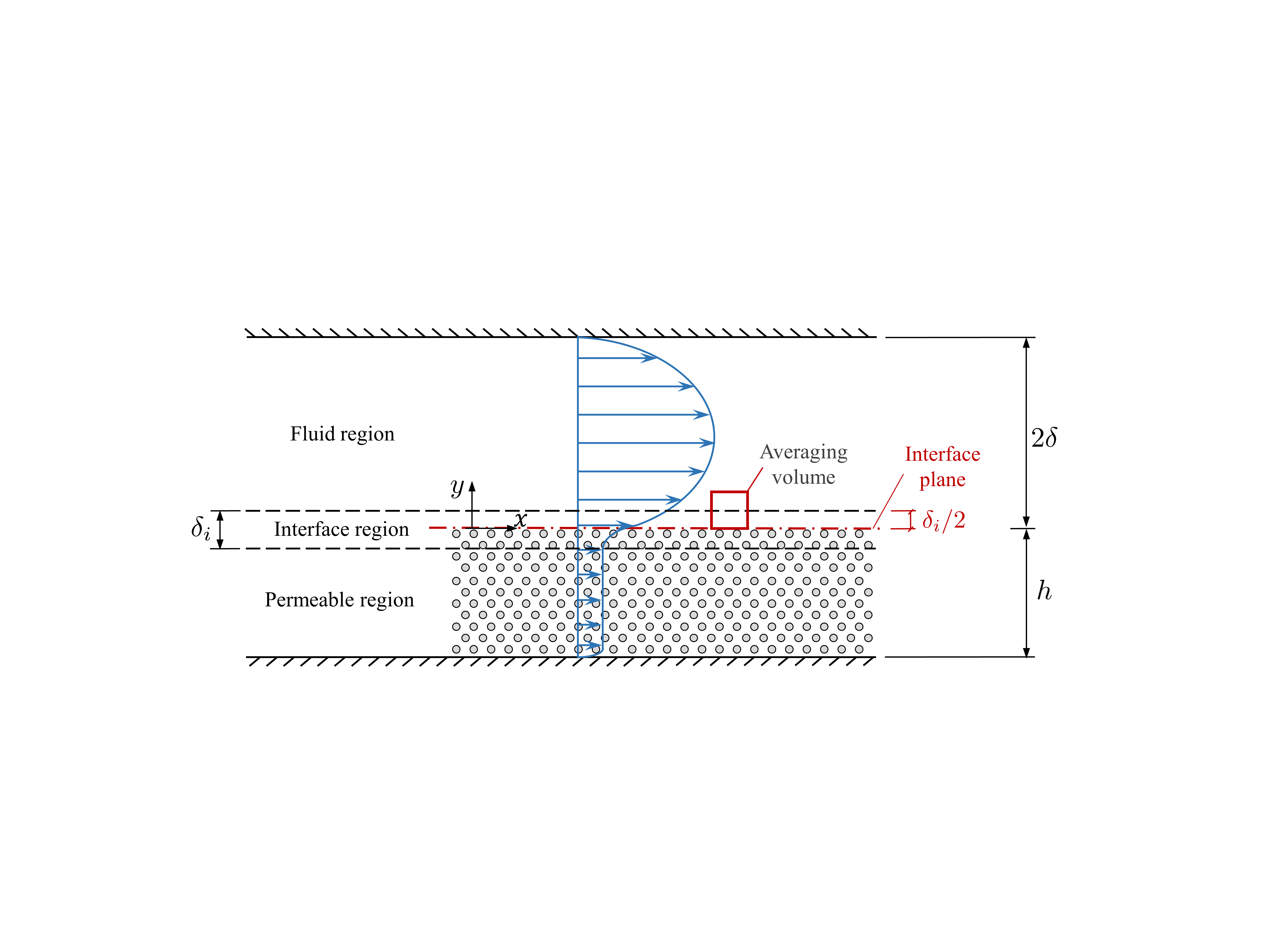}
  		\caption{Sketch of the channel of \cite{BreugemBoersma2006} used here for validation. The red dashed-dotted line corresponds to the location of the interface plane used in the present analysis for comparison with the analogous Brinkman model, BB\_Br.}
	    	\label{fig:BB_geom}	
\end{figure}

We validate the present Brinkman model with one of the cases studied by \cite{BreugemBoersma2006}, where the authors used the VANS equations within the permeable substrate. 
We consider their case E80, here referred to as BB\_E80, with a porosity $\epsilon =0.8$ -- which refers to the ratio between the void volume and the total volume of the substrate -- and an isotropic permeability $K^+ \approx 1$.
This permeability is of the same order of magnitude as our largest permeabilities $K_y^+$ and $K_z^+$ in the DNSs presented in \S\ref{sec:results_DNS}. The case BB\_E80 is compared to our Brinkman model, here referred to as BB\_Br, using approximately the same value of permeability, $K^+ \approx 1$.

To match the validation domain in \cite{BreugemBoersma2006}, we use an asymmetric channel of height $2 \delta$, delimited by an impermeable wall at the top and a permeable substrate at the bottom, as sketched in figure~\ref{fig:BB_geom}.
The thickness of the permeable layer is $h = 2 \delta$.

\cite{BreugemBoersma2006} used a VANS approach to model the flow within the substrate. At the interface with the free flow, they let the porosity, and hence the permeability, to evolve gradually from the inner value $\epsilon = 0.8$ to the free flow value $\epsilon = 1$ over a thin interfacial layer of thickness $\delta_i$. This corresponded to the averaging volumes in VANS capturing varying proportions of free flow and substrate, so that the volumes centred at the top of the interfacial layer did not contain any substrate, and vice versa, as illustrated in figure~\ref{fig:BB_geom}. This model is consistent with applying VANS on a setup with a sharp interface half-way through the interfacial layer. We set our reference plane $y=0$ at this height, so for comparison we represent the results from \cite{BreugemBoersma2006} in the same frame of reference. Note that in \cite{BreugemBoersma2006} the reference plane was at the top of the interfacial region instead. For a consistent comparison, the results from \cite{BreugemBoersma2006} have been rescaled with the friction velocity measured at our $y=0$ and with the bulk velocity integrated between that plane and the top impermeable smooth wall.

\begin{figure}
    \centering
        \includegraphics{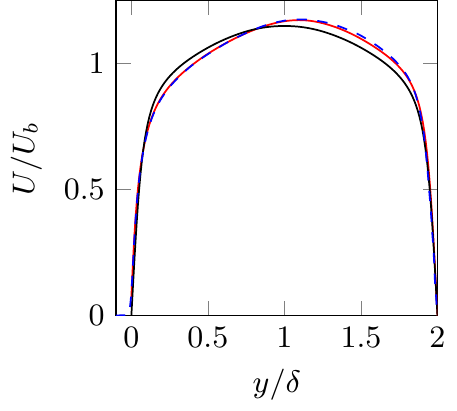}%
        \mylab{-3.4cm}{3.85cm}{(\aaa)}%
        \includegraphics{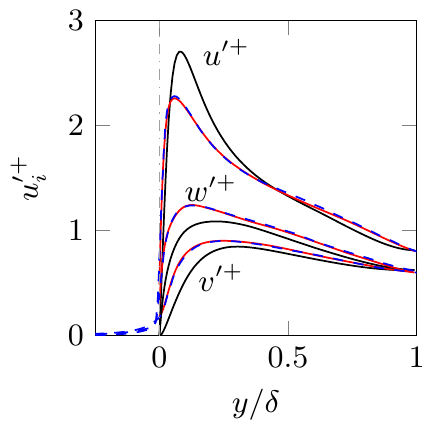}%
        \mylab{-3.4cm}{3.85cm}{(\bbb)}%
        \includegraphics{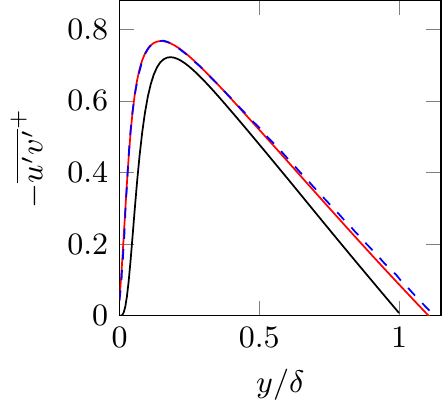}%
        \mylab{-3.22cm}{3.85cm}{(\ccc)}%
\caption{Comparison of a simulation from \cite{BreugemBoersma2006} using VANS (BB\_E80), \textcolor{blue}{\broken}, with a corresponding simulation using Brinkman's model (BB\_Br), \textcolor{red}{\full}. The curves from \cite{BreugemBoersma2006} are shifted by $\delta_i/2$ to match the substrate-channel interface in both setups. Black lines represent smooth-channel data for reference. (\aaa) Mean velocity profile, (\bbb) rms velocity fluctuations, (\ccc) Reynolds stress.}\label{fig:BB}
\end{figure}

Both BB\_Br and the original simulation of \cite{BreugemBoersma2006} were run at a constant mass flow rate started from a smooth channel, at $Re = 2750$ in the latter case and $Re = 2832$ in ours. Defining viscous units using the friction velocity at $y=0$, the initial friction Reynolds numbers were $\delta^+ = 176$ and  $\delta^+ = 180$, respectively, while in the final statistically-steady state they were $\delta^+ \approx 204$ and $\delta^+ \approx 203$, respectively.
Results from Breugem \& Boersma's VANS approach (BB\_E80) and Brinkman's model under study (BB\_Br) are compared in table~\ref{tab:BB} and figure~\ref{fig:BB}, all showing good agreement. 

This agreement between VANS and Brinkman's approach could be expected, given the similarities between the models for the values of the parameters considered.
VANS equations can be viewed as Brinkman's equation with the addition of the advective and temporal terms, with $\epsilon$ playing in the former the role that $\nu / \tilde{\nu}$ plays in the latter.
For small permeabilities, such as those under consideration, the advective terms can be neglected.
In addition, an order of magnitude analysis shows that the temporal term can also be neglected. This term is of the order $\sim O [u_{c}/t_c]$, where $t_c$ and $u_c$ denote a characteristic time and velocity, respectively. 
When the substrate is isotropic and highly-connected (i.e. $\tilde{\nu} \approx \nu$), both the Brinkman and Darcy terms are of the same order of magnitude, as the penetration length in an isotropic medium is of order $\sim \sqrt{K}$.
Comparing the temporal and the Brinkman terms then, we obtain
\begin{equation}
	\frac{u_c / t_c}{\nu u_c / K} \sim \frac{K}{\nu t_c} = \frac{K^+}{t_c^+}.
\end{equation} 
\noindent For the temporal term to be negligible, the characteristic time should satisfy $t_c^+ > K^+$.
Considering that the fastest-evolving turbulent structures near the wall are the quasi-streamwise vortices, with a radius $r^+ \sim 15$ and velocity $\sim u_{\tau}$, the smallest characteristic timescale would be $t_c^+ \sim 15$, and given that $K^+ \sim 1$, the condition $t_c^+ > K^+$ is satisfied. 
The flow within the permeable medium can then be assumed to be quasi-steady.
Additionally, for VANS and Brinkman's equation to be equivalent, the value of the porosity $\epsilon$ should be equal to the ratio $\nu / \tilde{\nu}$. In these simulations, these values differ slightly, $\epsilon = 0.8$ in BB\_E80 and $\nu / \tilde{\nu} = 1$ in our model.
Nonetheless, \cite{Rosti2015} reported that, for porosity values beyond $\epsilon \gtrsim 0.6$, a further increase of $\epsilon$ had no significant effect on the overlying flow, and the permeability $K$ was the only key parameter. This justifies the similarities between the results of the two models, even if the values of $\epsilon$ and $\nu /\tilde{\nu}$ are not exactly matching.

\section{Results and discussion for DNS}\label{sec:results_DNS} 

\begin{table}
 \begin{center}
  \begin{tabular}{lccccccccc}
    & Cases & $\sqrt{K_x^+}$ & $\sqrt{K_y^+}$ & $\sqrt{K_z^+}$ & $h^+$ & $U_b/U_{b_{sm}}$ & $\Delta U^+$ & $DR_{180}$  & $DR_{5000}$\\[2pt]
    \hline
    Smooth &              & ~~0 & 0~~~~~  & 0~~~~~ & ~0 & 1.0~~ & -     &   -    &   -    \\
    \hline
    \multirow{6}{*}{$\phi_{xy} = \sqrt{\dfrac{K_x}{K_y}} \approx 3.6$} & A1     & ~~0.71 & 0.20 & 0.20 & ~19.5 & 1.037 & ~~0.51 & ~~~5.64 & ~~~3.93 \\
       & A2     & ~~1.00 & 0.28  & 0.28 & ~28.1 & 1.045 & ~~0.68 & ~~~7.26 & ~~~5.08 \\
       & A3     & ~~1.42 & 0.39  & 0.39 & ~38.8 & 1.052 & ~~0.80 & ~~~8.44 & ~~~5.92 \\
       & A4     & ~~1.74 & 0.48  & 0.48 & ~48.1 & 1.041 & ~~0.54 & ~~~6.10 & ~~~4.25 \\
       & A5     & ~~2.45 & 0.68  & 0.68 & ~68.1 & 0.963 & ~-0.68 & ~~-7.38 & ~~-4.99 \\
       & A6    & ~~3.61 & 1.00  & 1.00 & 100.2 & 0.819 & ~-3.02 & ~-42.31 & ~-26.58 \\
       & A7    & ~~5.50 & 1.52  & 1.52 & 152.7 & 0.616 & ~-6.59 & -143.84 & ~-76.46 \\
       & A8   & ~~10.97 & 3.04  & 3.04 & 304.2 & 0.381 & -11.03 & -546.15 & -194.20 \\  
      \hline 
      \multirow{6}{*}{$\phi_{xy} = \sqrt{\dfrac{K_x}{K_y}} \approx 5.5$} & B1     & ~~1.00 & 0.18  & 0.18 & ~18.0  & 1.053 & ~0.84 & ~~~8.63 & ~~~6.06 \\
       & B2     & ~~1.79 & 0.32  & 0.32 & ~32.1  & 1.085 & ~1.29 & ~~12.71 & ~~~9.01 \\
	   & B3     & ~~2.12 & 0.39  & 0.39 & ~39.0  & 1.086 & ~1.31 & ~~12.93 & ~~~9.17 \\
       & B4     & ~~2.45 & 0.45  & 0.45 & ~45.0  & 1.070 & ~1.01 & ~~10.22 & ~~~7.20 \\
       & B5    & ~~3.61 & 0.66  & 0.66 & 65.7 & 0.979 & -0.46 & ~~-5.24 & ~~-3.56 \\
       & B6    & ~~5.48 & 1.00  & 1.00 & 100.0  & 0.792 & -3.66 & ~-56.35 & ~-34.47 \\
       & B7   & ~10.89 & 1.99  & 1.99 & 198.4  & 0.517 & -8.66 & -261.34 & -120.00 \\  
       \hline
       \multirow{6}{*}{$\phi_{xy} = \sqrt{\dfrac{K_x}{K_y}} \approx 11.4$} &  C1     & ~~1.00 & 0.09  & 0.09 & ~9.0 & 1.062 & ~0.98 & ~~~9.89 & ~~6.96 \\
       & C2     & ~~1.73 & 0.15  & 0.15 & 14.0 & 1.106 & ~1.67 & ~~16.01 & ~11.45 \\
       & C3     & ~~2.45 & 0.21  & 0.21 & 22.0 & 1.145 & ~2.24 & ~~20.63 & ~14.93 \\
       & C4    & ~~3.6 & 0.32  & 0.32 & 32.0 & 1.178 & ~2.84 & ~~25.10 & ~18.38 \\
       & C5    & ~~4.48 & 0.39 & 0.39 & 39.1 & 1.183 & ~2.87 & ~~25.34 & ~18.56 \\
       & C6    & ~~5.47 & 0.48  & 0.48 & 47.9 & 1.152 & ~2.34 & ~~21.38 & ~15.50 \\
       & C7   & ~10.89 & 0.96  & 0.96 & 95.6 & 0.898 & -2.21 & ~-29.35 & -18.92 \\   
       \hline
       \hline
    \multirow{4}{*}{$\dfrac{h}{\sqrt{K_x}} = 1.5$}
       & C$'$1     & ~2.45 & 0.21  & 0.21 & ~3.67 & 1.130 & ~2.00 & ~~18.74 & ~13.49 \\
       & C$'$2    & ~3.61 & 0.32  & 0.32 & ~5.40 & 1.171 & ~2.70 & ~~24.12 & ~17.62 \\
       & C$'$3    & ~5.49 & 0.48  & 0.48 & ~8.23 & 1.156 & ~2.40 & ~~21.87 & ~15.88 \\
       & C$'$4    & 10.84 & 0.95  & 0.95 & 16.26 & 0.962 & -0.90 & ~-10.84 & ~-7.27 \\
	\hline
    \multirow{5}{*}{$\dfrac{h}{\sqrt{K_x}} = 1.0$}
       & C$''$1    & ~3.61 & 0.32  & 0.32  & ~3.61 & 1.154 & ~2.42 & ~~22.02 & ~15.99 \\
       & C$''$2    & ~5.48 & 0.48  & 0.48 & ~5.51 & 1.163 & ~2.53 & ~~22.86 & ~16.64 \\
       & C$''$3    & ~7.01 & 0.62  & 0.62 & ~7.01 & 1.127 & ~1.90 & ~~17.93 & ~12.88 \\
       & C$''$4    & ~9.03 & 0.79  & 0.79 & ~9.03 & 1.066 & ~0.84 & ~~~8.62 & ~~6.05 \\
       & C$''$5    & 10.85 & 0.95  & 0.95 & 11.03 & 1.001 & -0.12 & ~~-1.32 & ~-0.91 \\
	\hline
	\multirow{6}{*}{$\dfrac{h}{\sqrt{K_x}} = 0.5$}
	   & C$'''$1   & ~~2.45 & 0.21  & 0.21 & 1.22 & 1.063 & ~0.93 & ~~9.46 & ~~6.65 \\
       & C$'''$2   & ~3.62 & 0.32  & 0.32 & 1.86 & 1.091 & ~1.36 & ~~13.35 & ~~9.48 \\
       & C$'''$3   & ~5.47 & 0.48  & 0.48 & 2.74 & 1.133 & ~2.04 & ~~19.11 & ~13.77 \\
       & C$'''$4   & ~7.01 & 0.62  & 0.62 & 3.50 & 1.153 & ~2.39 & ~~21.81 & ~15.83 \\
       & C$'''$5   & ~9.03 & 0.79  & 0.79 & 4.52 & 1.129 & ~1.95 & ~~18.34 & ~13.19 \\
       & C$'''$6   & 10.83 & 0.95  & 0.95 & 5.42 & 1.092 & ~1.30 & ~~12.88 & ~~9.13 \\
  \end{tabular}
  \caption{DNS parameters. $\protect\Ksqx$, $\protect\Ksqy$ and $\protect\Ksqz$ are the streamwise,
wall-normal and spanwise permeability lengths, $h^+$ is the thickness of the substrate, $\Delta U^+$ is the
shift of the velocity profile in the logarithmic region, and $DR_{180}$ and $DR_{5000}$ are the values of
drag reduction for $\delta^+=180$ and $\delta^+=5000$, respectively, obtained using expression~\eqref{eq:DR}.
The values $DR_{5000}$ have been calculated using the smooth-channel centreline velocity from \cite{LeeMoser2015}.
The first three substrate configurations A, B and C have thickness $h = 100 \sqrt{K_y}$ and different anisotropy ratios $\phi_{xy}$. The last three substrate configurations, C$'$, C$''$ and C$'''$, have $\phi_{xy} \approx 11.4$, same as substrate C, but different thickness $h / \sqrt{K_x}$.}\label{tab:cases}
 \end{center}
\end{table}


In this section, we present results from DNSs to investigate in detail the effect that permeable substrates have on the overlying flow and assess the validity of the predictions presented in \S\ref{sec:theory}.
We study three substrate configurations, given by three different anisotropy ratios $\phi_{xy} \approx 3.6$, $5.5$, $11.4$.
For our main set of simulations, the substrates have thickness $h = 100 \Ksqynp$, large
enough for the problem to become independent of it, and the same permeabilities in $y$ and $z$, $\phi_{zy}=1$.
An additional subset of simulations was conducted to explore the effect of a finite $h$ on the substrate
performance. For a given configuration (i.e. a fixed $\phi_{xy}$ and $h/\Ksqynp$), we vary proportionately the permeabilities in viscous units, $K_x^+$, $K_z^+$ and $K_y^+$, which is equivalent
to varying the viscous length. For each configuration, $\Ksqx$ varies between $0.7 - 11$. 
The simulations under study are summarised in table~\ref{tab:cases}, where each case is labelled with a
letter and a number. In the main set of simulations, the letter refers to the anisotropy
ratio $\phi_{xy}$ of the substrate and the number to the specific substrate, with fixed permeabilities in viscous units.
In the secondary set, an additional subscripts $'$, $''$ and $'''$ indicate decreasing substrate depth.

The virtual-origin model presented in \S\ref{sec:theory} is based on the idea that the near-wall cycle remains smooth-wall-like, other than by being displaced a depth $\ell_T$ towards the substrate.
Given that the origin perceived by turbulence is expected to be at $y =-\ell_T\approx-\sqrt{K_z}$,
throughout this section results are scaled taking that as the reference for the
wall-normal height. 
The friction velocity is obtained by extrapolating the total stresses to that height, $u_{\tau} = u_{\tau_{y=0}} \left( 1 + \sqrt{K_z}/\delta \right)^{1/2}$, and the effective half-height channel becomes $\delta' = \delta + \sqrt{K_z}$ \citep{RGM2018}, although the effect is negligible for the small values of $\sqrt{K_z}/\delta$ considered here.
Beyond the breakdown of the linear regime, the virtual-origin model begins to fail and the effect of the substrates can no longer be solely ascribed to a shift in origins.
Nonetheless, for the cases lying in the degraded regime, we still use the virtual origin
that would be valid in the linear regime, $y = -\sqrt{K_z}$, to measure $u_{\tau}$. In this
framework, any further effect can be interpreted as additive. The values of $\Delta U^+$
have been obtained using this $u_{\tau}$ and comparing with a smooth-wall velocity profile
with the origin shifted to $y = -\sqrt{K_z}$, although the effect of the shift on
$\Delta U^+$ is also negligible.


\subsection{Drag reduction curves}\label{subsec:DUp_curves}

\begin{figure}
    \centering
        \includegraphics{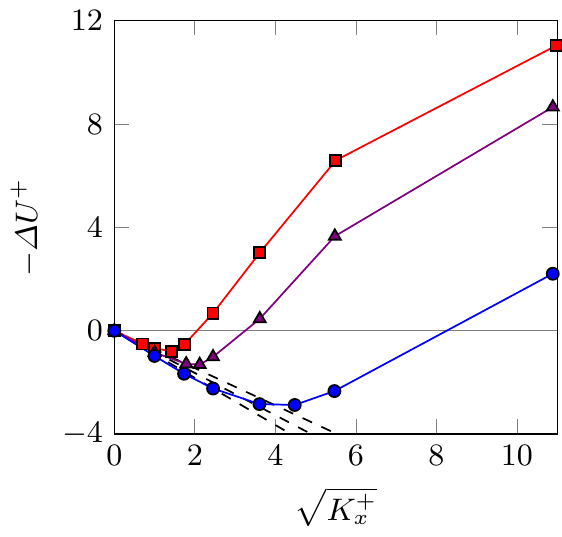}%
        \mylab{-4.45cm}{4.9cm}{(\aaa)}%
	\hspace{0.5cm}%
        \includegraphics{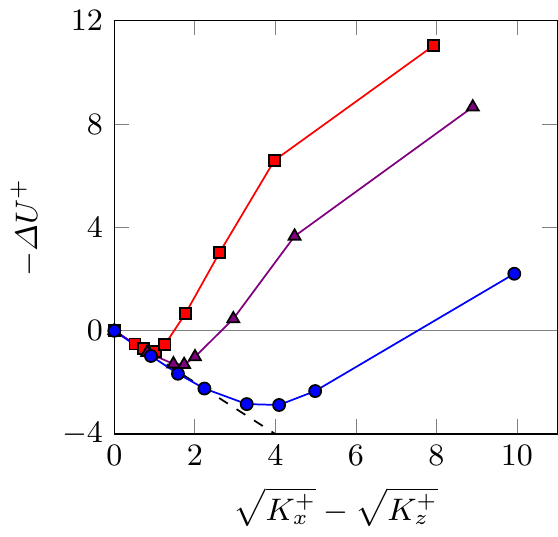}%
        \mylab{-4.45cm}{4.9cm}{(\bbb)}%
	
	\hspace{0.25cm}%
        \includegraphics{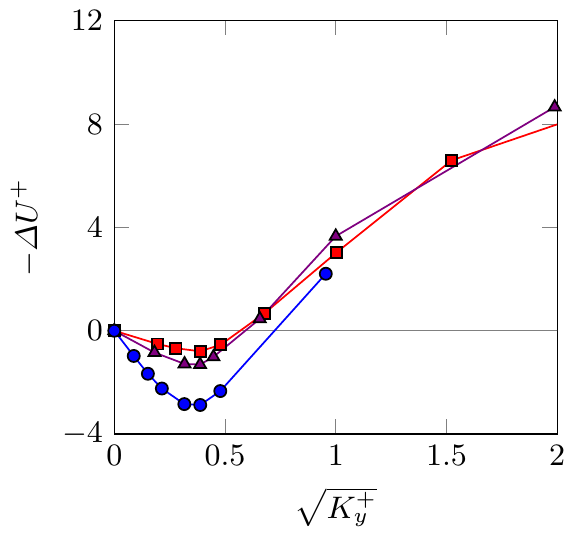}%
        \mylab{-4.6cm}{4.85cm}{(\ccc)}%
	\hspace{0.4cm}%
        \includegraphics{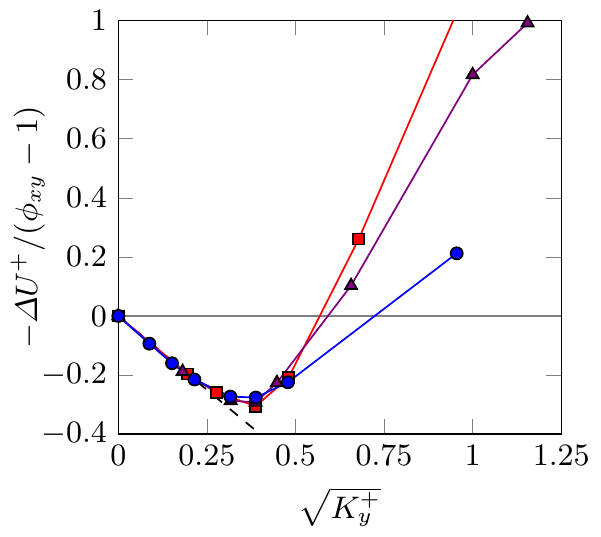}%
        \mylab{-4.8cm}{4.85cm}{(\ddd)}%
	\caption{Drag reduction curves for substrates with different anisotropy ratios. \protect\circlesolidblue, $\phi_{xy} \approx 11.4$; \protect\triangsolidviolet, $\phi_{xy} \approx 5.5$; and \protect\squaresolidred, $\phi_{xy} \approx 3.6$. The symbols correspond to DNSs listed in table~\ref{tab:cases}. $\Delta U^+$ is represented versus (\aaa) the streamwise permeability lengthscale, $\protect\Ksqx$; (\bbb) its predicted value in the linear regime, $\protect\Ksqx -\protect\Ksqz$; (\ccc) the wall-normal permeability lengthscale, $\protect\Ksqy$. (\ddd) $\Delta U^+$, reduced with its predicted slope, versus the wall-normal permeability lengthscale, $\protect\Ksqy$.}\label{fig:DNS_DU}
\end{figure}

\begin{figure}
    \centering
    \includegraphics{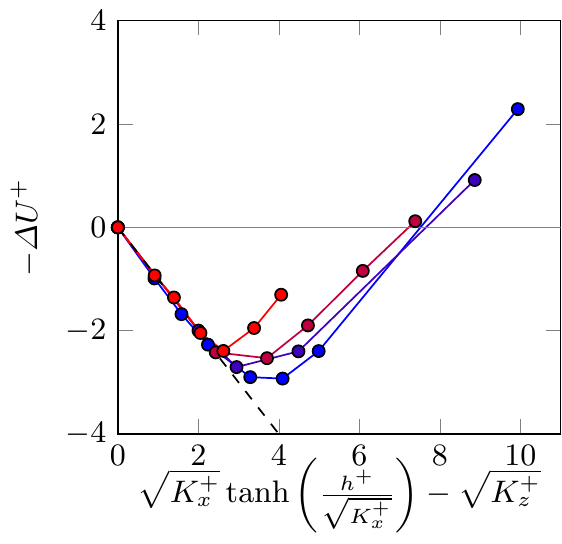}%
    \mylab{-4.4cm}{4.9cm}{(\aaa)}%
    \hspace{0.6cm}%
    \includegraphics{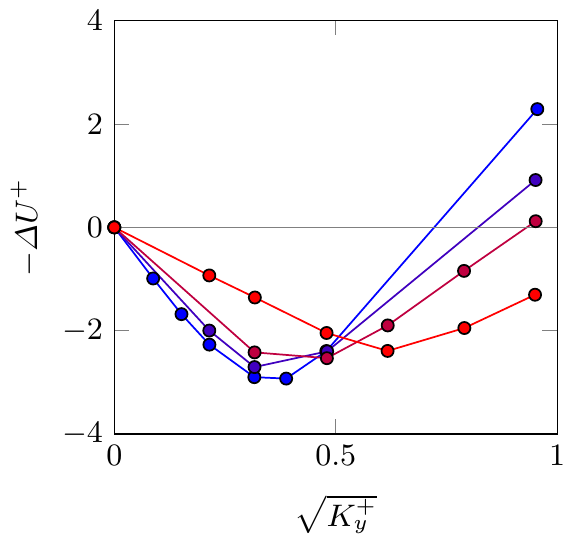}%
    \mylab{-4.57cm}{4.9cm}{(\bbb)}%
	
	\hspace{0.3cm}%
    \includegraphics{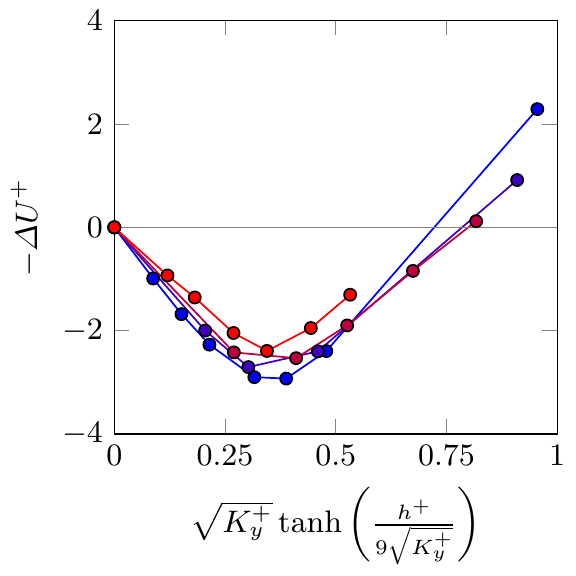}%
    \mylab{-4.6cm}{5.22cm}{(\ccc)}%
    \hspace{0.2cm}%
    \includegraphics{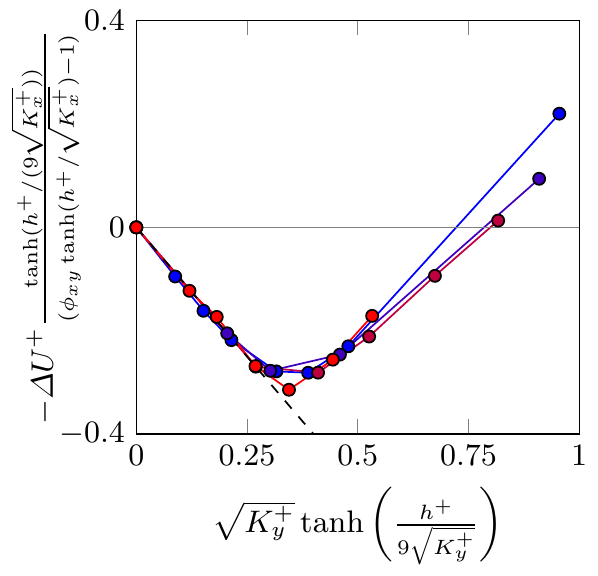}%
    \mylab{-4.68cm}{5.22cm}{(\ddd)}%
	\caption{Drag reduction curves for substrates with the same permeabilities but different substrate thickness. From
blue to red, representing decreasing thickness, cases C1-C7, C$'$1-C$'$7, C$''$1-C$''$7, and C$'''$1-C$'''$7, corresponding
to $h/\sqrt{K_x}=8.8$, $1.5$, $1.0$, and $0.5$. $\Delta U^+$ is represented versus (\aaa) its theoretical value in the linear regime; (\bbb) the wall-normal permeability lengthscale, $\sqrt{K_y^+}$; (\ccc) the fitted permeability lengthscale for the breakdown $\protect\KsqBrp$. (\ddd) $\Delta U^+$, reduced with its predicted linear slope, versus $\protect\KsqBrp$.
}
\label{fig:DNS_DU_h}
\end{figure}

The drag reduction curves obtained from the main set of DNSs are shown in
figure~\ref{fig:DNS_DU}. For small permeabilities, a linear
drag-reduction regime is observed. In \S\ref{subsec:DR}, we predicted $\Delta U^+$ in
this regime to be equal to the difference between the virtual origin for the mean flow,
$\ell_U^+$, and that perceived by turbulence, $\ell_T^+$. For the substrates under
consideration, these would be $\ell_U^+ \approx \Ksqx$ and $\ell_T^+ \approx
\Ksqz$, as given by equation~\eqref{eq:DR_K}. This prediction agrees well with the DNS
results, and the three substrate configurations exhibit roughly the same
initial unit slope in figure~\ref{fig:DNS_DU}(\bbb). The breakdown of the linear drag
reduction, however, occurs for different values of $\Ksqx-\Ksqz$ depending on the substrate.

In contrast, when the lengthscale is represented using $\Ksqy$ -- the
parameter predicted in \S\ref{subsec:LSA} to trigger the Kelvin-Helmholtz
instability -- the location of the breakdown coincides for all the curves, as shown in
figure~\ref{fig:DNS_DU}(\ccc). For all substrate configurations, the drag reduction
is maximum for $\Ksqy |_{opt} \approx 0.38$ and the drag becomes greater than for a
smooth wall for $\Ksqy \gtrsim 0.6$.

The common linear drag reduction behaviour, observed in figure~\ref{fig:DNS_DU}(b), and its common breakdown, observed in figure~\ref{fig:DNS_DU}(\ccc), are condensed in figure~\ref{fig:DNS_DU}(\ddd).
This is done by dividing $\Delta U^+$ from figure~\ref{fig:DNS_DU}(\ccc) by the slope for each curve expected from equation~\eqref{eq:DR_K2}, $\phi_{xy} - 1$.
Given that in this equation $\Delta U^+$ depends only on $\phi_{xy}$ and $\Ksqy$, the general collapse suggested by this figure could be used to predict the performance of permeable substrates different to those explored in this work.
Considering that the maximum  $\Delta U^+$ in figure~\ref{fig:DNS_DU}(\ddd) occurs
for $\Ksqy |_{opt} \approx 0.38$ and is approximately $80\%$ of that estimated by equation~\eqref{eq:DR_K2}, the maximum $\Delta U^+$ would depend only on the anisotropy ratio,
\begin{equation}
	\Delta U^+_{max} \approx 0.8 \times 0.38 \times \left( \phi_{xy} - 1 \right).
	\label{eq:DUmax}
\end{equation}
For substrates with different cross permeabilities, $\phi_{zy} \neq 1$, it follows from
equation~\eqref{eq:DR_K} that the maximum $\Delta U^+$ would be
$\Delta U^+_{max} \approx 0.8 \times 0.38 \times ( \phi_{xy} - \phi_{zy})$.

The secondary set of simulations aims to explore the effect of the substrate depth on $\Delta U^+$, and
to test if the performance could be improved by reducing the depth enough for it to become a parameter
in the problem. For this, the same substrate of cases C1-C7 is studied with depths $h/\sqrt{K_x}=1.5$,
$1.0$, and $0.5$. From equations~\eqref{eq:SL}, we can expect shallower substrates to have smaller $\ell_U^+$
and $\ell_T^+$, as the hyperbolic tangent terms become smaller than unity. This would reduce the slope of
the $\Delta U^+$ curve in the linear regime and be an adverse effect. However, a reduced depth would also
have the beneficial effect of making the substrate more robust to the onset of Kelvin-Helmholtz-like
rollers, as at a given Reynolds number (i.e. $\Ksqx$, $\Ksqy$) equation~\eqref{eq:Br_parameter_plus} would predict a smaller
$\KsqBr$. Note also that $\KsqBr$ is a parameter empirically fitted to the results from the
linear stability model, and that the actual results in \S\ref{subsec:LSA} show that shallower substrates have in fact a delayed
onset in terms of $\KsqBr$, as shown in figure~\ref{fig:sigmai}.

The results for $\Delta U^+$ for the shallow substrates of the secondary set of simulations are portrayed
in figure~\ref{fig:DNS_DU_h}, compared with the corresponding deep substrate from the main set, cases C1-C7. Given that
all our substrates have higher permeability in $x$, the first
terms to experience the effect of a finite $h$ in equations~\eqref{eq:SL} and \eqref{eq:Br_parameter_plus} are those where $h$
appears scaled with $\sqrt{K_x}$. Note that if we had considered values of $h$ small enough for $h/\sqrt{K_z}$
to be also small,
we would have $\ell_U^+ \approx \ell_T^+ \approx h^+$, which would yield no drag-reducing effect.
For the values of $h/\sqrt{K_x}$ studied, we have $h/\sqrt{K_y}=h/\sqrt{K_z}=
6$, $11$ and $17$, so the corresponding hyperbolic tangent terms in equations~\eqref{eq:SL} and \eqref{eq:Br_parameter_plus} are
still essentially unity. 
This can be appreciated for instance in figure~\ref{fig:DNS_DU_h}(\aaa), where
the predicted slope in the linear regime has been adjusted for the effect of $h^+$ on the streamwise slip,
$\ell_U^+ \approx \Ksqx \tanh (h^+/\Ksqx)$, but the spanwise slip remains $\ell_T^+ \approx \Ksqz$.
Figure~\ref{fig:DNS_DU_h}(\bbb), however, shows that $\Ksqy$ is no longer adequate to parametrise
the onset of the drag degradation. Panel (\ccc), in turn, suggests that a suitable alternative is
$\KsqBrp=\Ksqy \tanh (h^+/(9 \Ksqy))$, and that the optimum value is still $\KsqBrp\approx0.38$, as in
figure~\ref{fig:DNS_DU}. All the curves can be once more collapsed by reducing $\Delta U^+$ with its
predicted slope in the linear regime and expressing the Reynolds number in terms of $\KsqBrp$, as is
done in panel (\ddd). This suggests that the optimum performance for shallow substrates can also be
predicted and would be $\Delta U^+_{max} \approx 0.8 \times 0.38 \times
[\phi_{xy}\tanh(h/\sqrt{K_x}) - \phi_{zy}]/\tanh(h/9\Ksqynp)$. Note, however, that $\Delta U^+_{max}$
decreases slightly as the substrate depth is reduced, as can be appreciated in panel (\aaa), and that
even if there is a delay in the critical $\Ksqy$ in absolute terms, as observed in panel (\bbb), any gain in the relative width of the
`drag bucket' region -- the near-optimal range -- is insignificant, as is clear from panel (\ddd).

\subsection{Flow statistics}\label{subsec:stats}

\begin{figure}
  \centering
  \vspace{0.1cm}
  \includegraphics[width=0.95\textwidth]{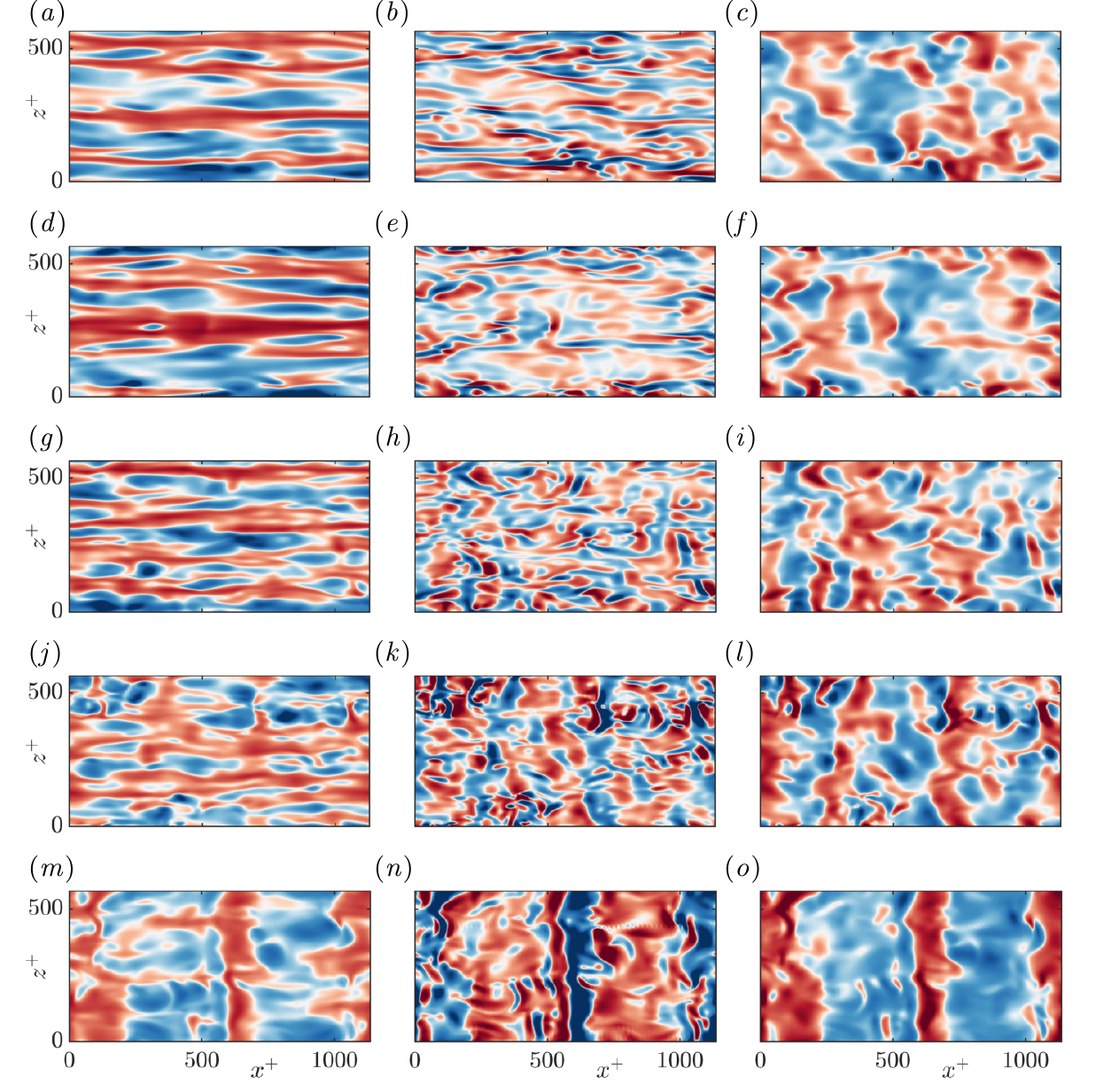}%
  \caption{Instantaneous realisations of $u^+$, $v^+$ and $p^+$ for a smooth channel and for substrates with $\phi_{xy} \approx 11.4$ at a $x$-$z$ plane $y^+ + \ell_T^+ \approx 2.5$. From left to right the columns are $u^+$, $v^+$ and $p^+$. From top to bottom, representing increasing permeabilities, (\aaa-\ccc) smooth wall, (\ddd-\fff) case C2, (e-g) case C4, (h-j) case C6 and (\mmm-\ooo) case C7. In all cases, red to blue corresponds to $(2.2 +  \sqrt{K_x^+} / 2) [-1, 1]$ for $u^+$, $(0.08 + 2/3 \sqrt{K_y^+}) [-1, 1]$ for $v^+$ and $(5+5 \sqrt[4]{K_y^+}) [-1, 1]$ for $p^+$. } \label{fig:inst}
\end{figure}

To explore the underlying mechanisms for the behaviour observed in the drag reduction curves, let us focus on a fixed substrate configuration, that is on one of the curves in figure~\ref{fig:DNS_DU}. Let us take the one with the anisotropy ratio $\phi_{xy} \approx 11.4$,
that is, simulations C1-C7.
The corresponding data for the other two substrate configurations can be found in Appendix~\ref{sec:appB}.
To illustrate how the overlying turbulence is modified at different points along the drag reduction curve, figure~\ref{fig:inst} shows instantaneous realisations of $u$, $v$ and $p$ in an $x$-$z$ plane immediately above the substrate-channel interface.
For small $\Ksqy$, the flow field resembles that observed over a smooth wall. This is shown in panels (\aaa-\ccc) and (\ddd-\fff), where the $u$-field displays the signature of near-wall
streaks, and the $v$-field that of quasi-streamwise vortices. 
As $\Ksqy$ increases beyond the linear regime, the flow begins to be altered, as shown in panels (\ggg-\lll). Some spanwise coherence emerges, becoming more prevalent for larger $\Ksqy$.
Eventually, the flow becomes strongly spanwise-coherent and no trace of the near-wall cycle remains, as shown for a drag-increasing case in panels (\mmm-\ooo).


\begin{figure}
    \centering
    \includegraphics{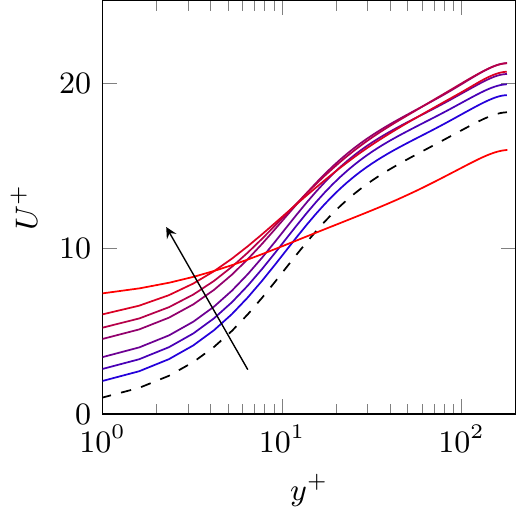}%
    \mylab{-4.0cm}{4.8cm}{(\aaa)}%
	\hspace{0.5cm}%
    \includegraphics{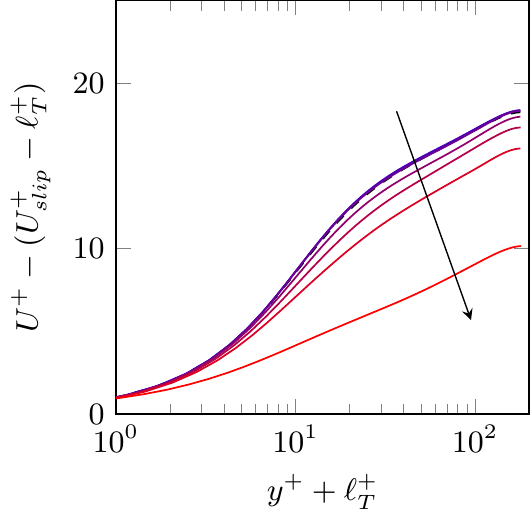}%
    \mylab{-4.0cm}{4.8cm}{(\bbb)}%
    \caption{Mean velocity profiles for a substrate configuration with $\phi_{xy} \approx 11.4$. Permeability values increase from blue to red and correspond to cases C1-C7. (\aaa) Profiles scaled with $u_{\tau}$ measured at the interface plane, $y^+= 0$. (\bbb) Profiles shifted by the linearly extrapolated virtual origin of turbulence, $\ell_T^+ = \protect\Ksqz$, and scaled with the corresponding $u_{\tau}$ at $y = -\ell_T$, where the value at the origin, i.e. the offset predicted from the linear theory, $\Delta U^+ = U_{slip}^+ - \ell_T^+$, has been subtracted. Black-dashed lines represent the smooth-channel case.}
    \label{fig:KxKy_130_mean}
    \vspace{0.3cm}
    \centering
	\includegraphics{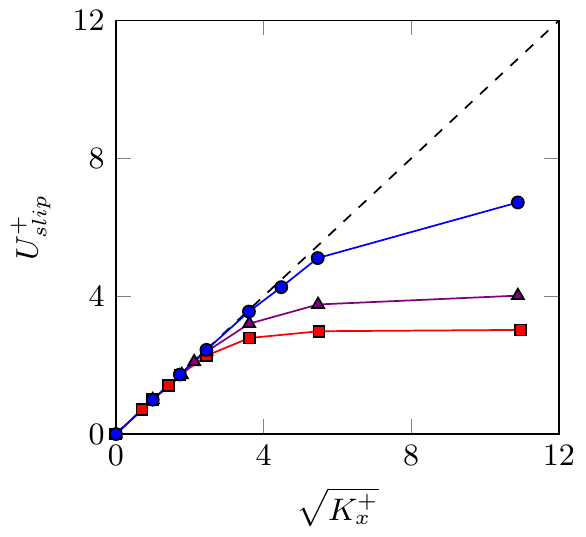}%
    \mylab{-4.0cm}{4.8cm}{(\bbb)}%
    \caption{Slip velocity at the substrate-channel interface, $U_{slip}^+$, versus $\protect\Ksqx$ for the three substrate configurations, \protect\circlesolidblue, $\phi_{xy} \approx 11.4$; \protect\triangsolidviolet, $\phi_{xy} \approx 5.5$; \protect\squaresolidred, $\phi_{xy} \approx 3.6$. The symbols correspond to DNS cases listed in table~\ref{tab:cases} and the dashed line to $U_{slip}^+ = \protect\Ksqx$. }
    \label{fig:Uslip_Kx}
\end{figure}

\begin{figure}
	\centering
    \includegraphics{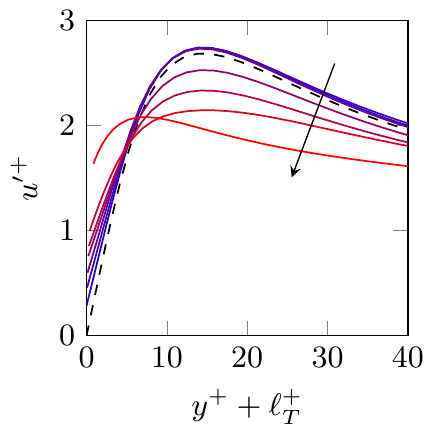}%
    \mylab{-3.45cm}{3.85cm}{(\aaa)}%
    \includegraphics{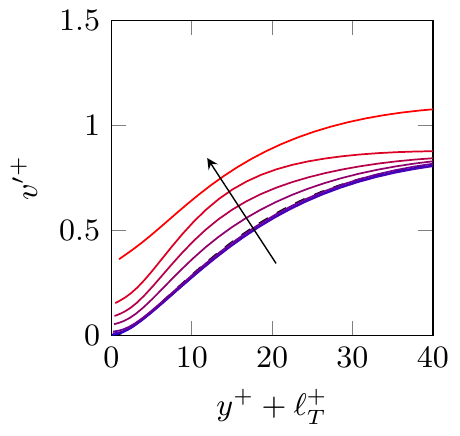}%
    \mylab{-3.45cm}{3.85cm}{(\bbb)}%
    \includegraphics{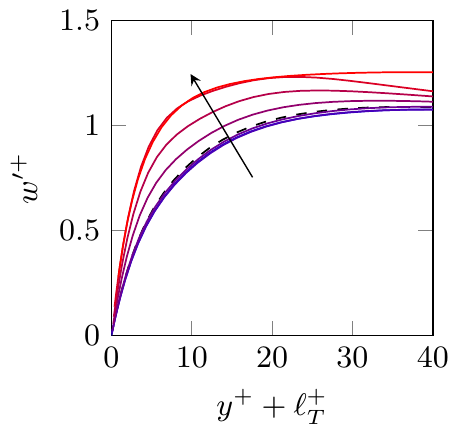}%
    \mylab{-3.45cm}{3.85cm}{(\ccc)}%
	
    \includegraphics{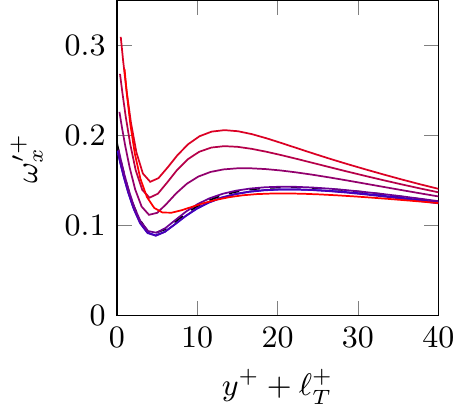}%
    \mylab{-3.45cm}{3.85cm}{(\ddd)}%
	\hspace{1.cm}%
    \includegraphics{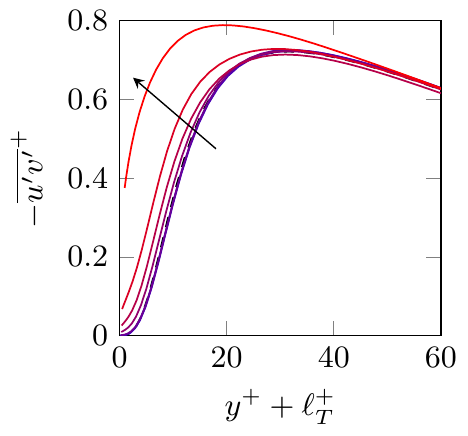}%
    \mylab{-3.48cm}{3.85cm}{(\eee)}%
\caption{One-point turbulent statistics for a substrate configuration with $\phi_{xy} \approx 11.4$. Permeability values increase from blue to red and correspond to cases C1-C7 scaled with the corresponding $u_{\tau}$ at $y = -\ell_T =- \sqrt{K_z}$, the linearly extrapolated virtual origin for turbulence. Black-dashed lines represent the smooth-channel case. Rms fluctuations of (\aaa) the streamwise velocity, (\bbb) the wall-normal velocity, (\ccc) the spanwise velocity, and (\ddd) the streamwise vorticity. (\eee) Reynolds stress.}\label{fig:KxKy_130}
\end{figure}

To assess quantitatively to what extent turbulence differs from that over smooth walls, we first focus
on the one-point statistics resulting from the DNSs, portrayed in figures~\ref{fig:KxKy_130_mean} and \ref{fig:KxKy_130}. The former shows the mean velocity profiles. In panel (\aaa) the results are represented with the
origin for the wall-normal height at the substrate-channel interface, $y^+ = 0$, as is typically done in the literature.
In this representation, the non-zero slip velocity at the interface, $U_{slip}^+$, is apparent at $y^+=0$, while far away from the wall the adverse effect of $\ell_T^+$ and the `roughness-like' shape of the
profile, that is the deviation from a smooth-wall-like shape, combine with $U_{slip}^+$ to yield the net
velocity offset. In this framework, the effect of $\ell_T^+$ and the deviation from the shape of a smooth-wall
profile cannot be easily disentangled.

If the velocity profiles are represented with the origin for the wall-normal
height at $y^+=-\ell_T^+$ and if turbulence remained smooth-wall like, the profiles could then be expected to be
like those for smooth walls, save for the offset given by equation \eqref{eq:DR_gg}. Subtracting that offset would
then give a collapse of all the velocity profiles, and any deviation can then be separately attributed to
modifications in the turbulence \citep{RGM2018}. In figure~\ref{fig:KxKy_130}(\bbb), the profiles are portrayed with the origin at $y^+=-\ell_T^+$ and with the offset subtracted from the
velocities.
For cases C1-C3, which lie in the linear regime, the resulting collapse is indeed good, but
beyond this regime the profiles deviate from the smooth-wall behaviour increasingly. Let us note that defining
$u_\tau$ at $y^+=-\ell_T^+$ implies that the wall-normal gradient of the mean profile at the interface is
no longer necessarily $d U^+/d y^+ |_{y^+=0} = 1$. 
This is because the stresses at that height
in viscous units sum slightly less than one, and more specifically, because a non-zero transpiration gives rise to a
Reynolds stress at the interface, so the viscous stress is no longer the only contribution to the total.
As
a result, $U_{slip}^+$ and $\ell_x^+$ do not strictly have equal value and cannot be used interchangeably.
For small values of $\Ksqy$, the Reynolds stress at the substrate-channel interface is negligible, 
so this effect is small and $U_{slip}^+ \approx \ell_x^+$. This is the case for the substrates lying on
the linear regime. However, as $\Ksqy$ increases, the Reynolds stress at the interface ceases to be
negligible, and $U_{slip}^+ = d U^+/d y^+ |_{y^+=0}\,\ell_x^+ < \ell_x^+$. 
This discrepancy between $U_{slip}^+$ and $\ell_x^+ \approx \Ksqx$ for the substrates under consideration is shown in figure~\ref{fig:Uslip_Kx}.
The effect is small for
the substrate of simulations C1-C7, but is significant for the substrates of B1-B7 and A1-A8, with results
portrayed in Appendix~\ref{sec:appB}. The effect is particularly intense
for the latter substrate, which reaches $\Ksqy\approx 3$ and experiences significant transpiration.
Although $U_{slip}^+$ and
$\ell_x^+$ represent essentially the same concept, the quantitative effect of the streamwise slip
is carried more accurately by $U_{slip}^+$, so the latter has been used for the velocity offset in
figure~\ref{fig:KxKy_130}(\bbb). Notice that this effect is negligible in slip-only simulations or other
idealised surfaces where zero transpiration is assumed \citep{Fairhall2018b}.

The observations on the agreement or deviation from smooth-wall data in the mean velocity profiles extend
also to the rms velocity fluctuations and streamwise vorticity, as well as the Reynolds shear stress, portrayed
in figures~\ref{fig:KxKy_130}(\aaa-\eee). For the cases in the linear regime, the agreement with smooth-wall
data is good. The only difference is a small deviation in the profile of $u'^+$ in the region immediately
above the interface. This deviation is caused by the streamwise velocity effectively tending to zero at
$y^+ = - \ell_U^+$, below the reference height $y^+=-\ell_T^+$, and essentially does not alter near-wall
dynamics \citep{ggMadrid}. Beyond the linear regime, the fluctuations of the streamwise velocity decrease in
intensity, while those of the transverse components increase. For rough surfaces, this is often associated with a decreased anisotropy of the fluctuating velocity \citep{Orlandi2006}.
The Reynolds stress behaves analogously, and the rms streamwise vorticity also becomes more intense, but
experiences a significant drop for the final case, C7. The snapshots of figure~\ref{fig:inst} could suggest
that this is caused by the eventual annihilation of the quasi-streamwise vortices of the near-wall cycle, as
the spanwise-coherent structures become prevalent.


\begin{figure}
    \centering
    \includegraphics{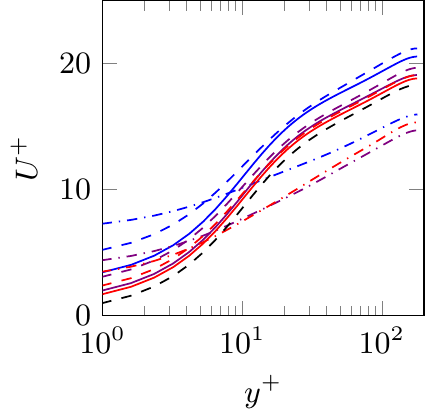}%
    \mylab{-3.2cm}{3.9cm}{(\aaa)}%
    \hspace{0.4cm}
    \includegraphics{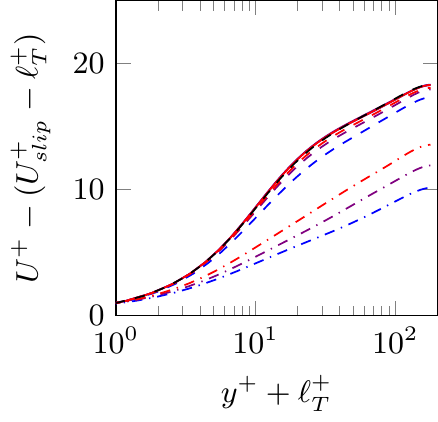}%
    \mylab{-3.2cm}{3.9cm}{(\bbb)}%
	
    \includegraphics{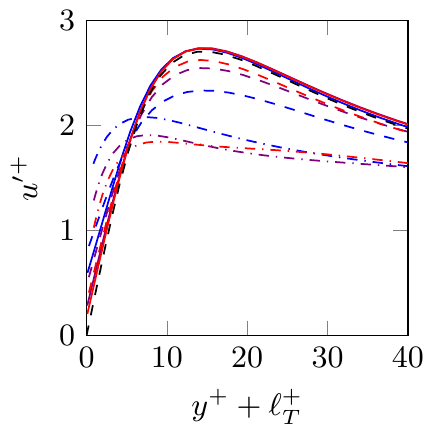}%
    \mylab{-3.45cm}{3.85cm}{(\ccc)}%
    \includegraphics{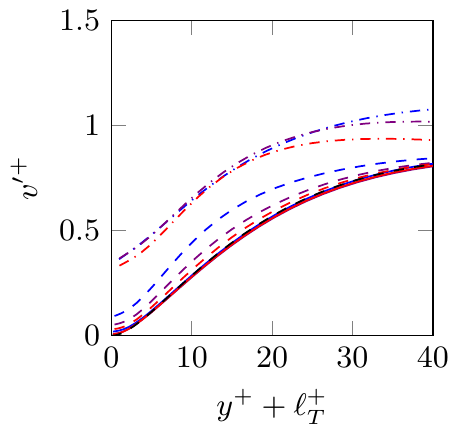}%
    \mylab{-3.45cm}{3.85cm}{(\ddd)}%
    \includegraphics{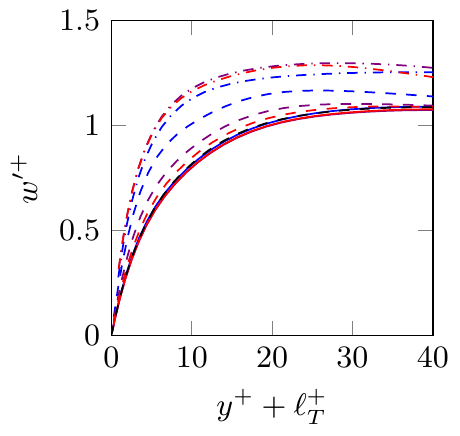}%
    \mylab{-3.45cm}{3.85cm}{(\eee)}%
	
    \includegraphics{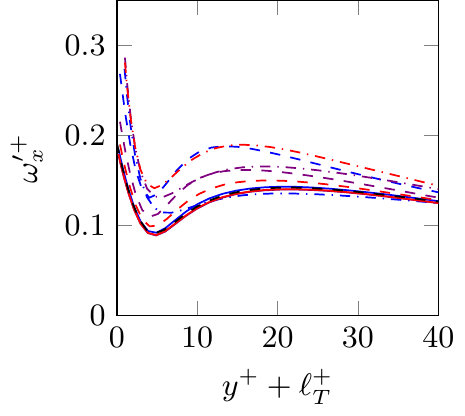}%
    \mylab{-3.45cm}{3.85cm}{(\fff)}%
	\hspace{0.1cm}%
    \includegraphics{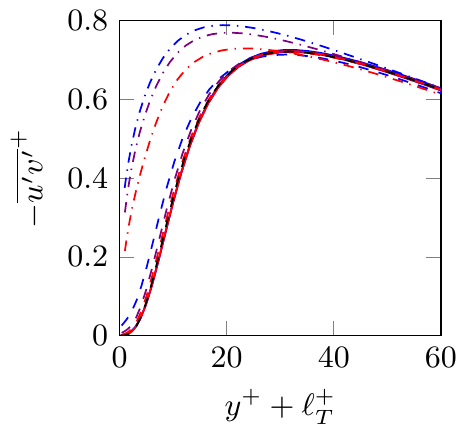}%
    \mylab{-3.48cm}{3.85cm}{(\ggg)}%
\caption{Turbulent statistics for different permeable substrates. Each linestyle is used for cases with approximately the same $\protect\Ksqy$ and $\protect\Ksqz$. \full, cases A1, B1 and C3, with $\protect\Ksqy \approx 0.2$; \broken, cases A3, B3 and C5, with $\protect\Ksqy \approx 0.4$; \chain, cases A6, B6 and C7, with $\protect\Ksqy \approx 1.0$. The colours represent substrate configurations with a fixed $\phi_{xy}$: red, $\phi_{xy} \approx 3.6$; purple, $\phi_{xy} \approx 5.5$; blue, $\phi_{xy} \approx 11.4$. Black lines correspond to the smooth-channel case. Variables are scaled with the corresponding $u_{\tau}$ at $y = -\sqrt{K_z}$, the linearly extrapolated virtual origin for turbulence.  (\aaa) Mean velocity profiles, (\bbb) mean velocity profiles shifted as in figure~\ref{fig:KxKy_130_mean}(\aaa). (\ccc), (\ddd) and (\eee) streamwise, wall-normal and spanwise rms velocity fluctuations. (\fff) Streamwise vorticity rms fluctuations. (\ggg) Reynolds stress.}\label{fig:stats_Ky}
\end{figure}

In the models proposed in \S\ref{sec:theory}, the streamwise, spanwise and wall-normal permeabilities
have separate effects. These models capture leading-order features, but in equations~\eqref{eq:BC3D}
the effect of the three permeabilities is coupled. This manifests in the DNS results and, although the
coupled effects are secondary, they become increasingly important for large permeabilities.

The leading-order effect of the substrate on the overlying turbulence is, as discussed above, set by the
transverse permeabilities. Although in the present study they are equal, it could be expected that
$\Ksqz$ governed the virtual-origin effect, while $\Ksqy$ governed the onset of spanwise-coherent dynamics.
However, once $\Ksqy$ becomes sufficiently large, $\Ksqx$ plays a secondary role by indirectly modulating the
transpiration. Quantitatively, this influence is embedded in equations~\eqref{eq:BC3D}. In essence, the
wall-normal flow that penetrates into the substrate is in a first instance impeded by $\Ksqy$, but from
continuity it eventually needs to traverse the substrate tangentially, being then impeded by $\Ksqx$,
before it leaves through the interface elsewhere. Thus, a large $\Ksqx$ amplifies the transpiration
effect of $\Ksqy$ or, rather, a small $\Ksqx$ limits it. This can be observed by comparing the three substrates
studied at roughly equal values of $\Ksqy$. As they have different anisotropy ratios, for the same $\Ksqy$
they have different $\Ksqx$. Examples are shown in figure~\ref{fig:stats_Ky}. The values
$\Ksqy\approx0.2$, $0.4$ and $1.0$ have been chosen to observe the secondary effect of $\Ksqx$ in the linear
regime, near the optimum drag reduction, and in the fully degraded regime, respectively. In the first case,
the effect of $\Ksqx$ is negligible. The only effect is essentially that of $\Ksqz$ setting the virtual origin,
and all the one-point statistics show good agreement with smooth wall data. The effect is still small near the optimum, for $\Ksqy\approx0.4$, but the modulation by $\Ksqx$ begins to manifest, amplifying the effects
of $\Ksqy$ already discussed above, such as the decreased anisotropy of the velocity fluctuations.
Nevertheless, the Reynolds stress curve, and thus the shape of the mean velocity profile, remain close
to those in the linear regime and for smooth walls. In the fully-degraded regime, $\Ksqy\approx1.0$, the modulating effect of $\Ksqx$ becomes stronger and results in a further degradation of the Reynolds stress,
the mean profile and the drag. The near-wall cycle is severely disrupted in this regime, and the main effect of
$\Ksqx$ on the velocity fluctuations is on $u'^+$ near the wall, directly through the increased streamwise slip.

In turn, $\Ksqy$ also has a secondary effect on the streamwise slip, through the
non-zero Reynolds stress at the interface discussed above.
Figure~\ref{fig:Uslip_Kx} illustrates how, for the same $\Ksqx$, which governs $U_{slip}^+$ to leading-order,
substrates with larger $\Ksqy$ have a smaller slip velocity.

While the analysis of the one-point statistics reveals variations in average intensities at different heights,
it cannot provide information on whether those variations are caused by contributions from lengthscales that
are not active over smooth walls, or from a change in the intensity of the typical lengthscales of canonical
wall turbulence. To investigate this, we analyse the spectral energy distribution of the fluctuating velocities.

\begin{figure}
  	\centering   	
  	\includegraphics[width=0.99\textwidth]{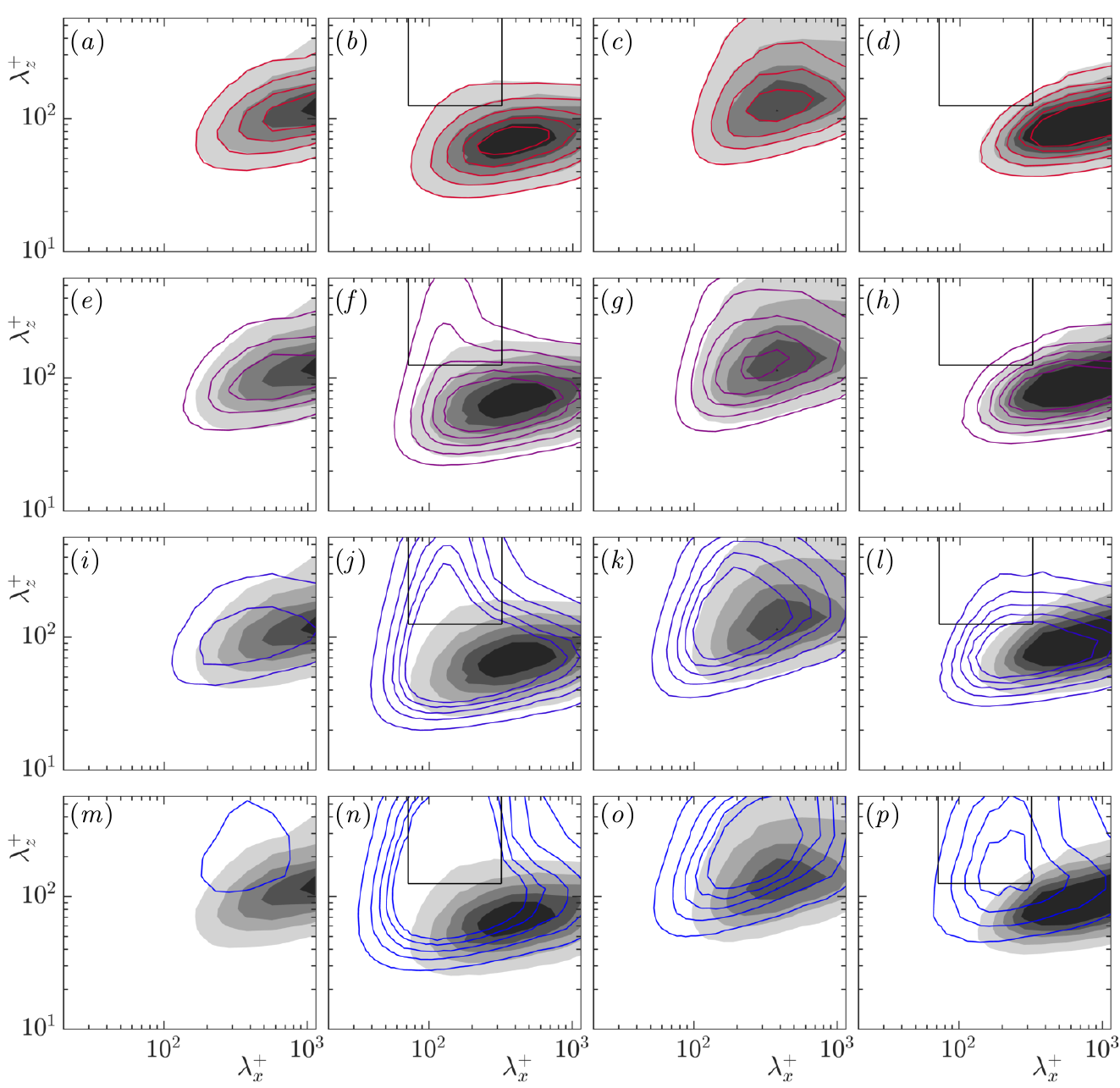}%
  	\caption{Premultiplied two-dimensional spectral densities for a substrate configuration with $\phi_{xy} \approx 11.4$ at a plane $y^+ + \ell_T^+ \approx 15.5$. First column, $k_x k_z E_{uu}$; second column, $k_x k_z E_{vv}$; third column, $k_x k_z E_{ww}$; fourth column, $k_x k_z E_{uv}$; with contour increments $0.3241$, $0.0092$, $0.0404$ and $0.0239$ in wall units, respectively. Shaded, smooth channel. Red contours, permeable cases: (\aaa-\ddd) case C2, (\eee-\hhh) case C4, (\iii-\lll) case C6, and (\mmm-\ppp) case C7. The box indicates the region of the spectrum considered in \S\ref{subsec:DU}.}\label{fig:spectra}
\end{figure}

As an example, spectral density maps of  $u^2$, $v^2$, $w^2$ and $uv$ are represented
at a height of roughly 15 wall units above the virtual origin for turbulence in figure~\ref{fig:spectra}.
For substrates in the linear regime, such as C2 in panels (\aaa-\ddd), the agreement in spectral distribution with
smooth-wall flows is excellent, as it was for the rms values, further supporting the idea that near-wall
turbulence remains essentially canonical. For substrate C4, which is just past the linear regime and has a
near-optimum $\Ksqy\approx0.32$, differences begin to appear in the spectral distributions, like additional
energy at slightly shorter streamwise wavelengths, but most notably the emergence of a spectral region with
high $v^2$ for large spanwise wavelengths, $\lambda_z^+\approx200-\infty$ and streamwise wavelengths
$\lambda_x^+\approx100-200$. This feature is consistent with the onset of spanwise-coherent structures
observed in figure~\ref{fig:inst}, and was also observed previously for riblets and connected to the
presence of Kelvin-Helmholtz-like rollers
\citep{Garcia-Mayoral2011}. The above effects become more intense for cases C6 and C7. For C6, which lies
in the degraded regime but still yields a net reduction in drag, energy appears in wavelengths as short as
$\lambda_x^+\approx50$, and the spanwise-coherent region spans a wider set of streamwise wavelengths,
$\lambda_x^+\approx60-350$, although there is still a trace of the spectral densities of smooth-wall
flow for long wavelengths, $\lambda_x^+\gtrsim500$, specially for $v^2$ and $w^2$. For case C7, which
gives a net drag increase, any residual trace of the spectral distribution for smooth-wall turbulence
has disappeared, and the range $\lambda_x^+\approx60-350$ becomes dominant in $v^2$.


\subsection{Contributions to $\Delta U^+$}\label{subsec:DU}

\begin{figure}
	\vspace{0.5cm}
    \centering
        \includegraphics{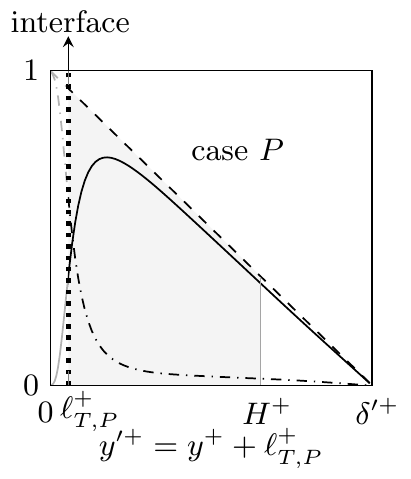}%
        \mylab{-4.3cm}{4.9cm}{(\aaa)}%
        \includegraphics{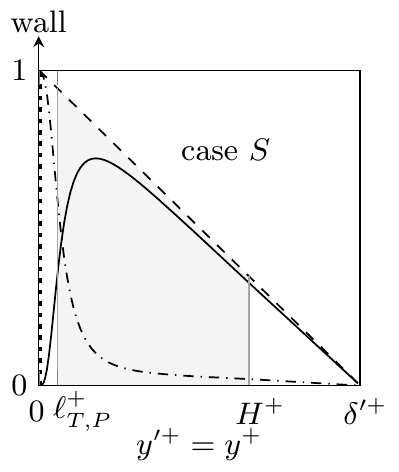}%
        \mylab{-4.2cm}{4.9cm}{(\bbb)}%
        \includegraphics{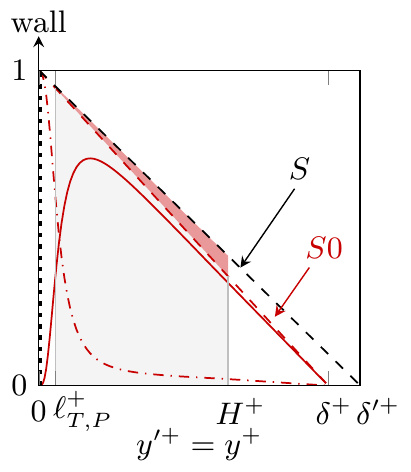}%
        \mylab{-4.3cm}{4.9cm}{(\ccc)}%
\caption{Sketch of stress curves taking the virtual origin of turbulence as reference. \chainII, viscous stress $ dU^+/dy^+$; \full,  $\overline{u'v'}^+$; \broken, total stress. (\aaa) Permeable case at a friction Reynolds number $\delta'^+ = \delta^+ + \ell_{T,P}^+$. (\bbb) Smooth-wall case at the same friction Reynolds number $\delta'^+$. (\ccc) Smooth-wall case at a different friction Reynolds number, $\delta^+$. The vertical black-dotted line indicates the substrate-channel interface in the permeable case and the wall in the smooth cases. The grey shaded area represents the integrated region in equation~\eqref{eq:DUDEC_PS1}. The red shaded area in (\ccc) shows the difference in the integrated area due to the difference in the friction Reynolds number.}\label{fig:DU_stress}
\end{figure}

The degradation of the drag reduction curves in figure~\ref{fig:DNS_DU} and the lack of collapse of the mean velocity profiles in figure~\ref{fig:KxKy_130_mean}(\bbb) show that there is an additional contribution to $\Delta U^+$ beyond the virtual-origins effect predicted in \S\ref{subsec:DR}.
To investigate this, we obtain an expression for $\Delta U^+$ by integrating the mean streamwise momentum equation for a permeable channel and comparing it with that for a smooth channel. This procedure follows closely \cite{MacDonald2016}, \cite{Abderrahaman2019} and \cite{Fairhall2018b}, and
is similar to that followed by \cite{Garcia-Mayoral2011}.  
The streamwise momentum equation is averaged in time and in the streamwise and spanwise directions, and integrated in the wall-normal direction, 
 \begin{equation}
    - \overline{u'v'}^{+} + \frac{d U^{+}}{dy^+} = \frac{\delta'^+ - y'^+}{\delta'^+}, \label{eq:DUDEC_1}\\
  \end{equation}
\noindent where the virtual origin of turbulence is taken as the reference for the wall-normal coordinate, i.e. $y'^+ = y^+ + \ell_T^+$, and it is also the height where $u_{\tau}$ is measured. The effective half-channel height or the effective friction Reynolds number is then $\delta'^{+} = \delta^+ + \ell_T^+$, as previously defined. In equation~\eqref{eq:DUDEC_1}, $\overline{u'v'}^+$ is the Reynolds stress, $d U^{+} /dy^+$ the viscous stress and the right-hand side represents the total stress.  
These three terms are represented in figure~\ref{fig:DU_stress}(\aaa). 

Integrating again between two heights, the viscous stress term gives the velocity $U^+$ at those two heights,
which can be compared to the corresponding equation for a smooth channel to obtain an expression for $\Delta U^+$.
The upper integration limit is then taken at an arbitrary height in the logarithmic region, $y'^+ = H^+$, so that the difference in $U^+$ yields $\Delta U^+$.   
For the lower limit, we set it at $y'^+ = \ell_{T,P}^+$, where $\ell_{T,P}^+$ refers to the virtual origin of turbulence for the permeable case,
since for that layout equation~\eqref{eq:DUDEC_1} is defined only above that height.
Integrating equation~\eqref{eq:DUDEC_1} from $y'^+ = \ell_{T,P}^+$, to an arbitrary height in the logarithmic region, $y'^+ = H^+$, yields
	\begin{equation}
    \int_{\ell_{T,P}^+}^{H^+} - \overline{u'v'}^{+} dy'^+ + U^+ (H^+) - U^+ (\ell_{T,P}^+) = H^+ - \ell_{T,P}^+ - \frac{H^{+2} - \ell_{T,P}^{+2}}{2 \delta'^+}. 
    \label{eq:DUDEC_2}
    \end{equation}
This equation applies not only to a permeable channel, but also to a smooth channel at the same Reynolds number, $\delta'^{+}$, as depicted in figure~\ref{fig:DU_stress}(\bbb). 
Note that for a smooth channel $y'^+ = y^+$, since the origin of turbulence is at the wall, but the lower integration limit can still be set at some height above the wall, $y'^+ = \ell_{T,P}^+$, with $\ell_{T,P}^+$ referring to the origin of the permeable case being compared. 

Subtracting equation~\eqref{eq:DUDEC_2} for the permeable case and for the smooth channel, the resulting expression for $\Delta U^+$ is,

\begin{equation}
\begin{split}
\Delta U^+ &= U_{P}^+ (H^+) - U_{S}^+ (H^+)  \\
&= \underbrace{U_{P}^+(\ell_{T,P}^+)}_{\let\scriptstyle\textstyle
    \substack{U_{slip}^+}} - U_{S}^+ (\ell_{T,P}^+)  \underbrace{-
\int_{\ell_{T,P}^+}^{H^+} \left[ \left(- \overline{u'v'}^{+}_{P} \right) - \left( -\overline{u'v'}^{+}_{S} \right) \right] dy'^+ ,}_{\let\scriptstyle\textstyle
    \substack{\mathcal{T}_{uv}}}
\end{split}
\label{eq:DUDEC_PS1}
\end{equation}

\noindent where subscript `$P$' denotes the permeable channel, and subscript `$S$' the reference smooth channel at the same friction Reynolds number $\delta'^+$. 
Equation~\eqref{eq:DUDEC_PS1} shows that $\Delta U^+$, defined as the difference in $U^+$ between a permeable and smooth channel measured at the same distance from their respective origins of turbulence, consists of the sum of three terms.

\begin{figure}
    \centering
    \includegraphics{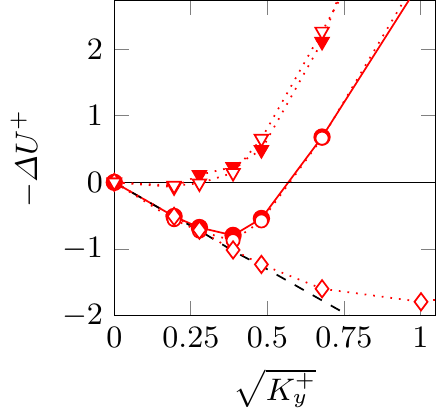}%
    	\mylab{-3.22cm}{3.9cm}{(\aaa)}%
    \includegraphics{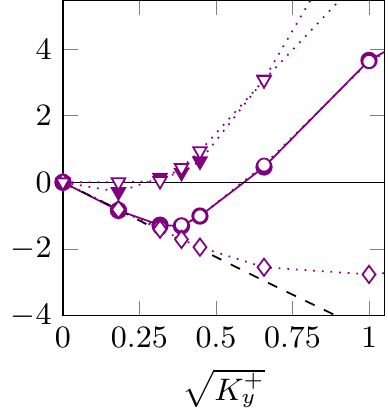}%
        \mylab{-3.2cm}{3.9cm}{(\bbb)}%
    \includegraphics{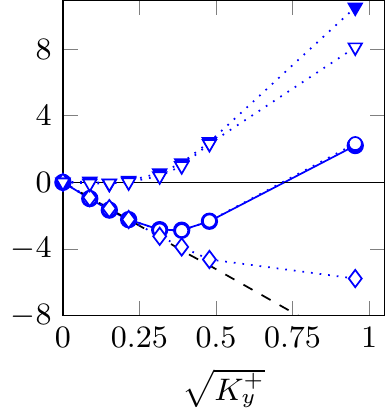}%
        \mylab{-3.2cm}{3.9cm}{(\ccc)}%
\caption{Different contributions to $\Delta U^+$ as a function of $\protect\Ksqy$ for (\aaa) substrates A1-A8, with $\phi_{xy} \approx 3.6$, (\bbb) substrates B1-B7, with $\phi_{xy} \approx 5.5$ and (\ccc) substrates C1-C7, with $\phi_{xy} \approx 11.4$. \protect\circlesolidblack, $\Delta U^+$ measured from the DNSs (same as in table~\ref{tab:cases}); \protect\dimonddot, contribution from the virtual-origin effect, $U_{slip}^+ - U_S^+ (\ell_T^+)$; \protect\triangdowndot, contribution from the additional Reynolds stress, $\mathcal{T}_{uv}$; \protect\triangdowndotblack, contribution from the additional Reynolds stress restricted to the spectral window $\lambda_x^+ \approx 70 - 320$ and $\lambda_z^+ \gtrsim 120$; \protect\circlesolid, $\Delta U^+$ calculated from equation~\eqref{eq:DUDEC_PS1}, as a sum of the contributions from the virtual-origin effect and the additional Reynolds stress.}\label{fig:DUDEC_PS}
\end{figure}

The first term, is the slip velocity of the permeable case at the substrate-channel interface, $U_{slip}^+$. This is a drag-reducing term, and for the cases lying in the linear regime it can be approximated to the virtual origin of the mean flow, $\ell_U^+$, since $dU_P^+/dy^+ |_{y'^+=\ell_{T,P}^+} \approx 1$.
The second term, $U_{S}^+ (\ell_{T,P}^+)$, is the mean velocity of the smooth channel measured at $y'^+ = \ell_{T,P}^+$.
It is a drag-increasing term, and if $\ell_{T,P}^+ \lesssim 5$, it can be accurately approximated as $U_{S}^+ (\ell_{T,P}^+) \approx \ell_{T,P}^+$.
This is essentially the same as the spanwise protrusion height of \cite{Luchini1991} and \cite{Luchini1996}, and the spanwise slip of superhydrophobic surfaces \citep{Min2004,BusseSandham2012}. 
The offset between these terms is then $U_{slip}^+ - U_S(\ell_{T,P}^+) \approx \ell_U^+ - \ell_{T,P}^+$ and represents the virtual-origin effect discussed throughout this paper. 
Note however that the exact contribution to $\Delta U^+$ involves velocities and not virtual origins as pointed out before.
The contribution of the offset between these two terms to $\Delta U^+$ is shown in figure~\ref{fig:DUDEC_PS}, where we can appreciate that the virtual origin approximation $\ell_U^+ - \ell_T^+$ is valid not only in the linear regime, but even slightly beyond the optimum. 

The third term, $\mathcal{T}_{uv}$, represents the additional Reynolds stress
induced by the permeable substrate. It is a drag-increasing term and its contribution to $\Delta U^+$ is also shown in figure~\ref{fig:DUDEC_PS}.
For the substrates lying in the linear regime, the Reynolds stress is smooth-wall-like, except for the displacement $\ell_T^+$ towards the interface, and the term $\mathcal{T}_{uv}$ is therefore zero.
The contribution of this term begins to be significant at the breakdown $\Ksqy |_{opt}$, and increases with $\Ksqy$ in the degraded region.
An increase in Reynolds stress is therefore responsible for the degradation of the drag-reducing behaviour of permeable substrates. 

The spectral energy distribution of the wall-normal velocity in figure~\ref{fig:spectra} shows the appearance of a new spectral region for large spanwise wavelengths centred around $\lambda_x^+ \approx 150$, which is associated to the large spanwise coherent structures observed in figure~\ref{fig:inst}.
To explore whether the additional Reynolds stress accounted for by $\mathcal{T}_{uv}$ is due to the energy accumulated in this spectral region, we define a spectral box with $\lambda_x^+ \approx 70 - 320$ and $\lambda_z^+ \gtrsim 120$, as that depicted in figure~\ref{fig:spectra}, and quantify its contribution to the additional Reynolds stress, as in \cite{Garcia-Mayoral2011}.
The values are also included in figure~\ref{fig:DUDEC_PS}, showing a close agreement with the whole $\mathcal{T}_{uv}$. This suggests that the new spanwise-coherent structures are indeed responsible for the degradation of the drag, as it was also observed for riblets in \cite{Garcia-Mayoral2011}.
In essence, these structures increase the turbulence mixing, increasing the local Reynolds stress, and consequently the global drag.


Note that equation~\eqref{eq:DUDEC_PS1} compares a permeable channel with a smooth one at the same friction Reynolds number.
Often, however, a reference smooth channel at exactly the same Reynolds number is not available. 
This is for instance the case for the simulations presented in this paper, where all the permeable cases are compared to the same smooth channel at a slightly different friction Reynolds number.
When the Reynolds numbers match exactly, the total stress, and thus the Reynolds stress, collapse sufficiently far away from the surface, as they approach zero-value at the centre of the channel. The contribution $\mathcal{T}_{uv}$ can then be entirely ascribed to wall effects, that is to the presence of the substrate. If the Reynolds numbers differ, however, there may be a significant contribution to $\mathcal{T}_{uv}$ far from the surface, which is a Reynolds-number effect, rather than a direct effect of the surface.
The same effect appears when comparing smooth channels at different friction Reynolds numbers, as illustrated in figure~\ref{fig:DU_stress}(\ccc).
To quantify this effect, we compare the smooth channel at Reynolds number $\delta'^+$ used for equation~\eqref{eq:DUDEC_PS1}, represented by a subscript `$S$', with another at a different Reynolds number $\delta^+$, represented by a subscript `$S0$'.
Subtracting the two integrated mean streamwise momentum equations, 
the universality of the near-wall mean velocity profile over smooth walls gives $U_S^+ (\ell_{T,P}^+) = U_{S0}^+ (\ell_{T,P}^+)$ and $U_S^+ (H^+) = U_{S0}^+ (H^+)$, yielding  

\begin{equation}
	\mathcal{T}_{Re} = - \int_{\ell_{T,P}^+}^{H^+} \left[ \left(- \overline{u'v'}^{+}_{S} \right) - \left(- \overline{u'v'}^{+}_{S0} \right) \right] dy'^+ = \frac{H^{+2} - \ell_{T,P}^{+2}}{2} \left( \frac{1}{\delta'^+} - \frac{1}{\delta^+} \right).
	\label{eq:DUDEC_PS2}
\end{equation}

When the break-up of equation~\eqref{eq:DUDEC_PS1} is applied to DNS results from a complex surface, $P$ in our case, and a smooth wall at a different Reynolds number, $S0$, the integral of the difference in Reynolds stresses would include both the surface and the Reynolds number effects. These, however, can be easily separated as 

\begin{equation}
\begin{split}
	 &- \int_{\ell_{T,P}^+}^{H^+} \left[ \left(- \overline{u'v'}^{+}_{P} \right) - \left(- \overline{u'v'}^{+}_{S0} \right) \right] dy'^+ = \\
	  & - \int_{\ell_{T,P}^+}^{H^+} \left[ \left(- \overline{u'v'}^{+}_{P} \right) - \left(- \overline{u'v'}^{+}_{S} \right) \right] dy'^+  -  \int_{\ell_{T,P}^+}^{H^+} \left[ \left(- \overline{u'v'}^{+}_{S} \right) - \left(- \overline{u'v'}^{+}_{S0} \right) \right] dy'^+ \\
	  &= \mathcal{T}_{uv} + \mathcal{T}_{Re}.
\end{split}
	\label{eq:DUDEC_PS3}
\end{equation}

\noindent Note that, from equation~\eqref{eq:DUDEC_PS2}, $\mathcal{T}_{Re}$ can be easily calculated \textit{a priori} as the area of the trapezoid formed between the total stress lines for $\delta^+$ and $\delta'^+$,
as highlighted in figure~\ref{fig:DU_stress}(\ccc).
$\mathcal{T}_{uv}$ can subsequently be obtained by subtracting $\mathcal{T}_{Re}$ from the integral of the difference in Reynolds stresses of cases $P$ and $S0$, as given by equation~\eqref{eq:DUDEC_PS3}, so that it only includes the effect of the surface.
This has been the procedure used to obtain the results shown in figure~\ref{fig:DUDEC_PS}, even though for the small values of $\ell_T^+$ considered, the Reynolds number effect, $\mathcal{T}_{Re}$, is negligible.

\subsection{Adjustment of the theoretical models}\label{subsec:DNS_LSA}

\begin{figure}
		\centering
		\includegraphics{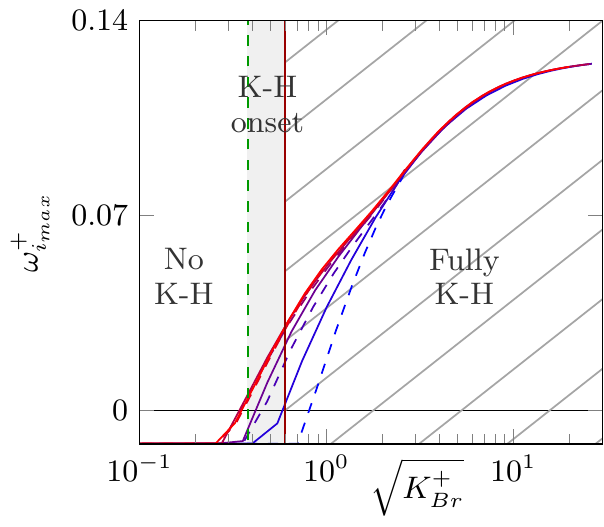}%
  		\mylab{-6.1cm}{5.05cm}{(\aaa)}%
  		\hspace{0.3cm}%
  		\includegraphics{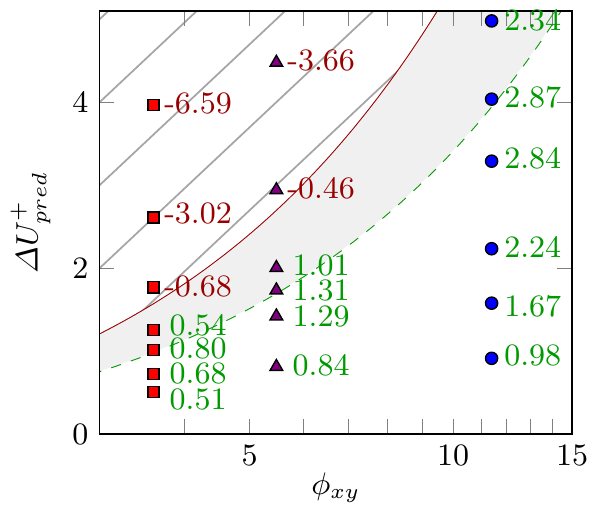}%
  		\mylab{-6.cm}{5.05cm}{(\bbb)}%
  		\caption{(\aaa) Amplification of the most unstable mode versus $\protect\KsqBr$, as in figure~\ref{fig:sigmai}, but with the threshold values for the onset of Kelvin-Helmholtz-like instability adjusted to $\protect\KsqBr \approx \protect\Ksqy = 0.38 - 0.6$. (\bbb) Predicted values of $\Delta U^+$ from the linear theory of equation~\eqref{eq:DR_K2} versus the anisotropy ratio $\phi_{xy}$, as in figure~\ref{fig:LSA}(\bbb), but with the adjusted thresholds for the degraded region. The green line corresponds approximately to the optimum $\Delta U^+$ ($ \protect\Ksqy |_{opt} \approx 0.38$); the red line corresponds approximately to zero $\Delta U^+$ ($ \protect\Ksqy |_{\Delta U^+=0} \approx 0.6$). The symbols represent the DNS cases studied and the values next to them are the actual $\Delta U^+$ measured from the DNSs. Cases beyond $\Delta U_{pred}^+ > 5$ are not displayed.}
	    \label{fig:DR_LSA_final}
\end{figure}

In \S\ref{sec:theory}, we presented theoretical models to estimate the drag-reducing behaviour for anisotropic permeable substrates, specifically, a linear drag-reduction model for small permeabilities given by equation~\eqref{eq:DR_K} and a threshold for the degradation of this linear regime based on the onset of Kelvin-Helmholtz rollers. 
The information obtained from the present DNSs can be used to assess the validity of the theoretical models summarised in \S\ref{sec:theory}, and if necessary adjust them, so that more accurate predictions can be made.

The drag reduction curves in figure~\ref{fig:DNS_DU} show that the linear regime is accurately represented by the offset between the virtual origins of the mean flow and that of turbulence, $\Ksqx - \Ksqz$, as predicted by equation~\eqref{eq:DR_K}.
As discussed above, $\Delta U^+$ in this regime would be more precisely given by the difference $U_{slip}^+ - U_{S}^+ (\ell_T^+)$, but the differences between $\Ksqx$ and $U_{slip}^+$, and between $\Ksqz$ and $U_{S}^+ (\ell_T^+)$ only become significant for larger permeabilities -- beyond the linear regime, as shown in figure~\ref{fig:DUDEC_PS}.

The DNS results and the discussion in \S\ref{subsec:DU} also support the idea that the degradation of the drag-reducing behaviour is caused by the formation of spanwise-coherent structures.
These are generally associated to a Kelvin-Helmholtz-like instability, as discussed in \S\ref{sec:theory}.
In that section, we predicted that the onset of these structures was governed by $\Ksqy$, the leading order term of $\KsqBr$ from equation~\eqref{eq:Br_parameter_plus}, as shown in figure~\ref{fig:sigmai}.
From this figure, we estimated an \textit{a priori} threshold for the onset of Kelvin-Helmholtz-like rollers in the range $\KsqBr \approx \Ksqy \approx 1 - 2.2$, beyond which equation~\eqref{eq:DR_K} would no longer be valid. 
The drag reduction curves in figure~\ref{fig:DNS_DU}(\ccc), however, show that the degradation sets in for lower values of $\Ksqy$ than initially hypothesised. The optimum value of $\Delta U^+$ occurs at $ \Ksqy |_{opt} \approx 0.38$, after which performance degrades, and drag becomes greater than that for smooth walls for $ \Ksqy |_{\Delta U^+ = 0}  \approx 0.6$. 
Adjusting figure~\ref{fig:sigmai} to account for these observed values, we obtain figure~\ref{fig:DR_LSA_final}(\aaa), which shows that the onset occurs as soon as the predicted amplification of the instability becomes positive.

In \S\ref{sec:theory}, combining the equation for the linear regime with the limiting values of $\Ksqy$, allowed us to design the parameter space for realisable drag reduction shown in figure~\ref{fig:LSA}(\bbb), which later served to select the DNS cases studied in \S\ref{sec:results_DNS}.
Using the limiting values of $\Ksqy$ observed in the DNSs ($\Ksqy\approx 0.38 - 0.6$), the adjusted prediction map for $\Delta U^+$ is shown in figure~\ref{fig:DR_LSA_final}(\bbb),
where the actual values of $\Delta U^+$ measured from DNSs are also shown.
This figure illustrates how the theoretical predictions compare to the actual results obtained from DNS.
In the linear regime, $\Delta U^+$ is well predicted by the theory.
At the optimum $\Delta U^+$ line, $\Ksqy \approx 0.38$, the exact value of $\Delta U^+$ is given by equation~\eqref{eq:DUmax}, that is, it is roughly $80 \%$ of the linear-regime prediction.
Beyond this line, the performance degrades, and for the line $\Ksqy \approx 0.6$, the drag reduction is fully negated.
Assuming that this behaviour holds for substrates with anisotropy ratios different to those studied in this work, 
figure~\ref{fig:DR_LSA_final}(\bbb),
which essentially contains the same information of figure~\ref{fig:DNS_DU_h}(\ddd), can be used to estimate their performance.

\section{Conclusions}\label{sec:conclusions}
   We have explored the ability of anisotropic permeable substrates to reduce turbulent skin friction. We have examined the effect of the streamwise, wall-normal and spanwise permeabilities in highly-connected substrates, and showed that streamwise-preferential substrates can reduce drag.

We have conducted a series of DNSs of turbulent channels delimited by permeable substrates, where the flow within the substrates was modelled using Brinkman's equation.
The resulting drag reduction curves obtained for different substrate configurations (different anisotropy ratios) are similar to the classical curves for riblets: they exhibit a linear drag reduction regime followed by a degradation of performance, eventually leading to an increase of drag.

We have observed that, in the linear regime of small permeabilities, the drag reduction is proportional to the difference between the virtual origin perceived by the mean flow and that perceived by turbulence, which for permeable substrates gives $\Delta U^+ \approx \sqrt{K_x^+} - \sqrt{K_z^+}$.
The drag-reducing ability of this technology results therefore from the streamwise-preferential configuration of the substrates, as in other complex surfaces \citep{RGM2018}. 
In this regime, the overlying turbulence remains smooth-wall-like, but shifted towards the substrate-channel interface by the origin perceived by turbulence, i.e. $\Ksqz$.

As permeabilities increase, the linear regime eventually breaks down. We observe that the breakdown is essentially governed by the wall-normal permeability, $K_y^+$, and occurs for $\Ksqy \approx 0.38$, independently of the substrate anisotropy. 
The breakdown can be attributed to the appearance of spanwise-coherent structures, associated to a Kelvin-Helmholtz-like instability. These structures appear to disrupt the near-wall cycle and modify the near-wall turbulence, increasing the Reynolds stress, and consequently, the drag.
As permeabilities increase, the drag-increasing, spanwise-coherent structures become prevalent in the flow, outweighing the drag-reducing effect of the streamwise slip and eventually leading to an increase of drag.


In order to predict the drag-reducing behaviour of anisotropic permeable substrates, we have established some theoretical models, which agree well with the behaviour observed from DNS results.
The linear regime is accurately described by the expression derived by \cite{Abderrahaman2017}, where $\Delta U^+ = \sqrt{K_x^+} - \sqrt{K_z^+}$. This assumes that the permeable medium is highly connected and that the substrate is sufficiently deep for the overlying turbulence not to perceive that its depth is finite, $h^+ \gtrsim 2 \Ksqx$. Beyond $\Ksqy \approx 0.38$, the formation of drag-increasing, Kelvin-Helmholtz rollers can be captured with a linear stability analysis.
These models provide design guidelines to produce a drag-reducing permeable substrate and give quantitative estimates as to how much drag reduction could be expected. 

Further work is nevertheless required to confirm these findings. Direct numerical simulations fully resolving the microstructure of the permeable substrates need to be conducted in order to set the region of validity of the current models, and to gain full understanding on the effect that these substrates have on the overlying turbulence. \\


GG was supported by an educational grant from Fundaci\'on Bancaria `la Caixa', Amelia Earhart Fellowship and an award from The Cambridge Commonwealth, European and International Trust.
Some simulations were run using the computational resources provided under EPSRC-UK Tier-2 grant EP/P020259/1 by CSD3, Cambridge.

\appendix

\section{Analytic solution of Brinkman's equation}\label{sec:appA}


The flow within the porous medium is approximated using Brinkman's equation~\eqref{eq:Br}, where $K_x$, $K_y$ and $K_z$ are the principal directions of the permeability tensor and are considered to be different. Together with the continuity equation, the system of equations is 

\begin{subequations}\label{eq:Br_ax_az}
	\begin{gather}
		\nu \left( \frac{\partial^2 u}{\partial x^2} + \frac{\partial^2 u}{\partial y^2} + \frac{\partial^2 u}{\partial z^2} \right) - \frac{\nu}{K_x} u - \frac{\partial p}{\partial x}=0, \label{subeq:Br_ax_az_1}\\
		\nu \left( \frac{\partial^2 v}{\partial x^2} + \frac{\partial^2 v}{\partial y^2} + \frac{\partial^2 v}{\partial z^2} \right) - \frac{\nu}{K_y} v - \frac{\partial p}{\partial y}=0, \label{subeq:Br_ax_az_2}\\
		\nu \left( \frac{\partial^2 w}{\partial x^2} + \frac{\partial^2 w}{\partial y^2} + \frac{\partial^2 w}{\partial z^2} \right) - \frac{\nu}{K_z} w - \frac{\partial p}{\partial z}=0, \label{subeq:Br_ax_az_3}\\
		\frac{\partial u}{\partial x} + \frac{\partial v}{\partial y} + \frac{\partial w}{\partial z} =0, \label{subeq:Br_ax_az_4}
	\end{gather}	
\end{subequations}
which can be solved analytically.
Here we restrict ourselves to the permeable substrate at the bottom of the channel, which extends from $y=-h$ to $y=0$ and we neglect the influence of a mean pressure gradient within the substrate, as discussed in \S\ref{subsec:Cf}.

In order to solve equation~\eqref{eq:Br_ax_az}, we reduce this system of partial differential equation (PDE) with three dependent variables into a single equation with a single dependent variable. We start by taking the divergence of the Brinkman equation~\eqref{subeq:Br_ax_az_1}-\eqref{subeq:Br_ax_az_3} and use the continuity equation~\eqref{subeq:Br_ax_az_4} to simplify, which yields

\begin{equation}
		\frac{1}{K_x} \frac{\partial u}{\partial x} + \frac{1}{K_y} \frac{\partial v}{\partial y} + \frac{1}{K_z} \frac{\partial w}{\partial z} + \frac{1}{\nu} \nabla^2 p = 0.
	\label{eq:Br_5} 
\end{equation}

\noindent We then take the $y$-derivative of equation~\eqref{eq:Br_5} and replace $\partial p / \partial y$ from equation~\eqref{subeq:Br_ax_az_2} to eliminate the pressure term. Using continuity again to remove the terms in $w$, the following equation is obtained

\begin{equation}
		\frac{\partial^2 u}{\partial x \partial y} \left( \frac{1}{K_x} - \frac{1}{K_z} \right) - \frac{1}{K_y} \left( \frac{\partial^2 v}{\partial x^2} + \frac{\partial^2 v}{\partial z^2} \right) - \frac{1}{K_z} \frac{\partial^2 v}{\partial y^2} + \nabla^4 v = 0,
	\label{eq:Br_6} 
\end{equation}

\noindent which has terms in $v$ and $u$ alone. 
To remove $u$, we take the $y$-derivative of equation~\eqref{subeq:Br_ax_az_1} and subtract the $x$-derivative of \eqref{subeq:Br_ax_az_2}.
The obtained expression is then differentiated with respect to $x$, yielding

\begin{equation}
		\left(\nabla^2 - \frac{1}{K_x} \right) \frac{\partial^2 u}{\partial x \partial y} - \left(\nabla^2 - \frac{1}{K_y} \right) \frac{\partial^2 v}{\partial x^2} = 0.
	\label{eq:Br_7} 
\end{equation}

\noindent Substituting for $\partial^2 u / \partial x \partial y$ from equation~\eqref{eq:Br_6}, a single equation for $v$ is obtained. This equation can be solved by expanding in Fourier series along $x$ and $z$, so that $v(x,y,z) = \hat{v}(y)\mathrm{e}^{i \alpha_x x} \mathrm{e}^{i \alpha_z z}$, where $\alpha_x$ and $\alpha_z$ are the wavenumbers, $i$ the imaginary unit, $i = \sqrt{-1}$. 
Differentiating in $x$ and $z$ becomes then multiplying by  $i \alpha_x$ and $i \alpha_z$, respectively, leading to the following ordinary differential equation (ODE)

\begin{multline}
		\Bigg\{ D^6 + D^4 \left[ - 3 \alpha^2 - \frac{1}{K_x} -\frac{1}{K_z} \right] + D^2 \bigg[ \frac{1}{K_y} \alpha^2 + \left( 2 \alpha^2 + \frac{1}{K_z} \right) \left( \alpha^2 + \frac{1}{K_x} \right) + \alpha^4 - \\ \alpha_{x}^2 \left( \frac{1}{K_x} - \frac{1}{K_z} \right) \bigg] + \left[ \left( \alpha^2 + \frac{1}{K_y} \right) \left( -\alpha^2 \left( \alpha^2 + \frac{1}{K_x} \right) + \alpha_{x}^2 \left( \frac{1}{K_x} - \frac{1}{K_z} \right) \right) \right] \Bigg\} \hat{v} = 0,
	\label{eq:Br_8} 
\end{multline}
\noindent where $D$ denotes $\partial / \partial y$ and $\alpha^2 = \alpha_{x}^2 + \alpha_{z}^2$. This is a sixth order equation, where all the derivatives are even, 
and the corresponding characteristic equation is a bicubic equation
\begin{equation}
		a_3 r^6 + a_2 r^4 +  a_1 r^2 +  a_0 = 0,
		\label{eq:Br_9} 
\end{equation}
\noindent where
\begin{equation*}
	\left\{ \begin{array}{l}
	\displaystyle
		a_3 = 1, \\[9pt]
	\displaystyle
		a_2 = - 3 \alpha^2 - \frac{1}{K_x} -\frac{1}{K_z},\\[9pt]
	\displaystyle
		a_1 = \frac{1}{K_y} \alpha^2 + \left( 2 \alpha^2 + \frac{1}{K_z} \right) \left( \alpha^2 + \frac{1}{K_x} \right) + \alpha^4 - \alpha_{x}^2 \left( \frac{1}{K_x} - \frac{1}{K_z} \right),\\[9pt]
	\displaystyle
		a_0 = \left( \alpha^2 + \frac{1}{K_y} \right) \left( -\alpha^2 \left( \alpha^2 + \frac{1}{K_x} \right) + \alpha_{x}^2 \left( \frac{1}{K_x} - \frac{1}{K_z} \right) \right).
	\end{array} \right.
\end{equation*}
\noindent This equation can be reduced to a cubic equation and then solved algebraically.
If the discriminant of equation~\eqref{eq:Br_9} is non-zero, i.e. $\Delta = 18 a_3 a_2 a_1 a_0 -4 a_2^3 a_0 + a_2^2 a_1^2 - 4 a_3 a_1 ^3 - 27 a_3^2 a_0^2 \neq 0$, there are 6 different roots. 
The roots of the original equation~\eqref{eq:Br_9} are denoted as $\pm r_1$, $\pm r_2$ and $\pm r_3$ and
the general solution for $\hat{v}$ is then
\begin{equation}
		\hat{v}(y) = A \mathrm{e}^{+r_1 y} + B \mathrm{e}^{-r_1 y} + C \mathrm{e}^{+r_2 y} + D \mathrm{e}^{-r_2 y} + E \mathrm{e}^{+r_3 y} + F\mathrm{e}^{-r_3 y}.
		\label{eq:Br_v}
\end{equation}

\noindent The constants $A$, $B$, $C$, $D$, $E$ and $F$ are determined once the boundary conditions are imposed and are a function of the geometry and the wavenumbers, $\alpha_x$ and $\alpha_z$.
Similar expressions for the pressure and the streamwise and spanwise velocities can be obtained by substitutions into equations~\eqref{subeq:Br_ax_az_2}, \eqref{eq:Br_6}, and the continuity equation \eqref{subeq:Br_ax_az_4}, respectively,
\begin{multline}
		\hat{p}(y) = \nu \bigg[ r_1 \left( A \mathrm{e}^{+r_1 y} - B \mathrm{e}^{-r_1 y} \right) + r_2 \left( C \mathrm{e}^{+r_2 y} - D  \mathrm{e}^{-r_2 y} \right) + r_3 \left( E \mathrm{e}^{+r_3 y} -  F \mathrm{e}^{-r_3 y} \right) \bigg]\\
		 - \nu \left( \alpha^2 + \frac{1}{K_y} \right) \bigg[ \frac{1}{r_1} \left( A \mathrm{e}^{+r_1 y} - B \mathrm{e}^{-r_1 y} \right) + \frac{1}{r_2} \left( C \mathrm{e}^{+r_2 y} - D \mathrm{e}^{-r_2 y} \right) + \\
		 \frac{1}{r_3} \left( E \mathrm{e}^{+r_3 y} - F \mathrm{e}^{-r_3 y} \right) \bigg],
		\label{eq:Br_p}
\end{multline}

\begin{multline}
		\hat{u}(y) = \mathrm{i} \frac{1}{1/K_x - 1/K_z} \frac{\alpha^2}{\alpha_x} \left( \frac{1}{K_y} + \alpha^2 \right) \bigg[ \frac{A}{r_1} \mathrm{e}^{+r_1 y} - \frac{B}{r_1} \mathrm{e}^{-r_1 y} + \frac{C}{r_2} \mathrm{e}^{+r_2 y} - \frac{D}{r_2} \mathrm{e}^{-r_2 y} \\
		+ \frac{E}{r_3} \mathrm{e}^{+r_3 y} - \frac{F}{r_3} \mathrm{e}^{-r_3 y} \bigg] 	- \mathrm{i}  \frac{1}{1/K_x - 1/K_z} \frac{1}{\alpha_x} \left( \frac{1}{K_z} + 2 \alpha^2 \right) \bigg[ A r_1 \mathrm{e}^{+r_1 y} \\
		- B r_1 \mathrm{e}^{-r_1 y} + C r_2 \mathrm{e}^{+r_2 y} - D r_2 \mathrm{e}^{-r_2 y} + E r_3 \mathrm{e}^{+r_3 y} -  F r_3 \mathrm{e}^{-r_3 y} \bigg] \\
		+ i \frac{1}{1/K_x - 1/K_z} \frac{1}{\alpha_x} \bigg[ A r_1^{3} \mathrm{e}^{+r_1 y} - B r_1^{3} \mathrm{e}^{-r_1 y} //
		+ C r_2^{3} \mathrm{e}^{+r_2 y} - D r_2^{3} \mathrm{e}^{-r_2 y} + E r_3^{3} \mathrm{e}^{+r_3 y} -  F r_3^{3} \mathrm{e}^{-r_3 y} \bigg],
		\label{eq:Br_u}
\end{multline}

\begin{equation}
		\hat{w}(y) = - \frac{\alpha_x}{\alpha_z} \hat{u} + \mathrm{i} \frac{1}{\alpha_z} \frac{d \hat{v}}{dy}.
		\label{eq:Br_w}
\end{equation}

To obtain $A$, $B$, $C$, $D$, $E$ and $F$, the boundary conditions need to be considered. The permeable substrate is delimited by an impermeable solid wall at the bottom, where no-slip and impermeability conditions are imposed, and by the free channel flow at the top, where continuity of the normal and tangential stresses holds, together with the continuity of the three velocity components. The boundary conditions at the substrate-channel interface have already been introduced in equation~\eqref{eq:BC_interface}. Expanding these boundary conditions in Fourier space, and assuming $\tilde{\nu} \approx \nu$, the continuity of the normal and tangential stresses at the interface simplifies to the continuity of pressure and wall-normal shear ($d\hat{u}/dy$ and $d\hat{w}/dy$), respectively. Thus, the boundary conditions for the permeable substrates are
	
\begin{subequations}\label{eq:BC_coating}
	\begin{gather}
	    \hat{u} = \hat{w} = \hat{v} = 0  \quad \text{at} \quad y=-h, \quad \text{and} \label{subeq:BC_coating1}\\
		\left. \nu \frac{d \hat{u}}{dy} \right|_{y=0^{-}} = \left. \nu \frac{d \hat{u}}{dy} \right|_{y=0^{+}}, \quad \left. \nu \frac{d \hat{w}}{dy} \right|_{y=0^{-}} = \left. \nu \frac{d \hat{w}}{dy} \right|_{y=0^{+}}, \quad \left. \hat{p} \right|_{y=0^{-}} = \left. \hat{p} \right|_{y=0^{+}} \quad \text{at} \quad y=0, \label{subeq:BC_coating2}
	\end{gather}	
\end{subequations}
\noindent where, at $y=0$, the plus and minus signs correspond to the substrate and fluid sides of the interface, respectively. 


By applying the above boundary conditions to equations~\eqref{eq:Br_v}, ~\eqref{eq:Br_u}, ~\eqref{eq:Br_w} and ~\eqref{eq:Br_p}, and particularising the solution at the substrate-channel interface, the velocities at the interface are
	
\begin{subequations}\label{eq:BC3D_ch}
	\begin{align}
		\left. \hat{u} \right|_{y=0^{-}} = \left. \mathcal{C}_{uu}(\alpha_x,\alpha_z) \frac{d \hat{u}}{dy} \right|_{y=0^{+}} + \left. \mathcal{C}_{uw}(\alpha_x,\alpha_z) \frac{d \hat{w}}{dy} \right|_{y=0^{+}} + \left. \mathcal{C}_{up}(\alpha_x,\alpha_z) \hat{p}\right|_{y=0^{+}},\label{subeq:BC3D_ch1}\\
		\left. \hat{w} \right|_{y=0^{-}} = \left. \mathcal{C}_{wu}(\alpha_x,\alpha_z) \frac{d \hat{u}}{dy} \right|_{y=0^{+}} + \left. \mathcal{C}_{ww}(\alpha_x,\alpha_z) \frac{d \hat{w}}{dy} \right|_{y=0^{+}} + \left. \mathcal{C}_{wp}(\alpha_x,\alpha_z) \hat{p}\right|_{y=0^{+}},\label{subeq:BC3D_ch2}\\
		\left. \hat{v} \right|_{y=0^{-}} = \left. \mathcal{C}_{vu}(\alpha_x,\alpha_z) \frac{d \hat{u}}{dy} \right|_{y=0^{+}} + \left. \mathcal{C}_{vw}(\alpha_x,\alpha_z) \frac{d \hat{w}}{dy} \right|_{y=0^{+}} + \left. \mathcal{C}_{vp}(\alpha_x,\alpha_z) \hat{p}\right|_{y=0^{+}} ,\label{subeq:BC3D_ch3}
	\end{align}	
\end{subequations}

\noindent where the coefficients $\mathcal{C}_{ij}(\alpha_x,\alpha_z)$ are a function of the wavenumbers, $\alpha_x$ and $\alpha_z$, and of the geometry of the substrate, i.e. $K_x$, $K_y$, $K_z$ and $h$. An equivalent analysis can be carried out for the upper permeable substrate. Considering the symmetry properties for each variable, this yields

\begin{subequations}\label{eq:BC3D_ch_top}
	\begin{align}
		\left. \hat{u} \right|_{y=(2 \delta)^{+}} = - \left. \mathcal{C}_{uu} (\alpha_x,\alpha_z) \frac{d \hat{u}}{dy} \right|_{y=(2 \delta)^{-}} - \left. \mathcal{C}_{uw}(\alpha_x,\alpha_z) \frac{d \hat{w}}{dy} \right|_{y=(2 \delta)^{-}} + \left. \mathcal{C}_{up}(\alpha_x,\alpha_z) \hat{p}\right|_{y=(2 \delta)^{-}},\label{subeq:BC3D_ch_top1}\\
		\left. \hat{w} \right|_{y=(2 \delta)^{+}} = - \left. \mathcal{C}_{wu}(\alpha_x,\alpha_z) \frac{d \hat{u}}{dy} \right|_{y=(2 \delta)^{-}} - \left. \mathcal{C}_{ww}(\alpha_x,\alpha_z) \frac{d \hat{w}}{dy} \right|_{y=(2 \delta)^{-}} + \left. \mathcal{C}_{wp}(\alpha_x,\alpha_z) \hat{p}\right|_{y=(2 \delta)^{-}},\label{subeq:BC3D_ch_top2}\\
		\left. \hat{v} \right|_{y=(2 \delta)^{+}} = \left. \mathcal{C}_{vu}(\alpha_x,\alpha_z) \frac{d \hat{u}}{dy} \right|_{y=(2 \delta)^{-}} + \left. \mathcal{C}_{vw}(\alpha_x,\alpha_z) \frac{d \hat{w}}{dy} \right|_{y=(2 \delta)^{-}} - \left. \mathcal{C}_{vp}(\alpha_x,\alpha_z) \hat{p}\right|_{y=(2 \delta)^{-}}. \label{subeq:BC3D_ch_top3}
	\end{align}	
\end{subequations}

\noindent When $\alpha_x = 0$ or $\alpha_z = 0$, Brinkman's equation simplifies and so does its solution. These cases are solved separately in \S\ref{sec:AppA_ax_0}, \ref{sec:AppA_0_az} and \ref{sec:AppA_mean}.



\subsection{Modes $\alpha_x \neq 0$, $\alpha_z=0$} \label{sec:AppA_ax_0}

When $\alpha_z = 0$, the $z$-derivatives become zero and the Brinkman equation for $w$, i.e. equation~\eqref{subeq:Br_ax_az_3}, decouples from the other two. The original system of equations simplifies then to
\begin{subequations}\label{eq:Br_ax_0}
	\begin{gather}
	    \nu \left( \frac{\partial^2 u}{\partial x^2} + \frac{\partial^2 u}{\partial y^2} \right) - \frac{\nu}{K_x} u - \frac{\partial p}{\partial x} = 0, \label{subeq:Br_ax_0_1}\\
	    \nu \left( \frac{\partial^2 v}{\partial x^2} + \frac{\partial^2 v}{\partial y^2} \right) - \frac{\nu}{K_y} v - \frac{\partial p}{\partial y} = 0, \label{subeq:Br_ax_0_2}\\
        \left( \frac{\partial^2 w}{\partial x^2} + \frac{\partial^2 w}{\partial y^2}\right) - \frac{1}{K_z} w = 0, \label{subeq:Br_ax_0_3}\\
        \frac{\partial u}{\partial x} + \frac{\partial v}{\partial y} = 0, \label{subeq:Br_ax_0_4}
    \end{gather}	
\end{subequations}

\noindent The velocities $u$ and $v$ can be solved with a procedure similar to that described above, while $w$ can be solved separately.

Taking the two-dimensional divergence of equations~\eqref{subeq:Br_ax_0_1} and \eqref{subeq:Br_ax_0_2} in the $(x,y)$ plane and using continuity yields
\begin{equation}
		\left( \frac{1}{K_y} - \frac{1}{K_x} \right) \frac{\partial v}{\partial y} + \frac{1}{\nu} \nabla_{xy}^2 p = 0.
	\label{eq:Br_ax_0_5} 
\end{equation}

\noindent Taking the $y$-derivative of equation~\eqref{subeq:Br_ax_0_2} and substituting $\partial v / \partial y$ from \eqref{eq:Br_ax_0_5} yields an equation in $\hat{p}$ alone. 
Taking the Fourier transform in $x$ leads to
\begin{equation}
		\Bigg\{ D^4 + \left[ - 2 \alpha_x^2 - \frac{1}{K_x} \right] D^2 + \alpha_x^2 \left[ \alpha_x^2 + \frac{1}{K_y} \right] \Bigg\} \hat{p} = 0.
	\label{eq:Br_ax_0_6} 
\end{equation}

\noindent The corresponding characteristic equation is biquadratic,
\begin{equation}
		m^4 + m^2 \left( -2 \alpha^2 - \frac{1}{K_x} \right) + \left( \alpha^4 + \frac{\alpha^2}{K_y} \right) = 0.
\end{equation}
	
\noindent Rewriting it as a second order equation, the roots of the characteristic equation are 
\begin{gather*}
		m_{1} = - m_{2} = \sqrt{\frac{2 \alpha^2 K_x + 1 + \sqrt{4 \alpha^2 K_x \left( 1-\frac{K_x}{K_y} \right) +1} }{2 K_x}},\\
		m_{3} = - m_{4} = \sqrt{\frac{2 \alpha^2 K_x + 1 - \sqrt{4 \alpha^2 K_x \left( 1-\frac{K_x}{K_y} \right) +1} }{2 K_x}}.\\
\end{gather*}

Except for the case in which $m_1 = m_3$, i.e. $\frac{K_x}{K_y} = \frac{1 + 4 \alpha^2 K_x}{4 \alpha^2 K_x}$, the expression for $\hat{p}$ becomes:

\begin{equation}
	\hat{p}(y) = A' \mathrm{e}^{m_1 y} + B' \mathrm{e}^{m_2 y} + C' \mathrm{e}^{m_3 y} + D' \mathrm{e}^{m_4 y},
	\label{eq:Br_ax_0_7}
\end{equation}

\noindent where $A'$, $B'$, $C'$ and $D'$ depend on the wavenumber $\alpha_x$ and the geometrical properties of the permeable medium, and are determined by imposing the boundary conditions -- impermeability and no slip conditions at $y=-h$, and continuity of pressure and $d \hat{u}/ dy$ at $y = 0$.
The general solutions for $\hat{v}$ and $\hat{u}$ can be obtained from equations~\eqref{eq:Br_ax_0_5} and \eqref{subeq:Br_ax_0_4}, respectively. 

In contrast, solving equation~\eqref{subeq:Br_ax_0_3} for $\hat{w}$ is straightforward. Expanding it in Fourier series gives
\begin{equation}
		\frac{\partial^2 \hat{w} }{\partial y^2} - \left( \alpha_x^2 + \frac{1}{K_z} \right) \hat{w} = 0,
	\label{eq:Br_ax_0_w1} 
\end{equation}

\noindent whose solution is
\begin{equation}
		\hat{w} = E'_{x0} \mathrm{e}^{y/ L_w} + F'_{x0} \mathrm{e}^{-y/ L_w},
	\label{eq:Br_ax_0_w1} 
\end{equation}
\noindent where $L_w = 1 / \sqrt{\alpha_x^2 + 1/K_z}$.
Applying now the boundary conditions for $\hat{w}$, $\hat{w}=0$ at the impermeable wall and continuity of $d \hat{w}/ dy$ at the interface, leads to
\begin{equation}
		\hat{w} = L_w \frac{\mathrm{e}^{ \left( y + h \right)/L_w } - \mathrm{e}^{- \left( y + h \right) / L_w }}{\mathrm{e}^{ h / L_w} + \mathrm{e}^{- h / L_w}} \left. \frac{d \hat{w}}{dy} \right|_{y=0^+}.
	\label{eq:Br_ax_0_w1} 
\end{equation}

\noindent Comparing to the expressions presented in \eqref{eq:BC3D_ch}, $\mathcal{C}_{ww}(\alpha_x,0)$ is the proportionality term in equation~\eqref{eq:Br_ax_0_w1} between $\hat{w}$ and its gradient, whereas $\mathcal{C}_{wu}(\alpha_x,0) = \mathcal{C}_{wp}(\alpha_x,0)= 0$. Also, from the general solutions for $\hat{u}$ and $\hat{w}$, we observe that $\mathcal{C}_{uw}(\alpha_x,0) = \mathcal{C}_{vw}(\alpha_x,0)= 0$, which was expected, as there is no coupling between $\hat{w}$, and the other two velocities, $\hat{u}$ and $\hat{v}$, for modes $(\alpha_x, 0)$.
Hence, in this case the 9 coefficients presented for the general interface conditions~\eqref{eq:BC3D_ch} are reduced to only 5.

\subsection{Modes $\alpha_x = 0$, $\alpha_z \neq 0$} \label{sec:AppA_0_az}

In this case, Brinkman's equation for $u$ decouples from the other two. 
For cases with the same permeability in $y$ and $z$ directions, $K_z = K_y$, the solution simplifies even more, to
\begin{subequations}\label{eq:Br_0_az}
	\begin{gather}
	    \left( \frac{\partial^2 u}{\partial y^2} + \frac{\partial^2 u}{\partial z^2}\right) - \frac{1}{K_x} u = 0, \label{subeq:Br_0_az1}\\
	    \nu \left( \frac{\partial^2 v}{\partial y^2} + \frac{\partial^2 v}{\partial z^2} \right) - \frac{\nu}{K_y} v - \frac{\partial p}{\partial y} = 0, \label{subeq:Br_0_az2}\\
	    \nu \left( \frac{\partial^2 w}{\partial y^2} + \frac{\partial^2 w}{\partial z^2} \right) - \frac{\nu}{K_y} w - \frac{\partial p}{\partial z} = 0, \label{subeq:Br_0_az3}\\
	    \frac{\partial v}{\partial y} + \frac{\partial w}{\partial z} = 0, \label{subeq:Br_0_az4}
	\end{gather}	
\end{subequations}

Taking the two-dimensional divergence of equations~\eqref{subeq:Br_0_az2} and ~\eqref{subeq:Br_0_az3} in the $(y,z)$ plane leads to a Laplace equation for the pressure. We then take the Fourier transform with respect to $z$ (i.e. $p(y,z) = \hat{p}(y)\mathrm{e}^{i \alpha_z z}$) to get

\begin{equation}
	\hat{p}(y) = A''_{0z} \mathrm{e}^{\alpha_z y} + B''_{0z} \mathrm{e}^{-\alpha_z y},
	\label{eq:Br_0_az_1}
\end{equation}

\noindent The general expressions for $\hat{v}$ and $\hat{w}$ can then be derived from the equations~\eqref{subeq:Br_0_az2} and ~\eqref{subeq:Br_0_az4}, respectively.

The streamwise velocity is solved similarly to $w$ in \S\ref{sec:AppA_ax_0}. We take the Fourier transform of equation~\eqref{subeq:Br_0_az1} in $z$, which gives
\begin{equation}
    \frac{\partial^2 \hat{u} }{\partial y^2} - \left( \alpha_z^2 + \frac{1}{K_x} \right) \hat{u} = 0.
    \label{eq:Br_0_az_u0}
\end{equation}

\noindent The solution, after applying the boundary conditions for $\hat{u}$, is

\begin{equation}
		\hat{u} = L_u \frac{\mathrm{e}^{\left( y + h \right) / L_u} - \mathrm{e}^{- \left( y + h \right)/ L_u}}{\mathrm{e}^{ h / L_u} + \mathrm{e}^{- h / L_u}} \left. \frac{d \hat{u}}{dy} \right|_{y=0^+},
	\label{eq:Br_ax_0_u1} 
\end{equation}

\noindent where $L_u = 1 / \sqrt{\alpha_z^2 + 1 / K_x}$. The proportionality coefficient relating $\hat{u}$ with its gradient is the coefficient $\mathcal{C}_{uu}(0,\alpha_z)$, i.e.

\begin{equation}
		\mathcal{C}_{uu}(0,\alpha_z) = L_u \frac{\mathrm{e}^{ \left( y + h \right) / L_u} - \mathrm{e}^{- \left( y + h \right) / L_u}}{\mathrm{e}^{ h / L_u} + \mathrm{e}^{- h / L_u}},
	\label{eq:Br_ax_0_u3} 
\end{equation}

\noindent and $\mathcal{C}_{up}(0,\alpha_z) = \mathcal{C}_{uw}(0,\alpha_z) = \mathcal{C}_{wu}(0,\alpha_z) = \mathcal{C}_{vu}(0,\alpha_z) = 0$.

\subsection{Mode $\alpha_x=0$, $\alpha_z=0$}\label{sec:AppA_mean}

Although the coefficients for the mean can be directly obtained from the expressions derived in \S\ref{sec:AppA_ax_0} and \S\ref{sec:AppA_0_az}, this case deserves further discussion. When $\alpha_x = \alpha_z =0$, the equations for the three velocities $u$, $v$ and $w$ decouple from each other and the velocities for mode zero become

\begin{subequations}\label{eq:Br_0_0}
	\begin{gather}	
		\hat{u} = \sqrt{K_x} \frac{\mathrm{e}^{ \left( y + h \right) / \sqrt{K_x}} - \mathrm{e}^{- \left( y + h \right) / \sqrt{K_x}}}{\mathrm{e}^{h / \sqrt{K_x}} + \mathrm{e}^{- h / \sqrt{K_x}}} \left. \frac{d \hat{u}}{dy} \right|_{y=0^+}, \label{subeq:Br_0_0_1} \\
		\hat{w} = \sqrt{K_z} \frac{\mathrm{e}^{\left( y + h \right) / \sqrt{K_z}} - \mathrm{e}^{- \left( y + h \right) / \sqrt{K_z}}}{\mathrm{e}^{ h / \sqrt{K_z}} + \mathrm{e}^{- h / \sqrt{K_z}}} \left. \frac{d \hat{w}}{dy} \right|_{y=0^+}, \label{subeq:Br_0_0_2} \\
		\hat{v} = 0. \label{subeq:Br_0_0_3}
	\end{gather}	
\end{subequations}

\noindent Equations~\eqref{subeq:Br_0_0_1} and \eqref{subeq:Br_0_0_2} are obtained from particularising equations~\eqref{eq:Br_ax_0_w1} and \eqref{eq:Br_ax_0_u1} for $\alpha_z = 0$ and $\alpha_x = 0$, respectively, while equation~\eqref{subeq:Br_0_0_2} is obtained from continuity, after applying the boundary condition that $\hat{v} = 0$ at $y= 0$. Particularising at $y=0$ and comparing to the general boundary conditions introduced in equation~\eqref{eq:BC3D_ch}, we have

\begin{subequations}\label{eq:Br_0_1}
	\begin{gather}	
		\left. \hat{u} \right|_{y=0} = \sqrt{K_x} \tanh \left( \frac{h}{\sqrt{K_x}}  \right) \left. \frac{d \hat{u}}{dy} \right|_{y=0^+} = \mathcal{C}_{uu} (0,0) \left. \frac{d \hat{u}}{dy} \right|_{y=0^+}, \label{subeq:Br_0_1_1} \\
		\left. \hat{w} \right|_{y=0} = \sqrt{K_z} \tanh \left( \frac{h}{\sqrt{K_z}}  \right) \left. \frac{d \hat{w}}{dy} \right|_{y=0^+} = \mathcal{C}_{ww} (0,0) \left. \frac{d \hat{w}}{dy} \right|_{y=0^+}, \label{subeq:Br_0_1_2} \\
		\left. \hat{v} \right|_{y=0} = 0, \label{subeq:Br_0_0_3}
	\end{gather}	
\end{subequations} 

\noindent where all the coefficients in equation~\eqref{eq:BC3D_ch} are zero except for $\mathcal{C}_{uu}$ and $\mathcal{C}_{ww}$, which relate the tangential velocities to their wall-normal gradient. These are the mean slip lengths $\ell_x^+$ and $\ell_z^+$ derived by \cite{Abderrahaman2017}.

\section{Turbulence statistics for permeable substrates}\label{sec:appB}

\begin{figure}
    \centering%
	\includegraphics{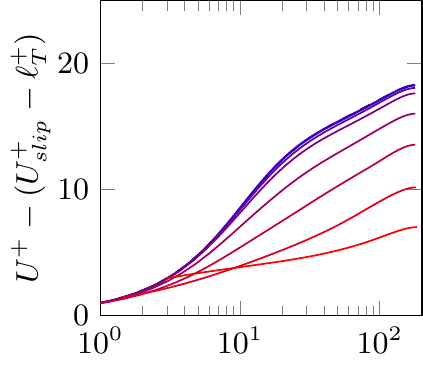}%
    \mylab{-3.2cm}{3.25cm}{(A.\aaa)}%
    \hspace{0.2cm}%
    \includegraphics{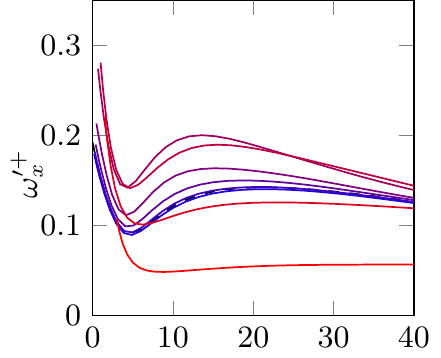}%
    \mylab{-3.45cm}{3.25cm}{(A.\bbb)}%
    \includegraphics{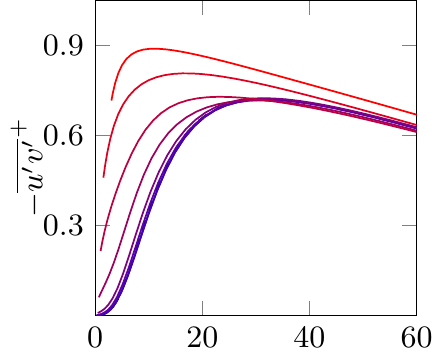}%
    \mylab{-3.45cm}{3.25cm}{(A.\ccc)}%
    
    \includegraphics{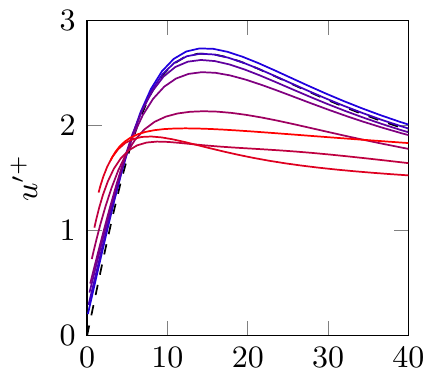}%
    \mylab{-3.45cm}{3.25cm}{(A.\ddd)}%
    \includegraphics{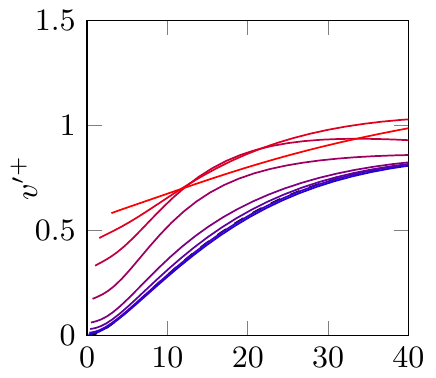}%
    \mylab{-3.45cm}{3.25cm}{(A.\eee)}%
    \includegraphics{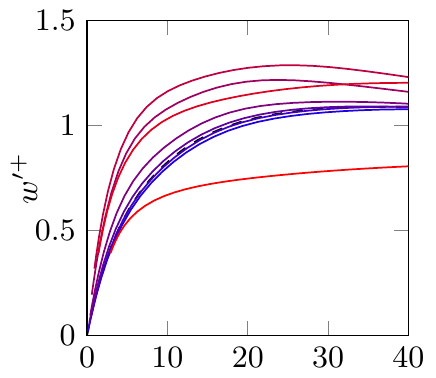}%
    \mylab{-3.45cm}{3.25cm}{(A.\fff)}%
    
	\centering%
	\vspace{0.7cm}
    \includegraphics{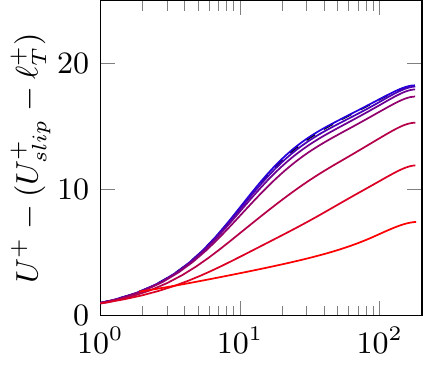}%
    \mylab{-3.2cm}{3.25cm}{(B.\aaa)}%
    \hspace{0.2cm}%
    \includegraphics{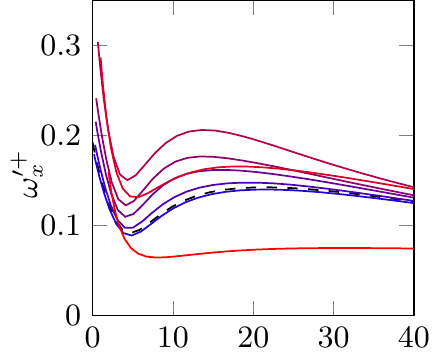}%
    \mylab{-3.45cm}{3.25cm}{(B.\bbb)}%
    \includegraphics{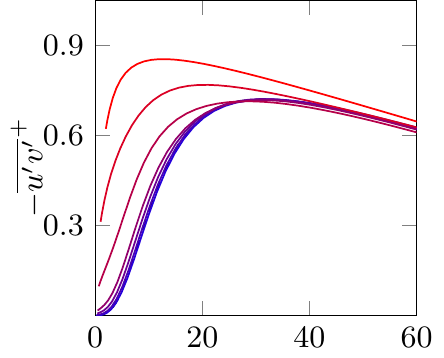}%
    \mylab{-3.45cm}{3.25cm}{(B.\ccc)}%
    
    \includegraphics{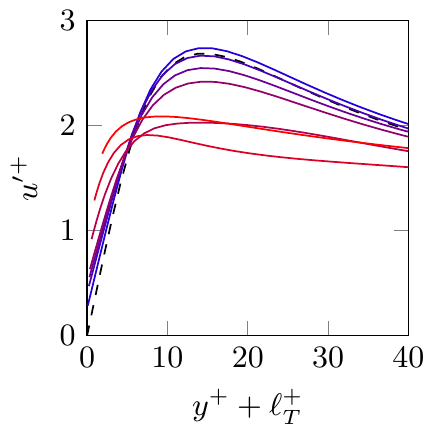}%
    \mylab{-3.45cm}{3.85cm}{(B.\ddd)}%
    \includegraphics{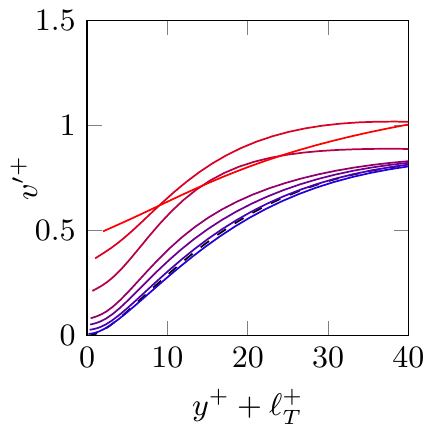}%
    \mylab{-3.45cm}{3.85cm}{(B.\eee)}%
    \includegraphics{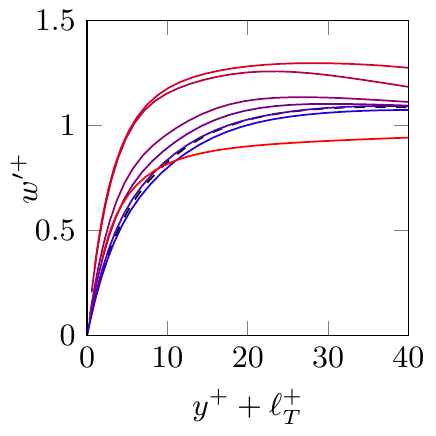}%
    \mylab{-3.45cm}{3.85cm}{(B.\fff)}%
\caption{One-point turbulent statistics for (A.\aaa -A.\fff) a substrate configuration with $\phi_{xy} \approx 3.6$, which corresponds to cases A1-A8; (B.\aaa -B.\fff) a substrate configuration with $\phi_{xy} \approx 5.5$, which corresponds to cases B1-B7. Permeability values increase from blue to red and profiles are scaled with the corresponding $u_{\tau}$ at $y = -\ell_T =- \sqrt{K_z}$, the linearly extrapolated virtual origin for turbulence. Black-dashed lines represent the smooth-channel case. (A.\aaa, B.\aaa) Mean velocity profiles shifted by $\ell_T^+$ and where the value at the origin, i.e. the offset predicted from the linear theory, $\Delta U^+ = U_{slip}^+ - \ell_T^+$, has been subtracted. Rms fluctuations of (A.\bbb, B.\bbb) the streamwise velocity, (A.\ccc, B.\ccc) the wall-normal velocity, (A.\ddd, B.\ddd) the spanwise velocity, and (A.\eee, B.\eee) the streamwise vorticity. (A.\fff, B.\fff) Reynolds stress.}\label{fig:KxKy_stats}
\end{figure}

In \S\ref{sec:results_DNS} results for only the permeable substrates with $\phi_{xy} \approx 11.4$ are discussed. In this appendix, the flow statistics for the other two substrate configurations are presented.
The mean velocity profiles and the turbulence fluctuations for configurations with $\phi_{xy} \approx 5.5$ and $\phi_{xy} \approx 3.6$ are compiled in figure~\ref{fig:KxKy_stats}.

   
\newpage      
\bibliography{porosity}
\bibliographystyle{jfm}

\end{document}